\documentclass[12pt]{article}
\usepackage{amsfonts}
\usepackage{amssymb}
\usepackage{amsthm}
\usepackage{amsmath}
\usepackage{bm}
\begin{document}

\renewcommand{\theequation}{\thesection.\arabic{equation}}

\title{The multi-dimensional Hamiltonian Structures in the Whitham 
method.}

\author{A.Ya. Maltsev.}

\date{
\centerline{\it L.D. Landau Institute for Theoretical Physics}
\centerline{\it 142432 Chernogolovka, pr. Ak. Semenova 1A, 
maltsev@itp.ac.ru}}

\maketitle

\begin{abstract}
 We consider the averaging of local field-theoretic 
Poisson brackets in the multi-dimensional case. As a result, we 
construct a local Poisson bracket for the regular Whitham system in the
multidimensional situation. The procedure is based on the
procedure of averaging of local conservation laws and follows 
the Dubrovin - Novikov scheme of the bracket averaging suggested
in one-dimensional case. However, the features of the phase space 
of modulated parameters in higher dimensions lead to a different 
natural class of the averaged brackets in comparison with the
one-dimensional situation. Here we suggest a direct procedure
of construction of the bracket for the Whitham system for $d > 1$
and discuss the conditions of applicability of the corresponding
scheme. At the end, we discuss canonical forms of the
averaged Poisson bracket in the  multidimensional case. 
\end{abstract}

\section{Introduction.}

In this paper we consider special
features of the Hamiltonian formulation of 
the Whitham method (\cite{whith1,whith2,whith3}) in the multi-dimensional
situation. Thus, we will consider here slow modulations of periodic or 
quasiperiodic $m$-phase solutions of nonlinear systems
\begin{equation}
\label{insyst}
F^{i}(\bm{\varphi}, \bm{\varphi}_{t}, \bm{\varphi}_{x^{1}}, 
\dots, \bm{\varphi}_{x^{d}}, \dots )
\,\, = \,\, 0
\,\,\,\,\,\,\,\, , \,\,\,\,\, i = 1, \dots, n \,\,\, , \,\,\,
\bm{\varphi} = (\varphi^{1}, \dots, \varphi^{n})
\end{equation}
with $d$ spatial dimensions.

 The corresponding $m$-phase solutions will be represented here
in the form
\begin{equation}
\label{phasesol}
\varphi^{i} ({\bf x}, t) \,\, = \,\, \Phi^{i} \left( 
{\bf k}_{1}({\bf U})\, x^{1} \, + \, \dots \, + \,
{\bf k}_{d}({\bf U})\, x^{d} 
\, + \, \bm{\omega}({\bf U})\, t \, + \, \bm{\theta}_{0}, \,
{\bf U} \right)
\end{equation}
where the
functions ${\bf k}_{q} ({\bf U})$ and $\bm{\omega}({\bf U})$
play the role of the ``wave numbers'' and
``frequencies'' of $m$-phase solutions, while the parameters
$\bm{\theta}_{0}$ represent the ``initial phase shifts''.
We are going to consider here systems with $d$ spatial
variables $(x^{1}, \dots, x^{d})$ and one time variable $t$. 
The parameters ${\bf U} = (U^{1}, \dots, U^{N})$ 
can be chosen in an arbitrary way, we just assume that they do not
change under shifts of the initial phases of solutions
$\bm{\theta}_{0}$. We can write 
${\bf k}_{p}({\bf U}) = (k^{1}_{p}({\bf U}), \dots, k^{m}_{p}({\bf U}))$,
$\bm{\omega}({\bf U}) = (\omega^{1}({\bf U}), \dots,
\omega^{m}({\bf U}))$ for the wave numbers and the frequencies
of $m$-phase solutions of (\ref{insyst}).

 The functions $\Phi^{i}(\bm{\theta})$ are supposed to be 
$2\pi$-periodic with respect to each $\theta^{\alpha}$, 
$\alpha = 1, \dots, m$, and satisfy the system
\begin{equation}
\label{PhaseSyst0}
F^{i} \left( {\bf \Phi}, \, \omega^{\alpha}
{\bf \Phi}_{\theta^{\alpha}}, \,
k^{\beta_{1}}_{1} {\bf \Phi}_{\theta^{\beta_{1}}}, \dots,
k^{\beta_{d}}_{d} {\bf \Phi}_{\theta^{\beta_{d}}},
\dots \right) \,\,\, \equiv
\,\,\, 0
\,\,\,\,\,\,\,\, , \,\,\,\,\, i = 1, \dots, n
\end{equation}

 The functions ${\bf \Phi}(\bm{\theta}, {\bf U})$, having 
``zero initial phase shift'', can in fact be chosen in an arbitrary
(smooth) way for each value of ${\bf U}$, such that the corresponding set 
of $m$-phase solutions of (\ref{insyst}) can be represented in the 
form (\ref{phasesol}).  

 Consider now a set $\Lambda$ of functions
$\bm{\Phi}(\bm{\theta} + \bm{\theta}_{0}, {\bf U})$,
depending smoothly on the parameters ${\bf U}$ and satisfying  
system  (\ref{PhaseSyst0}) for all ${\bf U}$. In the Whitham approach the 
parameters ${\bf U}$ and $\bm{\theta}_{0}$
become slowly varying functions of ${\bf x}$ and $t$:
${\bf U} = {\bf U}({\bf X}, T)$, 
$\bm{\theta}_{0} = \bm{\theta}_{0}({\bf X}, T)$,
where ${\bf x} = (x^{1}, \dots, x^{d})$, \linebreak
${\bf X} = (X^{1}, \dots, X^{d})$,
$X^{q} = \epsilon x^{q}$, $T = \epsilon t$
($\epsilon \rightarrow 0$).

In the simplest case
(see \cite{luke}) we try to construct asymptotic solutions
of system (\ref{insyst}) in the form
\begin{equation}
\label{whithsol}
\varphi^{i}(\bm{\theta}, {\bf X}, T, \epsilon) 
\,\,\, = \,\,\, \sum_{k\geq0}
\Psi^{i}_{(k)} \left( {{\bf S}({\bf X},T) \over \epsilon} +
\bm{\theta}, \, {\bf X}, \, T \right) \,\, \epsilon^{k}   
\end{equation}
with $2\pi$-periodic in $\bm{\theta}$ functions
$\bm{\Psi}_{(k)}$.  The function
${\bf S}({\bf X},T) = (S^{1}({\bf X},T), \dots, S^{m}({\bf X},T))$
is called the ``modulated phase'' of solution
(\ref{whithsol}). We have to put now
$$F^{i} \left( \bm{\varphi}, \epsilon \, \bm{\varphi}_{T},
\epsilon \, \bm{\varphi}_{X^{1}}, \dots, 
\epsilon \, \bm{\varphi}_{X^{d}}, 
\dots \right) \,\,\, = \,\,\, 0
\,\,\,\,\,\,\,\, , \,\,\,\,\, i = 1, \dots, n $$
and try to find a sequence of all terms of (\ref{whithsol}).

 The main term in the series (\ref{whithsol}) is given then by the
modulated $m$-phase solution of system (\ref{insyst}) and
for the construction of the corresponding asymptotic solution
the functions ${\bf U}({\bf X}, T)$ must satisfy some system of
differential equations (the Whitham system). Indeed, assume now that 
the function
$\bm{\Psi}_{(0)}(\bm{\theta}, {\bf X}, T)$ 
belongs to the family $\Lambda$
of $m$-phase solutions of (\ref{insyst})
for all ${\bf X}$ and $T$. We have then
\begin{equation}
\label{psi0}
\bm{\Psi}_{(0)} (\bm{\theta}, {\bf X}, T) \,\,\, = \,\,\,   
\bm{\Phi} \left( \bm{\theta} + \bm{\theta}_{0}({\bf X}, T), 
{\bf U}({\bf X} ,T) \right)
\end{equation}
and
$$S^{\alpha}_{T}({\bf X},T) \, = \, 
\omega^{\alpha}({\bf U}({\bf X}, T)) \,\,\, ,
\,\,\,\,\, S^{\alpha}_{X^{p}}({\bf X},T) \, = \, 
k^{\alpha}_{p} ({\bf U}({\bf X}, T)) $$
as follows after the substitution of (\ref{whithsol}) into system
(\ref{insyst}).

 In the simplest case the functions 
$\bm{\Psi}_{(k)} (\bm{\theta}, {\bf X}, T)$ 
are determined from the linear systems
\begin{equation}
\label{ksyst}
{\hat L}^{i}_{j[{\bf U}, \bm{\theta}_{0}]}({\bf X},T) \,\,
\Psi_{(k)}^{j} (\bm{\theta}, {\bf X}, T) \,\,\, = \,\,\,
f_{(k)}^{i} (\bm{\theta}, {\bf X}, T)
\end{equation}
where ${\hat L}^{i}_{j[{\bf U}, \bm{\theta}_{0}]}({\bf X}, T)$
is a linear operator defined by the linearization of system
(\ref{PhaseSyst0}) on the solution (\ref{psi0}). The solubility
conditions of systems (\ref{ksyst}) in the space of periodic functions
can be written as the conditions of orthogonality of the functions
${\bf f}_{(k)} (\bm{\theta}, {\bf X}, T)$ to all the ``left eigenvectors''
(the eigenvectors of the adjoint operator) of the operator
${\hat L}^{i}_{j[{\bf U}, \bm{\theta}_{0}]}({\bf X}, T)$ corresponding
to zero eigenvalue.

 We should say, however, that the solubility conditions of 
systems (\ref{ksyst}) can actually be quite complicated in general
multi-phase case, since the eigenspaces of the operators
${\hat L}_{[{\bf U}, \bm{\theta}_{0}]}$
and ${\hat L}^{\dagger}_{[{\bf U}, \bm{\theta}_{0}]}$
on the space of $2\pi$-periodic functions can be rather nontrivial
in the multi-phase situation. As a result, the determination of the 
next corrections from systems (\ref{ksyst}) is impossible in general 
multiphase situation and the corrections to the main approximation 
(\ref{psi0}) have more complicated and rather nontrivial form 
(\cite{dobr1,dobr2,DobrKrichever}).

 These difficulties do not arise commonly in the single-phase
situation ($m = 1$) where the behavior of eigenvectors of
${\hat L}_{[{\bf U}, \bm{\theta}_{0}]}$
and ${\hat L}^{\dagger}_{[{\bf U}, \bm{\theta}_{0}]}$, as a rule,
is quite regular. The solubility conditions of system 
(\ref{ksyst}) for $k = 1$
\begin{equation}
\label{1syst}
{\hat L}^{i}_{j[{\bf U}, \bm{\theta}_{0}]}({\bf X}, T) \,\,
\Psi_{(1)}^{j} (\bm{\theta}, {\bf X}, T) \,\,\, = \,\,\,
f_{(1)}^{i} (\bm{\theta}, {\bf X}, T)
\end{equation}
with the relations
$$k_{p T} \, = \, \omega_{X^{p}} \,\,\,\,\, ,
\,\,\,\,\, k_{p X^{l}} \, = \, k_{l X^{p}} 
\,\,\,\,\,\,\,\, , \,\,\,\,\,\,\,\,\,\, p, l = 1, \dots, d $$ 
define in this case the Whitham system 
for the single-phase solutions of (\ref{insyst}) which plays the
central role in considering the slow modulations.

 For the multi-phase solutions the Whitham system is usually defined
by the orthogonality conditions of the right-hand part of (\ref{1syst})
to the maximal set of ``regular'' left eigenvectors corresponding to
zero eigenvalues which are defined for all values of ${\bf U}$
and depend smoothly on ${\bf U}$. The construction of asymptotic series
(\ref{whithsol}) in the multi-phase case is impossible in general
situation (see \cite{dobr1,dobr2,DobrKrichever}). 
Nevertheless, the ``regular'' Whitham system defined in the way
described above and the leading term of the expansion
(\ref{whithsol}) play the major role in consideration of modulated
solutions also in this case, representing the main approximation for
the corresponding modulated solutions. The corrections to the main
term have in general more complicated form than (\ref{whithsol}),
but they also tend to zero in the limit
$\epsilon \rightarrow 0$ (\cite{dobr1,dobr2,DobrKrichever}).

 Some incomplete list of the classical papers devoted to the 
foundations of the Whitham method can be given by
\cite{AblBenny, dm, DobrMaslMFA, dobr1, dobr2, DobrKrichever,
dn1, dn2, dn3, ffm, Hayes, krichev1, luke, Newell, theorsol, Nov, 
whith1, whith2, whith3}. 
We will be interested here only in Hamiltonian aspects of 
the multi-dimensional Whitham method. In many aspects the article 
uses the technique and methods developed in the paper 
\cite{DNMultDim}, applied to the multi-dimensional case. However, as we 
will see, the features of the phase space of modulated parameters in 
higher dimensions lead to a different natural class of the averaged 
brackets in comparison with the one-dimensional situation.

 Let us use for simplicity the notation $\Lambda$ both for the family 
of the functions $\bm{\Phi} (\bm{\theta} + \bm{\theta}_{0}, {\bf U})$
and the corresponding family of $m$-phase solutions of system
(\ref{insyst}), such that we will denote by $\Lambda$ both the 
parameter-dependent families of the 
$2\pi$-periodic in all $\theta^{\alpha}$ functions 
$\bm{\Phi} (\bm{\theta} + \bm{\theta}_{0}, {\bf U})$ and
$\bm{\varphi}_{[{\bf U}, \bm{\theta}_{0}]} ({\bf x}) = \bm{\Phi}
({\bf k}_{q} ({\bf U}) \, x^{q} + \bm{\theta}_{0}, {\bf U})$. 
We will assume
everywhere below that the family $\Lambda$ represents a smooth family
of $m$-phase solutions of system (\ref{insyst}) in the sense  
discussed above.

 It is generally assumed that the parameters $k^{\alpha}_{p}$,
$\omega^{\alpha}$ are independent on the family
$\Lambda$, such that the full family of the $m$-phase solutions
of (\ref{insyst}) depends on $N = (d + 1) m + s$, ($s \geq 0$)
parameters $U^{\nu}$ and $m$ initials phase shifts
$\theta^{\alpha}_{0}$. In this case it is convenient to represent
the parameters ${\bf U}$ in the form
${\bf U} = ({\bf k}_{1}, \dots, {\bf k}_{d}, \bm{\omega}, {\bf n})$, 
where ${\bf k}_{p}$
represent the wave numbers, $\bm{\omega}$ - the frequencies
of the $m$-phase solutions  and ${\bf n} = (n^{1}, \dots, n^{s})$ -
some additional parameters (if any).

 It is easy to see that the functions
$\bm{\Phi}_{\theta^{\alpha}} (\bm{\theta} + \bm{\theta}_{0},
{\bf k}_{1}, \dots, {\bf k}_{d},
\bm{\omega}, {\bf n})$, $\alpha = 1, \dots, m$, \linebreak
$\bm{\Phi}_{n^{l}} (\bm{\theta} + \bm{\theta}_{0},
{\bf k}_{1}, \dots, {\bf k}_{d}, 
\bm{\omega}, {\bf n})$, $l = 1, \dots, s$,
belong to the kernel of the operator
${\hat L}^{i}_{j[{\bf k}_{1}, \dots, {\bf k}_{d},
\bm{\omega},{\bf n},\bm{\theta}_{0}]}$.
In the regular case we assume that the set
of the functions ($\bm{\Phi}_{\theta^{\alpha}}$, $\bm{\Phi}_{n^{l}}$)
represents the maximal linearly independent set of the kernel
vectors of the operator ${\hat L}$, regularly depending on
the parameters 
$({\bf k}_{1}, \dots, {\bf k}_{d}, \bm{\omega}, {\bf n})$. 
For the construction
of the ``regular'' Whitham system we have to require the following
property of regularity and completeness of the family of $m$-phase
solutions of system (\ref{insyst}):

\vspace{0.2cm}

{\bf Definition 1.1.}

{\it We call a family $\Lambda$ a complete regular family of
$m$-phase solutions of system (\ref{insyst}) if:

1) The values ${\bf k}_{p} = (k^{1}_{p}, \dots, k^{m}_{p})$,
$\bm{\omega} = (\omega^{1}, \dots, \omega^{m})$
represent independent parameters on the family $\Lambda$, 
such that the total set of parameters of the $m$-phase solutions
can be represented in the form
$({\bf U}, \bm{\theta}_{0}) =
({\bf k}_{1}, \dots, {\bf k}_{d},  
\bm{\omega}, {\bf n}, \bm{\theta}_{0})$;

2) The functions
$\bm{\Phi}_{\theta^{\alpha}} (\bm{\theta} + \bm{\theta}_{0},
{\bf k}_{1}, \dots, {\bf k}_{d},
\bm{\omega}, {\bf n})$,
$\bm{\Phi}_{n^{l}} (\bm{\theta} + \bm{\theta}_{0},
{\bf k}_{1}, \dots, {\bf k}_{d},
\bm{\omega}, {\bf n})$
are linearly independent and give the maximal linearly independent
set among the kernel vectors of the operator
${\hat L}^{i}_{j[{\bf k}_{1}, \dots, {\bf k}_{d},
\bm{\omega},{\bf n},\bm{\theta}_{0}]}$,
smoothly depending on the parameters
$({\bf k}_{1}, \dots, {\bf k}_{d}, \bm{\omega}, {\bf n})$ 
on the whole set of parameters;

3) The operator
${\hat L}^{i}_{j[{\bf k}_{1}, \dots, {\bf k}_{d},
\bm{\omega},{\bf n},\bm{\theta}_{0}]}$
has exactly $m + s$ linearly independent left eigenvectors
with zero eigenvalue
$$\bm{\kappa}^{(q)}_{[{\bf U}]} (\bm{\theta} + \bm{\theta}_{0})
\,\,\, = \,\,\,
\bm{\kappa}^{(q)}_{[{\bf k}_{1}, \dots, {\bf k}_{d},
\bm{\omega}, {\bf n}]}
(\bm{\theta} + \bm{\theta}_{0})
\,\,\,\,\,\,\,\,\,\, , \,\,\,\,\,\,\,\,\,\,
q = 1, \, \dots , \, m + s $$ 
among the vectors smoothly depending  on the parameters
$({\bf k}_{1}, \dots, {\bf k}_{d}, \bm{\omega}, {\bf n})$ 
on the whole set of parameters.
}

\vspace{0.2cm}

 By definition, we will call the regular Whitham system for a
complete regular family of $m$-phase solutions of (\ref{insyst})
the conditions of orthogonality of the discrepancy
${\bf f}_{(1)}(\bm{\theta}, {\bf X}, T)$ to the functions
$\bm{\kappa}^{(q)}_{[{\bf U}({\bf X},T)]}
(\bm{\theta}\, + \, \bm{\theta}_{0}({\bf X},T))$
\begin{equation}
\label{ortcond}
\int_{0}^{2\pi}\!\!\!\!\!\dots\int_{0}^{2\pi}
\kappa^{(q)}_{[{\bf U}({\bf X},T)]\, i}
(\bm{\theta}\, + \, \bm{\theta}_{0}({\bf X},T)) \,\,
f^{i}_{(1)} (\bm{\theta},{\bf X},T) \,\,
{d^{m} \theta \over (2\pi)^{m}} \,\,\, = \,\,\, 0
\,\,\,\,\,\,\,\,\,\, , \,\,\,\,\,\,\,\,\,\,
q = 1, \, \dots , \, m + s
\end{equation}
with the compatibility conditions
\begin{equation}
\label{comcond}
k^{\alpha}_{p T} \, = \, \omega^{\alpha}_{X^{p}} \,\,\,\,\, ,
\,\,\,\,\, k^{\alpha}_{p X^{l}} \, = \, k^{\alpha}_{l X^{p}}
\,\,\,\,\,\,\,\, , \,\,\,\,\,\,\,\, \alpha = 1, \dots, m
\,\,\, , \,\,\,\,\, p, l = 1, \dots, d
\end{equation}

 System (\ref{ortcond}) - (\ref{comcond}) gives
$m d (d + 1) / 2 \, + \, (m \, + \, s) $
conditions at every ${\bf X}$ and $T$ for the parameters of the
zero approximation $\bm{\Psi}_{(0)} (\bm{\theta},{\bf X},T)$.

\vspace{0.2cm}

{\bf Lemma 1.1.}

{\it Under the regularity conditions formulated above the
orthogonality conditions (\ref{ortcond}) do not contain the functions
$\theta^{\alpha}_{0}({\bf X},T)$ and give constraints only to the
functions $U^{\nu}({\bf X},T)$, having the form
$$C^{(q)}_{\nu} ({\bf U}) \, U^{\nu}_{T} \,\, - \,\, 
D^{(q)p}_{\nu} ({\bf U}) \, U^{\nu}_{X^{p}} \,\, = \,\, 0
\,\,\,\,\,\,\,\,\,\, , \,\,\,\,\,\,\,\, q = 1, \dots, m + s$$
(with some functions $C^{(q)}_{\nu} ({\bf U})$,
$D^{(q)p}_{\nu} ({\bf U})$, $\nu = 1, \dots, N$,
$p = 1, \dots, d$).
}

\vspace{0.2cm}

Proof.

Let us write down the part 
${\bf f}^{\prime}_{(1)}$ of the function ${\bf f}_{(1)}$,
which contains the derivatives $\theta^{\beta}_{0T}({\bf X},T)$    
and $\theta^{\beta}_{0X^{p}}({\bf X},T)$. We have
$$f^{\prime i}_{(1)}(\bm{\theta},{\bf X},T) \, = \, - \,
{\partial F^{i} \over \partial \varphi^{j}_{t}}
\left(\bm{\Psi}_{(0)}, \, \dots \right) \,
\Psi^{j}_{(0)\theta^{\beta}} \, \theta^{\beta}_{0T} \, - \,
{\partial F^{i} \over \partial \varphi^{j}_{x^{p}}}
\left(\bm{\Psi}_{(0)}, \, \dots \right) \,
\Psi^{j}_{(0)\theta^{\beta}} \, \theta^{\beta}_{0X^{p}} \, - $$
$$- {\partial F^{i} \over \partial \varphi^{j}_{tt}} \!
\left(\bm{\Psi}_{(0)},  \dots \right) \, 2 
\omega^{\alpha} ({\bf X},T) 
\Psi^{j}_{(0)\theta^{\alpha}\theta^{\beta}} 
\theta^{\beta}_{0T} \, - 
\, {\partial F^{i} \over \partial \varphi^{j}_{x^{p}x^{l}}} \! 
\left(\bm{\Psi}_{(0)}, \, \dots \right) \, 2 
k^{\alpha}_{p} ({\bf X},T) 
\Psi^{j}_{(0)\theta^{\alpha}\theta^{\beta}} 
\theta^{\beta}_{0X^{l}} \, -  \dots $$
 
 Let us choose the parameters ${\bf U}$ in the form
$${\bf U} = (k^{1}_{p}, \dots, k^{m}_{p}, \omega^{1}, 
\dots, \omega^{m}, n^{1}, \dots, n^{s})$$

 We can write then
$$f^{\prime i}_{(1)}(\bm{\theta},{\bf X},T) \, = \, \left[
- \, {\partial \over \partial \omega^{\beta}} \,
F^{i} \left( \bm{\Phi} (\bm{\theta} + \bm{\theta}_{0}, {\bf U}),
\dots \right) \, + \, {\hat L}^{i}_{j} \, 
{\partial \over \partial \omega^{\beta}} \,
\Phi^{j} (\bm{\theta} + \bm{\theta}_{0}, {\bf U}) \right] \,
\theta^{\beta}_{0T} \, + \, $$
$$+ \, \left[ - \, {\partial \over \partial k^{\beta}_{p}} \,
F^{i} \left( \bm{\Phi} (\bm{\theta} + \bm{\theta}_{0}, {\bf U}),
\dots \right) \, + \, {\hat L}^{i}_{j} \,
{\partial \over \partial k^{\beta}_{p}} \,
\Phi^{j} (\bm{\theta} + \bm{\theta}_{0}, {\bf U}) \right] \,
\theta^{\beta}_{0X^{p}} $$

 The total derivatives $\partial F^{i} / \partial \omega^{\beta}$
and $\partial F^{i} / \partial k^{\beta}_{p}$ are identically equal to
zero on $\Lambda$ according to (\ref{PhaseSyst0}). We have then
$$\int_{0}^{2\pi}\!\!\!\!\!\dots\int_{0}^{2\pi}
\kappa^{(q)}_{[{\bf U}({\bf X},T)]\, i}
(\bm{\theta}\, + \, \bm{\theta}_{0}({\bf X},T)) \,
f^{\prime i}_{(1)} (\bm{\theta},{\bf X},T) \,
{d^{m} \theta \over (2\pi)^{m}} \,\,\, \equiv \,\,\, 0 $$
since 
$\bm{\kappa}^{(q)}_{[{\bf U}({\bf X},T)]} 
(\bm{\theta} + \bm{\theta}_{0}({\bf X},T))$ 
are left eigenvectors of ${\hat L}$ with the zero eigenvalue.

{\hfill Lemma 1.1 is proved.}

\vspace{0.2cm}

 Let us note that the statement of Lemma 1.1 was present 
in the Whitham approach from the very beginning 
(see \cite{whith1, whith2, whith3, luke}).
In fact, under various assumptions it can be also shown that the
additional phase shifts 
$\theta^{\alpha}_{0}({\bf X},T)$ can be always
absorbed by the functions 
$S^{\alpha} ({\bf X},T)$ after a suitable
correction of initial data (see e.g. \cite{Haberman1, Haberman2,
MaltsevJMP, DobrMinenkov}). It should be noted, however, that
the corresponding phase shift can play rather important role
in the weakly nonlinear case
(\cite{Newell}, see also \cite{MaltsevAmerMath, DobrMinenkov}).

\vspace{0.2cm}

 Conditions (\ref{ortcond}) together
with the compatibility conditions
$$k^{\alpha}_{p T} \, = \, \omega^{\alpha}_{X^{p}} 
\,\,\,\,\,\,\,\, , \,\,\,\,\,\,\,\, \alpha = 1, \dots, m
\,\,\, , \,\,\,\,\, p = 1, \dots, d $$
give $(m +s) + m d = m (d + 1) + s$ restrictions on the functions
${\bf U} ({\bf X}, T)$ which is exactly equal to the number of the
parameters $U^{\nu}$.

 Let us call the system
\begin{equation}
\label{RegWhithSyst}
\begin{array}{c}
C^{(q)}_{\nu} ({\bf U}) \, U^{\nu}_{T} \,\, = \,\,
D^{(q)p}_{\nu} ({\bf U}) \, U^{\nu}_{X^{p}} \,\,\, , \,\,\,\,\,
q \, = \, 1, \dots, m+s
\cr
\cr
k^{\alpha}_{p T} \,\, = \,\, \omega^{\alpha}_{X^{p}}
\,\,\,\,\,\,\,\, , \,\,\,\,\,\,\,\, \alpha = 1, \dots, m
\,\,\, , \,\,\,\,\, p = 1, \dots, d
\end{array}
\end{equation}
the evolutionary part of a regular Whitham system for a complete 
regular family $\Lambda$, while the restrictions
\begin{equation}
\label{InConstraints}
k^{\alpha}_{p X^{l}} \, = \, k^{\alpha}_{l X^{p}}
\,\,\,\,\,\,\,\, , \,\,\,\,\,\,\,\, \alpha = 1, \dots, m
\,\,\, , \,\,\,\,\, p, l = 1, \dots, d
\end{equation}
will be considered as additional constraints for the evolutionary
system (\ref{RegWhithSyst}). It is easy to see that the constraints
(\ref{InConstraints}) are compatible with the evolutionary
system (\ref{RegWhithSyst}) in the sense that the restrictions
(\ref{InConstraints}) are conserved by system (\ref{RegWhithSyst})
being imposed at the initial time.

 In generic case the derivatives $U^{\nu}_{T}$ can be expressed in
terms of $U^{\mu}_{X^{p}}$ from system (\ref{RegWhithSyst})
and the evolutionary part of a regular Whitham system can be written 
in the form 
\begin{equation}
\label{WhithamSystem}
U^{\nu}_{T} \,\, = \,\, V^{\nu p}_{\mu} ({\bf U}) \, U^{\mu}_{X^{p}} 
\end{equation}

 The Hamiltonian properties of systems (\ref{WhithamSystem})
are very well developed in the one-dimensional situation.
The general theory of the one-dimensional
systems (\ref{WhithamSystem}),
which are Hamiltonian with respect to local Poisson brackets of
Hydrodynamic type (Dubrovin - Novikov brackets) was constructed by
B.A. Dubrovin and S.P. Novikov. Let us give here a brief description
of the Dubrovin - Novikov Hamiltonian structures and of the properties
of the corresponding systems (\ref{WhithamSystem}).

 The Dubrovin - Novikov bracket on the space of fields
$(U^{1}(X), \dots, U^{N}(X))$ has the form
\begin{equation}
\label{DNbr}
\{U^{\nu}(X), U^{\mu}(Y)\} \,\, = \,\, g^{\nu\mu}({\bf U}) \,
\delta^{\prime}(X-Y) \, + \,
b^{\nu\mu}_{\gamma}({\bf U}) \, U^{\gamma}_{X} \,
\delta (X-Y) \,\,\,\,\, , \,\,\, \nu, \mu = 1, \dots, N
\end{equation}
with the Hamiltonian operator
$${\hat J}^{\nu\mu} \,\, = \,\,
g^{\nu\mu}({\bf U}) \, {d \over dX} \, + \,
b^{\nu\mu}_{\gamma}({\bf U}) \, U^{\gamma}_{X} $$

 As was shown by B.A. Dubrovin and S.P. Novikov
(\cite{dn1,DubrNovDAN84,dn2,dn3}), 
the expression (\ref{DNbr}) with non-degenerate
tensor $g^{\nu\mu}({\bf U})$ defines a Poisson bracket on the space
of fields ${\bf U}(X)$ if and only if:

1) Tensor $g^{\nu\mu}({\bf U})$ gives a symmetric flat pseudo-Riemannian
metric with upper indices on the space of parameters
$(U^{1}, \dots, U^{N})$;

2) The values
$$\Gamma^{\nu}_{\mu\gamma} = - g_{\mu\lambda} \,\,
b^{\lambda\nu}_{\gamma}$$
where
$g^{\nu\lambda}({\bf U}) \, g_{\lambda\mu}({\bf U}) = \delta^{\nu}_{\mu}$,
represent the Christoffel symbols for the corresponding metric
$g_{\nu\mu}({\bf U})$.

 Every Dubrovin - Novikov
bracket with non-degenerate tensor $g^{\nu\mu}({\bf U})$ can be written
in the canonical form (\cite{dn1,DubrNovDAN84,dn2,dn3}):
$$  \{n^{\nu}(X), n^{\mu}(Y)\} \,\, = \,\, \epsilon^{\nu}
\delta^{\nu\mu} \, \delta^{\prime}(X-Y) \,\,\,\,\, , \,\,\,
\epsilon^{\nu} = \pm 1 $$
after the transition to the flat coordinates
$n^{\nu} = n^{\nu}({\bf U})$ for the metric $g_{\nu\mu}({\bf U})$.

 The functionals
$$N^{\nu} \,\, = \,\, \int_{-\infty}^{+\infty} 
n^{\nu}(X) \, dX \,\,\,\,\, , \,\,\,\,\,\,\,\,
P \,\, = \,\, \int_{-\infty}^{+\infty} {1 \over 2}
\sum_{\nu=1}^{N} \epsilon^{\nu}  (n^{\nu})^{2} (X) \, dX$$
represent the annihilators and the momentum functional of
the bracket (\ref{DNbr}) respectively.

 The Hamiltonian functions in the theory of brackets (\ref{DNbr})
are represented usually by the functionals of Hydrodynamic Type, i.e.
$$H  \,\,\, = \,\,\,  \int_{-\infty}^{+\infty}
h({\bf U}) \, dX $$

 Another important form of the Dubrovin - Novikov bracket is the
diagonal form. It corresponds to the case when the coordinates
$U^{\nu}$ represent the diagonal coordinates for the metric
$g_{\nu\mu}({\bf U})$. This form of the
Dubrovin - Novikov bracket is connected with the integration
theory of systems of Hydrodynamic Type which can be written in the
diagonal form
\begin{equation}
\label{DiagSystHT}
U^{\nu}_{T} \,\,\, = \,\,\,
V^{\nu} ({\bf U}) \, U^{\nu}_{X}
\end{equation}

 According to conjecture of S.P. Novikov, all the systems of
Hydrodynamic Type having form (\ref{DiagSystHT}) and Hamiltonian
with respect to any bracket (\ref{DNbr}) are integrable. This
conjecture was proved by S.P. Tsarev (\cite{tsarev,tsarev2})
who suggested a method of integration of these systems. In fact,
the method of Tsarev is applicable to a wider class of diagonalizable
systems of hydrodynamic type which was called by Tsarev
``semi-Hamiltonian''. As it turned out later,
the class of ``semi-Hamiltonian systems'' contains also the
systems Hamiltonian with respect to generalizations of the Dubrovin -
Novikov bracket - the weakly nonlocal Mokhov - Ferapontov bracket
(\cite{mohfer1}) and the Ferapontov bracket (\cite{fer1, fer2}).
Various aspects of the weakly nonlocal brackets of Hydrodynamic
Type are discussed in
\cite{mohfer1, fer1, fer2, fer3, fer4, Pavlov1, PhysD}.

 B.A. Dubrovin and S.P. Novikov (\cite{dn1,dn2,dn3}) suggested also
the procedure for constructing bracket (\ref{DNbr})
for the Whitham system in the one-dimensional
case. The original system has in this scheme the evolutionary
form
\begin{equation}
\label{OneDimEvSyst}
\varphi^{i}_{t} \,\, = \,\, F^{i}
(\bm{\varphi}, \bm{\varphi}_{x}, \bm{\varphi}_{xx}, \dots)
\end{equation}
and is Hamiltonian with respect to a local field-theoretic bracket
\begin{equation}
\label{LocTheorFieldBr}
\{\varphi^{i}(x) , \varphi^{j}(y)\} \,\, = \,\, \sum_{k \geq 0}
B^{ij}_{(k)} (\bm{\varphi}, \bm{\varphi}_{x}, \dots)
\, \delta^{(k)}(x-y)
\end{equation}
with the local Hamiltonian of the form
\begin{equation}
\label{hamfun}
H \,\, = \,\, \int P_{H}
 (\bm{\varphi}, \bm{\varphi}_{x}, \dots) \, dx
\end{equation}

 Method of B.A. Dubrovin and S.P. Novikov is based on
the existence of $N$ (equal to the number of parameters
$U^{\nu}$ of the family $\Lambda$) local integrals
\begin{equation}
\label{integ}
I^{\nu} \,\, = \,\, \int
P^{\nu}(\bm{\varphi}, \bm{\varphi}_{x},\dots) \, dx
\end{equation}
which commute with Hamiltonian (\ref{hamfun}) and with each other
\begin{equation}
\label{invv}
\{I^{\nu} , H\} = 0 \,\,\,\,\, , \,\,\,\,\,
\{I^{\nu} , I^{\mu}\} = 0
\end{equation}

 We have then
$$P^{\nu}_{t} (\bm{\varphi}, \bm{\varphi}_{x},\dots) \,\, \equiv \,\,
Q^{\nu}_{x} (\bm{\varphi}, \bm{\varphi}_{x},\dots) $$
for some functions $Q^{\nu} (\bm{\varphi}, \bm{\varphi}_{x},\dots)$,
while the calculation of the pairwise Poisson brackets
of the densities $P^{\nu}$ gives
\begin{equation}
\label{PnuPmuskob}
\{P^{\nu}(x), P^{\mu}(y)\} \,\, = \,\, \sum_{k\geq 0}
A^{\nu\mu}_{k}(\bm{\varphi}, \bm{\varphi}_{x}, \dots) \,
\delta^{(k)}(x-y)
\end{equation}
where
$$A^{\nu\mu}_{0}(\bm{\varphi}, \bm{\varphi}_{x}, \dots)
\,\, \equiv \,\,
\partial_{x} Q^{\nu\mu}(\bm{\varphi}, \bm{\varphi}_{x}, \dots) $$
according to (\ref{invv}).

 Using now the procedure $\langle \dots \rangle$ of the averaging
of any expression $f (\bm{\varphi}, \bm{\varphi}_{x}, \dots)$ over
the phase variables on the space of the $m$-phase solutions of
(\ref{OneDimEvSyst}) we can write the corresponding Dubrovin-Novikov
bracket on the space of functions ${\bf U}(X)$ in the form:
\begin{equation}
\label{dubrnovb}
\{U^{\nu}(X), U^{\mu}(Y)\} \,\, = \,\, \langle A^{\nu\mu}_{1}\rangle
({\bf U}) \,\, \delta^{\prime}(X-Y) \,\, + \,\,
{\partial \langle Q^{\nu\mu} \rangle \over
\partial U^{\gamma}} \,\, U^{\gamma}_{X} \,\, \delta (X-Y)
\end{equation}
where $U^{\nu} \equiv \langle P^{\nu}(x) \rangle$.

 The Whitham system is written now in the conservative form
$$\langle P^{\nu} \rangle_{T} \,\, = \,\,
\langle Q^{\nu} \rangle_{X} \,\,\,\,\, , \,\,\,\,\,\,\,\,
\nu = 1, \dots, N $$
and is Hamiltonian with respect
to the Dubrovin - Novikov bracket (\ref{dubrnovb}) with the
Hamiltonian
\begin{equation}
\label{UsrHamFunc}
H^{av} \,\, = \,\, \int_{-\infty}^{+\infty}
\langle P_{H} \rangle \left( {\bf U} (X) \right) \,\, d X
\end{equation}

 The proof of the Jacobi identity for bracket
(\ref{dubrnovb}) was suggested in \cite{izvestia}
under certain assumptions about the family of $m$-phase solutions 
of (\ref{EvInSyst}). Besides that, it was
shown in \cite{MalPav} that the Dubrovin - Novikov procedure is
compatible with the procedure of averaging of local Lagrangian
functions when carrying out of both the procedures is possible.
Let us note also that the generalization of the Dubrovin - Novikov
procedure for the weakly nonlocal case was proposed in \cite{malnloc2}.

 In paper \cite{Pavlov2} all the local brackets
(\ref{DNbr}) for the Whitham equations for KdV, NLS, and SG
equations were found. Besides that, in paper \cite{Alekseev}
the hierarchies of the weakly nonlocal Hamiltonian structures
for the Whitham systems for KdV were represented.
 
 The most detailed discussion and justification
of the Dubrovin - Novikov procedure separately for the single-phase and
the multiphase cases can be found in \cite{DNMultDim}. In particular, it
was shown that the justification of the procedure is in fact insensitive
to the appearance of ``resonances'' which can arise in the multi-phase
case, which is the basis for its widespread use in the multiphase
situation.

 However, as we will see, the Hamiltonian structure of system
(\ref{RegWhithSyst}) under the constraints (\ref{InConstraints})
should have another form in the case $d > 1$. Thus, we can not 
introduce a Poisson bracket on the whole space of fields 
${\bf U}({\bf X})$ in the general case. Instead, the Poisson bracket
on the submanifold given by constraints (\ref{InConstraints}) should
be considered.

 In this paper we will represent the Hamiltonian structure for
system (\ref{RegWhithSyst})-(\ref{InConstraints}) using the functions
$S^{\alpha} ({\bf X})$ as a part of coordinates on the submanifold
given by constraints (\ref{InConstraints}). As a result, the general 
form of the Poisson bracket will be different from that given by
(\ref{DNbr}). However, as we will see, an analogue of the 
Dubrovin - Novikov procedure under the same requirements as in the
one-dimensional case can be used also in the multi-dimensional 
situation. Finally, the whole space of parameters will be considered
in the form
$$\left( S^{1}({\bf X}), \dots, S^{m}({\bf X}), \, U^{1}({\bf X}),
\dots, U^{m+s}({\bf X}) \right) $$
where $S^{\alpha}_{X^{l}} \, \equiv \, k^{\alpha}_{l} ({\bf X})$
and $U^{1}({\bf X}), \dots, U^{m+s}({\bf X})$ - are some additional
necessary parameters, while the corresponding Poisson bracket will
be written as
\begin{multline}
\label{AdvDNbr}
\left\{ S^{\alpha}({\bf X}), S^{\beta}({\bf Y}) \right\} 
\,\, = \,\, 0 \,\,\,\,\, , \,\,\,\,\,
\left\{ S^{\alpha}({\bf X}), U^{\lambda}({\bf Y}) \right\}
\,\, = \,\, \omega^{\alpha\lambda}({\bf X}) \,\, 
\delta ({\bf X} - {\bf Y}) \,\,\, , \\
\left\{ U^{\lambda}({\bf X}), U^{\rho}({\bf Y}) \right\} \,\, = \,\,
\sum_{p=1}^{d} g^{\lambda\rho p}({\bf X}) \,\,
\delta_{X^{p}} ({\bf X} - {\bf Y}) \,\,\, + \,\,\,
\left( \sum_{p=1}^{d} \left[ \gamma^{\lambda\rho p}({\bf X})
\right]_{X^{p}} \right) \, \delta ({\bf X} - {\bf Y})
\end{multline}

 We can see that the form of bracket (\ref{AdvDNbr}) combines
in fact the Dubrovin - Novikov form and the form connected with the
action-angle variables for the multi-dimensional brackets. Here we
discuss also the possibility of the pure canonical representation
of bracket (\ref{AdvDNbr}) after some change of coordinates. As
we will see, this possibility will 
depend in fact on some special 
features of the space of the additional parameters 
$(U^{1}({\bf X}), \dots, U^{m+s}({\bf X}))$.

 In Chapter 2 we consider the Hamiltonian formulation of the Whitham
method and discuss the regularity conditions necessary for the
application of the analogue of Dubrovin - Novikov procedure in the
multi-dimensional case.

 In Chapter 3 we suggest the averaging procedure for the Poisson
bracket in the multi-dimensional case and give the necessary
justification of it's application.

 In Chapter 4 we discuss the canonical forms of the averaged
Poisson brackets and demonstrate the averaging procedure using
the simple example of the nonlinear wave equation in $d$ spatial
dimensions.

\section{Hamiltonian formulation of the Whitham method.}
\setcounter{equation}{0}

 In this paper we are going to consider the Hamiltonian formulation
of the Whitham equations. Let us consider then another approach to 
the construction of the regular Whitham system which
is connected with the method of averaging of conservation laws.
According to further consideration of the Hamiltonian structure
of the Whitham equations we will assume now that system (\ref{insyst})
is written in an evolutionary form
\begin{equation}
\label{EvInSyst}
\varphi^{i}_{t} \,\, = \,\,
F^{i} \left( \bm{\varphi}, \, \bm{\varphi}_{\bf x}, \,
\bm{\varphi}_{\bf xx},
\dots \right)
\end{equation}

 The families of the $m$-phase solutions of (\ref{EvInSyst})
are defined then by solutions of the system
\begin{equation}
\label{EvPhaseSyst}   
\omega^{\alpha} \, \varphi^{i}_{\theta^{\alpha}} \,\, = \,\,
F^{i} \left( \bm{\varphi}, \,
k^{\beta_{1}}_{1} \, \bm{\varphi}_{\theta^{\beta_{1}}}, \dots,
k^{\beta_{d}}_{d} \, \bm{\varphi}_{\theta^{\beta_{d}}}, 
\dots \right)
\end{equation}
on the space of $2\pi$-periodic in each $\theta^{\alpha}$ functions
$\bm{\varphi} (\bm{\theta})$.

 We will assume that the conservation laws of system (\ref{EvInSyst})
have the form
$$P^{\nu}_{t} \left( \bm{\varphi}, \, \bm{\varphi}_{\bf x}, \,
\bm{\varphi}_{\bf xx}, \dots \right) \,\,\, = \,\,\, 
Q^{\nu 1}_{x^{1}} \left( \bm{\varphi}, \,
\bm{\varphi}_{\bf x}, \, \bm{\varphi}_{\bf xx}, \dots \right) \, + \,
\dots \, + \, Q^{\nu d}_{x^{d}} \left( \bm{\varphi}, \,
\bm{\varphi}_{\bf x}, \, \bm{\varphi}_{\bf xx}, \dots \right) $$
such that the values
$$I^{\nu} \,\, = \,\, \int
P^{\nu} \left( \bm{\varphi}, \, \bm{\varphi}_{\bf x}, \, 
\bm{\varphi}_{\bf xx}, \dots \right) \, d^{d} x $$
represent translationally invariant conservative quantities
for system (\ref{EvInSyst}) in the case of
the rapidly decreasing at infinity functions  
$\bm{\varphi} ({\bf x})$. We can also define the conservation laws 
for system (\ref{EvInSyst}) in the periodic case with the periods
$K_{1}$, $\dots$, $K_{d}$
$$I^{\nu} \,\, = \,\, {1 \over K_{1} \dots K_{d}} \, 
\int_{0}^{K_{1}} \!\!\! \dots \int_{0}^{K_{d}} 
P^{\nu} \left( \bm{\varphi}, \, \bm{\varphi}_{\bf x}, \, 
\bm{\varphi}_{\bf xx}, \dots \right) \, d^{d} x $$
or in the quasiperiodic case 
$$I^{\nu} \,\, = \,\, \lim_{K\rightarrow\infty} \,\,\,
{1 \over (2K)^{d}} \,\, \int_{-K}^{K} \! \dots \int_{-K}^{K} \,
P^{\nu} \left( \bm{\varphi}, \, \bm{\varphi}_{\bf x}, \,
\bm{\varphi}_{\bf xx},
\dots \right) \, d^{d} x $$

 It is natural also to define the variation derivatives of the
functionals $I^{\nu}$ with respect to the variations of
$\bm{\varphi} ({\bf x})$ having the same periodic or quasiperiodic
properties as the original functions. Easy to see then that the
standard Euler - Lagrange expressions for the variation derivatives
can be used in this case.

 Let us write the functionals $I^{\nu}$ in the general form
\begin{equation}
\label{Integrals}
I^{\nu} \,\, = \,\, \int
P^{\nu} \left( \bm{\varphi}, \, \bm{\varphi}_{\bf x}, \, 
\bm{\varphi}_{\bf xx},
\dots \right) \, d^{d} x
\end{equation}
assuming the appropriate definition in the corresponding situations.

 Let us define a quasiperiodic function $\bm{\varphi} ({\bf x})$
with fixed wave numbers
$({\bf k}_{1}, \dots, {\bf k}_{d})$ as a function 
$\bm{\varphi} ({\bf x})$ on $\mathbb{R}^{d}$ coming from a smooth 
periodic function $\bm{\varphi} (\bm{\theta})$ on the torus
$\mathbb{T}^{m}$:
$$\bm{\varphi} ({\bf k}_{1} x^{1} + \dots + {\bf k}_{d} x^{d} +
\bm{\theta}_{0})  \,\,\,\,\, \rightarrow \,\,\,\,\, 
\bm{\varphi} (x^{1}, \dots, x^{d}) $$

 Let us define the functionals
\begin{equation}
\label{Jnu}
J^{\nu} \,\, = \,\, \int_{0}^{2\pi}\!\!\!\!\!\dots\int_{0}^{2\pi}
P^{\nu} \left( \bm{\varphi}, \,
k^{\beta_{1}}_{1} \, \bm{\varphi}_{\theta^{\beta_{1}}}, \dots,
k^{\beta_{d}}_{d} \, \bm{\varphi}_{\theta^{\beta_{d}}},
\dots \right) \, {d^{m} \theta \over (2\pi)^{m}}
\end{equation}
on the space of $2\pi$-periodic in $\bm{\theta}$ functions.

 It's not difficult to see that the functions
\begin{equation}
\label{VarDer}
\zeta^{(\nu)}_{i [{\bf U}]} (\bm{\theta} + \bm{\theta}_{0})
\,\, = \,\, \left. 
{\delta J^{\nu} \over \delta \varphi^{i} (\bm{\theta})} 
\right|_{\bm{\varphi}(\bm{\theta}) = \bm{\Phi}(\bm{\theta} + 
\bm{\theta}_{0}, {\bf U})}
\end{equation}
represent left eigenvectors of the operator
${\hat L}^{i}_{j[{\bf U}, \bm{\theta}_{0}]}$ with zero
eigenvalues, regularly depending on parameters ${\bf U}$
on a fixed smooth family $\Lambda$.

 Indeed, the operator 
${\hat L}^{i}_{j[{\bf U}, \bm{\theta}_{0}]}$ 
is defined in this case by the distribution
$$L^{i}_{j[{\bf U}, \bm{\theta}_{0}]}
(\bm{\theta}, \bm{\theta}^{\prime}) \,\, = \,\,
\delta^{i}_{j} \, \omega^{\alpha} \, \delta_{\theta^{\alpha}}
(\bm{\theta} - \bm{\theta}^{\prime}) \, - \, \left.
{\delta F^{i} ( \bm{\varphi}, \,
k^{\beta_{1}}_{1} \, \bm{\varphi}_{\theta^{\beta_{1}}}, \dots,
k^{\beta_{d}}_{d} \, \bm{\varphi}_{\theta^{\beta_{d}}},
\dots ) \over
\delta \varphi^{j} (\bm{\theta}^{\prime})} 
\right|_{\bm{\varphi}(\bm{\theta}) = \bm{\Phi}(\bm{\theta} +  
\bm{\theta}_{0}, {\bf U})} $$

 We have 
$$\int_{0}^{2\pi}\!\!\!\!\!\dots\int_{0}^{2\pi}
{\delta J^{\nu} \over \delta \varphi^{i} (\bm{\theta})} \,
\left( \omega^{\alpha} \, \varphi^{i}_{\theta^{\alpha}} \, - \,  
F^{i} ( \bm{\varphi}, \,
k^{\beta_{1}}_{1} \, \bm{\varphi}_{\theta^{\beta_{1}}}, \dots,
k^{\beta_{d}}_{d} \, \bm{\varphi}_{\theta^{\beta_{d}}},
\dots ) \right)
\, {d^{m} \theta \over (2\pi)^{m}} \,\, \equiv \,\, 0 $$
for any translationally invariant integral of (\ref{EvInSyst}).
Taking the variation derivative of this relation with respect to
$\varphi^{j} (\bm{\theta}^{\prime})$ 
on $\Lambda$ we get the required statement.

 Thus, we can write
\begin{equation}
\label{SviazZetaKappa}
\zeta^{(\nu)}_{i [{\bf U}]} (\bm{\theta}) \,\,\, = \,\,\,
\sum_{q=1}^{m+s} \, 
c^{\nu}_{q} ({\bf U}) \,\, \kappa^{(q)}_{i[{\bf U}]}
(\bm{\theta}) \,\,\,\,\,\,\,\,\,\, , \,\,\,\,\,\,\,\,\,\,
\nu = 1, \dots, N 
\end{equation}
with some smooth functions $c^{\nu}_{q} ({\bf U})$
on a complete regular family $\Lambda$.

 For our consideration of the regular Whitham system we will need
a sufficient number of the first integrals (\ref{Integrals}), such
that the values of the functionals $J^{\nu}$ on $\Lambda$ represent
the full set of parameters $U^{\nu} = J^{\nu}|_{\Lambda}$. Thus,
we will require here the presence of $N = m (d + 1) + s$ independent
integrals $I^{\nu}$, $\nu = 1, \dots, m (d + 1) + s$. 

 Coming back to the definition of a complete regular family of
$m$-phase solutions of system (\ref{EvInSyst}) we can see that in
the case of a complete regular family $\Lambda$ the number of linearly
independent vectors (\ref{VarDer}) on $\Lambda$ is always finite.
More precisely, if $N = m (d + 1) + s$ is the number of parameters of
$m$-phase solutions of (\ref{EvInSyst}) (excluding the initial phase 
shifts) then for a complete regular family of $m$-phase solutions
we require the presence of exactly $m + s = N - m d$ left eigenvectors 
$\bm{\kappa}^{(q)}_{[{\bf U}]} (\bm{\theta} + \bm{\theta}_{0})$
with zero eigenvalues, regularly depending on parameters
(in accordance with the number of the vectors 
$\bm{\Phi}_{\theta^{\alpha}}$, $\bm{\Phi}_{n^{l}}$). 
Thus, according to Definition 1.1, we assume
here that the number of linearly independent vectors
defined by formula (\ref{VarDer}) does not exceed
$m + s = N - m d$ for a complete regular family $\Lambda$.

 Let us note that the conditions on the variation derivatives
of $J^{\nu}$ formulated above do not contradict to the condition 
that the values $J^{\nu}$ ($\nu = 1, \dots, N$) can be chosen as 
the parameters $U^{\nu}$ on the family of $m$-phase solutions. 
Indeed, the definition of $J^{\nu}$ (\ref{Jnu}) explicitly includes 
the additional $m d$ functions $k^{\alpha}_{p}$, which provide
the necessary functional independence of the values of $J^{\nu}$ on
$\Lambda$. In other words, we can use the Euler - Lagrange
expressions for the variation derivatives of $I^{\nu}$ 
only on subspaces with fixed wave numbers 
$({\bf k}_{1}, \dots, {\bf k}_{d})$.
The variation of the wave numbers gives linearly growing
variations which do not allow to use the Euler - Lagrange
expressions.

 Let us make an agreement that we will always assume here that
the Jacobian of the coordinate transformation
$$\left({\bf k}_{q}, \, \bm{\omega}, \, {\bf n} \right) \,\,
\rightarrow \,\, \left( U^{1}, \, \dots, \, U^{N} \right) $$
is different from zero on $\Lambda$ whenever we say that the
values $U^{\nu} ({\bf k}_{q}, \bm{\omega}, {\bf n})$ represent
a complete set of parameters on ${\Lambda}$ 
(excluding the initial phase shifts). 

 Under the conditions formulated above let us 
prove here the following proposition:

\vspace{0.2cm}

{\bf Proposition 2.1.}

{\it 
Let ${\Lambda}$ be a complete regular family of $m$-phase solutions
of system (\ref{EvInSyst}). Let the values $(U^{1}, \dots, U^{N})$
of the functionals $(J^{1}, \dots, J^{N})$ (\ref{Jnu}) give a
complete set of parameters on $\Lambda$ excluding the initial phase 
shifts. Then:

1) The set of the vectors
$$\left\{ \bm{\Phi}_{\omega^{\alpha}} 
(\bm{\theta} + \bm{\theta}_{0}, 
\, {\bf k}_{q}, \, \bm{\omega}, \, {\bf n}) \, ,
\,\,\, 
\bm{\Phi}_{n^{l}}
(\bm{\theta} + \bm{\theta}_{0}, 
\, {\bf k}_{q}, \, \bm{\omega}, \, {\bf n}) \, , 
\,\,\,\,\,\,\,\, \alpha = 1, \dots, m, \,\,\, l = 1, \dots, s
\right\} $$
is linearly independent on $\Lambda$;

2) The variation derivatives
$\zeta^{(\nu)}_{i [{\bf U}]} (\bm{\theta} + \bm{\theta}_{0})$,
given by (\ref{VarDer}), generate the full space of the regular
left eigenvectors of the operator
${\hat L}^{i}_{j[{\bf U},\bm{\theta}_{0}]}$ with zero eigenvalues
on the family ${\Lambda}$.
}

\vspace{0.2cm}

 Proof.

 Indeed, we require that the rows given by the derivatives
$$\left( {\partial U^{1} \over \partial \omega^{\alpha}}, 
\, \dots \, ,
{\partial U^{N} \over \partial \omega^{\alpha}} \right) \,\,\,\,\, ,
\,\,\,\,\, 
\left( {\partial U^{1} \over \partial n^{l}}, \, \dots \, ,
{\partial U^{N} \over \partial n^{l}} \right) $$
are linearly independent on $\Lambda$. Using the expressions
$${\partial U^{\nu} \over \partial \omega^{\alpha}} \,\, = \,\,
\int_{0}^{2\pi}\!\!\!\!\!\dots\int_{0}^{2\pi} 
\zeta^{(\nu)}_{i [{\bf U}]} (\bm{\theta}) \,\,
\Phi^{i}_{\omega^{\alpha}} (\bm{\theta}, \, {\bf U}) \,\,
{d^{m} \theta \over (2\pi)^{m}} \,\,\,\,\, , \,\,\,\,\,\,\,\,
\alpha = 1, \dots, m $$
$${\partial U^{\nu} \over \partial n^{l}} \,\, = \,\,
\int_{0}^{2\pi}\!\!\!\!\!\dots\int_{0}^{2\pi}
\zeta^{(\nu)}_{i [{\bf U}]} (\bm{\theta}) \,\,
\Phi^{i}_{n^{l}} (\bm{\theta}, \, {\bf U}) \,\,
{d^{m} \theta \over (2\pi)^{m}} \,\,\,\,\, , \,\,\,\,\,\,\,\,
l = 1, \dots, s $$
on $\Lambda$, we get that the set 
$\{\bm{\Phi}_{\omega^{\alpha}}, \, \bm{\Phi}_{n^{l}} \}$ is
linearly independent on $\Lambda$ and the number of linearly
independent variation derivatives (\ref{VarDer}) is not less
than $m + s$.

 We obtain then that the variation derivatives (\ref{VarDer}) 
generate in this case a space of 
regular left eigenvectors of the operator
${\hat L}^{i}_{j[{\bf U},\bm{\theta}_{0}]}$ with zero eigenvalues
of dimension $(m + s)$.

{\hfill Proposition 2.1 is proved.}

\vspace{0.2cm}

 The following lemma will be very important in our further 
considerations.

\vspace{0.2cm}

{\bf Lemma 2.1.}

{\it Let the values $U^{\nu}$ of the functionals $J^{\nu}$
on a complete regular family of $m$-phase solutions $\Lambda$ 
be functionally independent and give a complete set of parameters
(excluding initial phase shifts) on $\Lambda$, such that we have 
$k^{\alpha}_{p} = k^{\alpha}_{p} (U^{1}, \dots, U^{N})$.
Then the functionals  $k^{\alpha}_{p} (J^{1}, \dots, J^{N})$
have zero variation derivatives on $\Lambda$.
}

\vspace{0.2cm}

Proof.

 As we have seen, the conditions of the Lemma imply the existence 
of $m d$ independent relations
\begin{equation}
\label{intder}
\sum_{\nu=1}^{N} \lambda^{\tau}_{\nu}({\bf U}) \left.
{\delta J^{\nu} \over \delta \varphi^{i} (\bm{\theta})}
\right|_{\bm{\varphi}(\bm{\theta}) \,\, = \,\, \bm{\Phi}(\bm{\theta} 
+ \bm{\theta}_{0}, {\bf U})} \,\, \equiv \,\, 0
\,\,\,\,\, ,
\,\,\,\,\, \tau = 1, \dots, m d
\end{equation}
on $\Lambda$.

 For the corresponding coordinates $U^{\nu}$ on $\Lambda$ this
implies the relations
$$\sum_{\nu=1}^{N} \lambda^{\tau}_{\nu}({\bf U}) \, dU^{\nu} 
\,\, = \,\, \sum_{p=1}^{d} \sum_{\beta=1}^{m} 
\mu^{(\tau)}_{(\beta p)}({\bf U}) \, dk^{\beta}_{p}({\bf U}) $$
for some matrix $\mu^{(\tau)}_{(\beta p)}({\bf U})$.

 Let us consider the matrix $\mu^{(\tau)}_{(\beta p)}({\bf U})$
as a $m d \times m d$ matrix giving a linear transformation
${\hat \mu}: \mathbb{R}^{md} \rightarrow \mathbb{R}^{md}$
between the spaces with bases parametrized by the pairs
$(\beta p)$ and the index $\tau$ respectively.
Since $U^{\nu}$ provide coordinates on $\Lambda$
the matrix $\mu^{(\tau)}_{(\beta p)}({\bf U})$ has the full rank
and, therefore, invertible. We can then write
$$d k^{\beta}_{p} \,\, = \,\, \sum_{\tau=1}^{md}
({\hat \mu}^{-1})^{(\beta p)}_{(\tau)}({\bf U}) \,
\sum_{\nu=1}^{N} \lambda^{(\tau)}_{\nu}({\bf U}) \,
dU^{\nu}$$

 The assertion of the Lemma follows then from (\ref{intder}).

{\hfill Lemma 2.1 is proved.}

\vspace{0.2cm}

 Let us consider now the system
\begin{equation}
\label{ConsWhitham}
\langle P^{\nu} \rangle_{T} \,\, = \,\, 
\langle Q^{\nu 1} \rangle_{X^{1}} \, + \, \dots \, + \,
\langle Q^{\nu d} \rangle_{X^{d}}
\,\,\,\,\,\,\,\, , \,\,\,\,\,\,\,\, 
\nu = 1, \dots , N = m (d + 1) + s
\end{equation}
on the space of the functions ${\bf U} ({\bf X})$,
where $\langle \dots \rangle$ denotes the averaging operation
on $\Lambda$ defined by the formula
$$\langle f (\bm{\varphi},  \bm{\varphi}_{\bf x}, \dots ) \rangle
\,\, \equiv \,\, \int_{0}^{2\pi}\!\!\!\!\!\dots\int_{0}^{2\pi}
 f \left( \bm{\Phi}, \, 
k_{1}^{\beta_{1}} \bm{\Phi}_{\theta^{\beta_{1}}},
\dots,  k_{d}^{\beta_{d}} \bm{\Phi}_{\theta^{\beta_{d}}}, \dots
\right) \, {d^{m} \theta \over (2\pi)^{m}} $$

 Let us prove here the following lemma about the connection between
systems (\ref{ConsWhitham}) and (\ref{RegWhithSyst}).

\vspace{0.2cm}

{\bf Lemma 2.2.}

{\it Let the values $U^{\nu}$ of the functionals $J^{\nu}$
on a complete regular family of $m$-phase solutions $\Lambda$
be functionally independent and give a complete set of parameters
on $\Lambda$ excluding the initial phase shifts.
Then on the space of functions ${\bf U} ({\bf X})$ satisfying the
system of constraints
\begin{equation}
\label{KKConstraints}
{\partial k^{\alpha}_{p} ({\bf U} ({\bf X})) \over \partial X^{l}}
\,\,\, = \,\,\,
{\partial k^{\alpha}_{l} ({\bf U} ({\bf X})) \over \partial X^{p}}
\end{equation}
system (\ref{ConsWhitham}) is equivalent to 
the evolutionary part (\ref{RegWhithSyst}) of the regular Whitham
system.
}

\vspace{0.2cm}

 Proof.

 Let us introduce the
functions
\begin{equation}
\label{PiFunc}
\Pi^{\nu (l_{1} \dots l_{d})}_{i} 
(\bm{\varphi}, \bm{\varphi}_{\bf x}, \dots )
\,\, \equiv \,\,
{\partial P^{\nu} (\bm{\varphi}, \bm{\varphi}_{\bf x}, \dots )
\over \partial \varphi^{i}_{l_{1} x^{1} \dots l_{d} x^{d}} }
\,\,\,\,\, , \,\,\,\,\,\,\,\, l_{1}, \dots , l_{d} \, \geq \, 0
\end{equation}

 Using the expression for the evolution of the densities
$P^{\nu} (\bm{\varphi}, \epsilon \bm{\varphi}_{\bf X}, \dots )$
we can write the following identities
$$P^{\nu}_{t} (\bm{\varphi}, \epsilon \bm{\varphi}_{\bf X}, \dots )
\,\,\, = \,\,\, \sum_{l_{1}, \dots , l_{d}} 
\epsilon^{l_{1} + \dots + l_{d}} \,\, 
\Pi^{\nu (l_{1} \dots l_{d})}_{i}
(\bm{\varphi}, \epsilon \bm{\varphi}_{\bf X}, \dots ) \,
\left( F^{i} (\bm{\varphi}, \epsilon \bm{\varphi}_{\bf X}, \dots )
\right)_{l_{1} X^{1} \dots l_{d} X^{d}} \,\,\, \equiv $$
\begin{equation}
\label{PnuQnu}  
\equiv \,\,\, \epsilon \, Q^{\nu 1}_{X^{1}} 
(\bm{\varphi}, \epsilon \bm{\varphi}_{\bf X}, \dots )
\, + \, \dots \, + \, \epsilon \, Q^{\nu d}_{X^{d}}                      
(\bm{\varphi}, \epsilon \bm{\varphi}_{\bf X}, \dots )
\end{equation}

 To calculate the values 
$\epsilon \langle Q^{\nu 1} \rangle_{X^{1}} \, + \, \dots \, + \,
\epsilon \langle Q^{\nu d} \rangle_{X^{d}}$ 
let us put now 
\begin{equation}
\label{VarphiSubs}
\varphi^{i} (\bm{\theta}, {\bf X}) \,\, = \,\, \Phi^{i}
\left( {{\bf S} ({\bf X}) \over \epsilon} + \bm{\theta}, 
{\bf U}({\bf X}) \right)
\end{equation}
where $S^{\alpha}_{X^{p}} = k^{\alpha}_{p}({\bf U}({\bf X}))$.

 The operators $\epsilon \, \partial / \partial X^{p}$ acting on the 
functions (\ref{VarphiSubs}) can be naturally represented as a sum
of $k^{\alpha}_{p} \, \partial / \partial \theta^{\alpha}$ and the terms
proportional to $\epsilon$. So, any expression
$f (\bm{\varphi}, \epsilon \bm{\varphi}_{\bf X}, \dots )$ on the
submanifold (\ref{VarphiSubs}) can be naturally represented in the 
form
$$f (\bm{\varphi}, \epsilon \bm{\varphi}_{\bf X}, \dots ) \,\, = \,\, 
\sum_{l \geq 0} \epsilon^{l} \, f_{[l]} \left[ \bm{\Phi}, {\bf U}
\right] $$
where $f_{[l]} [\bm{\Phi}, {\bf U}]$ are smooth functions of
$(\bm{\Phi}, \bm{\Phi}_{\theta^{\alpha}}, \bm{\Phi}_{U^{\nu}}, \dots)$
and $({\bf U}, {\bf U}_{\bf X}, {\bf U}_{\bf XX}, \dots) $, 
polynomial in the derivatives 
$({\bf U}_{\bf X}, {\bf U}_{\bf XX}, \dots)$,
and having degree $l$ in terms of the total number of derivations of 
${\bf U}$ w.r.t. ${\bf X}$. The functions
$\bm{\Phi}$ appear in $ f_ {[l]} $ with the phase shift
${\bf S}({\bf X}) / \epsilon$ according to (\ref{VarphiSubs}).
However, the common phase shift is not important for the integration 
with respect to $\bm{\theta}$, so let us assume below that the phase 
shift ${\bf S}({\bf X}) / \epsilon$ is omitted after taking all the 
differentiations with respect to ${\bf X}$.

 According to (\ref{PnuQnu}) and (\ref{EvPhaseSyst}) we can write
$$\epsilon \, \langle Q^{\nu 1} \rangle_{X^{1}} \, + \, \dots \, + \,
\epsilon \, \langle Q^{\nu d} \rangle_{X^{d}} \,\, = \,\, 
\epsilon \, \int_{0}^{2\pi}\!\!\!\!\!\dots\int_{0}^{2\pi} \left(
Q^{\nu 1}_{X^{1} [1]} \, + \, \dots \, + \,
Q^{\nu d}_{X^{d} [1]} \right) 
\, {d^{m} \theta \over (2\pi)^{m}} \,\, = $$
$$= \,\, \epsilon \, \int_{0}^{2\pi}\!\!\!\!\!\dots\int_{0}^{2\pi}
\sum_{l_{1}, \dots, l_{d}} 
\left( \Pi^{\nu (l_{1} \dots l_{d})}_{i \, [0]} \, 
F^{i}_{l_{1} X^{1} \dots l_{d} X^{d} \, [1]} \, + \, 
\Pi^{\nu (l_{1} \dots l_{d})}_{i \, [1]} \, 
F^{i}_{l_{1} X^{1} \dots l_{d} X^{d} \, [0]} \right)
{d^{m} \theta \over (2\pi)^{m}} \,\, = $$
$$= \,\,\, \epsilon \, \int_{0}^{2\pi}\!\!\!\!\!\dots\int_{0}^{2\pi}
\sum_{l_{1}, \dots, l_{d}} \left(
\Pi^{\nu (l_{1} \dots l_{d})}_{i \, [0]} \,\, 
k^{\gamma^{1}_{1}}_{1} \dots k^{\gamma^{1}_{l_{1}}}_{1} \,\, \dots \,\,
k^{\gamma^{d}_{1}}_{d} \dots k^{\gamma^{d}_{l_{d}}}_{d} \,\,\,  
F^{i}_{[1] \, \theta^{\gamma^{1}_{1}}\dots\theta^{\gamma^{1}_{l_{1}}}
\, \dots \, \theta^{\gamma^{d}_{1}}\dots\theta^{\gamma^{d}_{l_{d}}}} 
\,\,\, +  \right. $$
$$+ \,\,\, \left. 
\Pi^{\nu (l_{1} \dots l_{d})}_{i \, [0]} \, 
\left( \omega^{\beta} \, \Phi^{i}_{\theta^{\beta}} 
\right)_{l_{1} X^{1} \dots l_{d} X^{d} \, [1]} \,\, + \,\,
\Pi^{\nu (l_{1} \dots l_{d})}_{i \, [1]} \,       
\left( \omega^{\beta} \, \Phi^{i}_{\theta^{\beta}}
\right)_{l_{1} X^{1} \dots l_{d} X^{d} \, [0]} \right)
\, {d^{m} \theta \over (2\pi)^{m}} \,\,\, =$$

$$= \, \epsilon  \int_{0}^{2\pi}\!\!\!\!\!\dots\int_{0}^{2\pi}
\! \sum_{l_{1}, \dots, l_{d}} \! \left(
k^{\gamma^{1}_{1}}_{1} \dots k^{\gamma^{1}_{l_{1}}}_{1} \,\, \dots \,\,
k^{\gamma^{d}_{1}}_{d} \dots k^{\gamma^{d}_{l_{d}}}_{d} \,\,
(-1)^{l_{1} + \dots + l_{d}} \,\,\,
\Pi^{\nu (l_{1} \dots l_{d})}_{i \, [0] \,
\theta^{\gamma^{1}_{1}}\dots\theta^{\gamma^{1}_{l_{1}}}
\, \dots \, \theta^{\gamma^{d}_{1}}\dots\theta^{\gamma^{d}_{l_{d}}}}
\,\, F^{i}_{[1]} \,\, +  \right. $$
$$+ \,\,\, \omega^{\beta}_{X^{1}} \,\,
\Pi^{\nu (l_{1} \dots l_{d})}_{i[0]} \,\, l_{1} \,\,
\Phi^{i}_{\theta^{\beta} \, (l_{1}-1) X^{1} \dots l_{d} X^{d} \, [0]}
\,\,\, + \, \dots \, + \,\,\, 
\omega^{\beta}_{X^{d}} \,\,
\Pi^{\nu (l_{1} \dots l_{d})}_{i[0]} \,\, l_{d} \,\,
\Phi^{i}_{\theta^{\beta} \, l_{1} X^{1} \dots (l_{d}-1) X^{d} \, [0]}
\,\,\, +$$
$$\left. + \,\,\, \omega^{\beta} \,\,
\Pi^{\nu (l_{1} \dots l_{d})}_{i[0]} \,\,
\Phi^{i}_{\theta^{\beta} \, l_{1} X^{1} \dots l_{d} X^{d} \, [1]}
\,\,\, + \,\,\, \omega^{\beta} \,\,
\Pi^{\nu (l_{1} \dots l_{d})}_{i[1]} \,\,
\Phi^{i}_{\theta^{\beta} \, l_{1} X^{1} \dots l_{d} X^{d} \, [0]}
\, \right) \, {d^{m} \theta \over (2\pi)^{m}} $$

 The last two terms in the above expression represent the integral
of the value
$$\omega^{\beta} \, \sum_{l_{1}, \dots, l_{d}}
\left( \Pi^{\nu (l_{1} \dots l_{d})}_{i} \,
\Phi^{i}_{\theta^{\beta} \, l_{1} X^{1} \dots l_{d} X^{d}} 
\right)_{[1]} \,\,\, \equiv \,\,\,
\omega^{\beta} \, \partial P^{\nu}_{[1]} / \partial \theta^{\beta}$$
and are equal to zero.

 It is not difficult to see also that for arbitrary dependence of
parameters ${\bf U}$ of $T$, the derivative of the average
$\langle P^{\nu} \rangle$ w.r.t. $T$ can be written as:
$$\langle P^{\nu} \rangle_{T} \, = 
\int_{0}^{2\pi}\!\!\!\!\!\dots\int_{0}^{2\pi} \!\!
\sum_{l_{1}, \dots, l_{d}}
\Pi^{\nu (l_{1} \dots l_{d})}_{i \, [0]} \, 
\left( 
k^{\gamma^{1}_{1}}_{1} \dots k^{\gamma^{1}_{l_{1}}}_{1} \, \dots \,
k^{\gamma^{d}_{1}}_{d} \dots k^{\gamma^{d}_{l_{d}}}_{d} \,\,
\Phi^{i}_{\theta^{\gamma^{1}_{1}}\dots\theta^{\gamma^{1}_{l_{1}}}
\, \dots \, \theta^{\gamma^{d}_{1}}\dots\theta^{\gamma^{d}_{l_{d}}}}
\right)_{T}  {d^{m} \theta \over (2\pi)^{m}} $$

 Now, we can write the relations
$\,\, \langle P^{\nu} \rangle_{T} \, - \, 
\langle Q^{\nu 1} \rangle_{X^{1}}
- \dots - \langle Q^{\nu d} \rangle_{X^{d}} \, = \, 0  \,\,\, $ 
as
\begin{multline*}
\int_{0}^{2\pi}\!\!\!\!\!\dots\int_{0}^{2\pi} \left(
\zeta^{(\nu)}_{i[{\bf U}({\bf X})]} (\bm{\theta}) \, \left[
\Phi^{i}_{T} (\bm{\theta}, {\bf U}({\bf X})) \, - \, 
F^{i}_{[1]} (\bm{\theta}, {\bf X}) \right] \right. \,\, + \\  \\
+ \,\, ( k^{\beta}_{1 T} \, - \, \omega^{\beta}_{X^{1}} ) \!
\sum_{l_{1}, \dots, l_{d}} 
\Pi^{\nu (l_{1} \dots l_{d})}_{i \, [0]} \,\, l_{1} \,
k^{\gamma^{1}_{1}}_{1} \dots k^{\gamma^{1}_{l_{1}-1}}_{1} \dots 
k^{\gamma^{d}_{1}}_{d} \dots k^{\gamma^{d}_{l_{d}}}_{d} \,\,
\Phi^{i}_{\theta^{\beta}
\theta^{\gamma^{1}_{1}}\dots\theta^{\gamma^{1}_{l_{1}-1}}
\, \dots \, \theta^{\gamma^{d}_{1}}\dots\theta^{\gamma^{d}_{l_{d}}}}
\,\, + \,\, \dots  \\
\left. 
+ \, ( k^{\beta}_{d T} - \omega^{\beta}_{X^{d}} ) \!\! 
\sum_{l_{1}, \dots, l_{d}} \!
\Pi^{\nu (l_{1} \dots l_{d})}_{i \, [0]} \, l_{d} \,
k^{\gamma^{1}_{1}}_{1} \! \dots k^{\gamma^{1}_{l_{1}}}_{1} \dots 
k^{\gamma^{d}_{1}}_{d} \! \dots k^{\gamma^{d}_{l_{d}-1}}_{d} \,
\Phi^{i}_{\theta^{\beta}
\theta^{\gamma^{1}_{1}}\dots\theta^{\gamma^{1}_{l_{1}}} \, \dots \, 
\theta^{\gamma^{d}_{1}}\dots\theta^{\gamma^{d}_{l_{d}-1}}} \!
\right) \!\! {d^{m} \theta \over (2\pi)^{m}} = 0
\end{multline*}
where the values $\zeta^{(\nu)}_{i[{\bf U}({\bf X})]} (\bm{\theta})$
are given by (\ref{VarDer}).

 Consider the convolution (in $\nu$) of the above expression
with the values $\partial k^{\alpha}_{p} / \partial U^{\nu}$.
The expressions
$${\partial k^{\alpha}_{p} \over \partial U^{\nu}} ({\bf U}({\bf X}))
\,\,\, \zeta^{(\nu)}_{i[{\bf U}({\bf X})]} (\bm{\theta}) $$
are identically equal to zero according to Lemma 2.1.

 From the other hand we have
\begin{multline*}
{\partial k^{\alpha}_{p} \over \partial U^{\nu}} \,
\int_{0}^{2\pi} \!\!\!\!\! \dots \int_{0}^{2\pi} 
\sum_{l_{1}, \dots, l_{d}} l_{q} \,
k^{\gamma^{1}_{1}}_{1} \dots k^{\gamma^{1}_{l_{1}}}_{1}
\,\, \dots \,\,
k^{\gamma^{q}_{1}}_{q} \dots k^{\gamma^{q}_{l_{q}-1}}_{q}
\,\, \dots \,\,
k^{\gamma^{d}_{1}}_{d} \dots k^{\gamma^{d}_{l_{d}}}_{d}   
\,\, \times \\
\times \,\, \Phi^{i}_{\theta^{\beta}
\theta^{\gamma^{1}_{1}}\dots\theta^{\gamma^{1}_{l_{1}}} \, \dots \,
\theta^{\gamma^{q}_{1}}\dots\theta^{\gamma^{q}_{l_{q}-1}} \, \dots \,
\theta^{\gamma^{d}_{1}}\dots\theta^{\gamma^{d}_{l_{d}}} } \,\,\,\,\,
\Pi^{\nu (l_{1} \dots l_{d})}_{i \, [0]} 
\,\,\,\,\, {d^{m} \theta \over (2\pi)^{m}} \,\,\, = 
\end{multline*}
\begin{equation}
\label{kalphakbetaUnu}
= \,\, \left. \left( 
{\partial k^{\alpha}_{p} \over \partial U^{\nu}} \,
{\partial \over \partial k^{\beta}_{q}} \,
J^{\nu}[\bm{\varphi}, {\bf k}_{1}, \dots , {\bf k}_{d}] \right)
\right|_{\bm{\varphi}(\bm{\theta}) = \bm{\Phi}(\bm{\theta}, {\bf U})} 
\,\,\,\,\, = \,\,\,\,\,   
\delta^{\alpha}_{\beta} \,\, \delta^{q}_{p}
\end{equation}
since the variations of the functions $\bm{\Phi}$
are insignificant for the values of $k^{\alpha}_{p}$ 
according to Lemma 2.1.

 We get then that conditions (\ref{ConsWhitham}) imply
the relations $k^{\alpha}_{p T} = \omega^{\alpha}_{X^{p}}$, 
which are the second part of system (\ref{RegWhithSyst}). 

 Now the conditions 
$$\int_{0}^{2\pi}\!\!\!\!\!\dots\int_{0}^{2\pi} 
\zeta^{(\nu)}_{i[{\bf U}({\bf X})]} (\bm{\theta}) \, \left[
\Phi^{i}_{T} (\bm{\theta}, {\bf U}({\bf X})) \, - \,
F^{i}_{[1]} (\bm{\theta}, {\bf X}) \right] \, 
{d^{m} \theta \over (2\pi)^{m}} \,\, = \,\, 0 $$
express the conditions of orthogonality of the vectors
(\ref{VarDer}) to the function
$- \bm{\Phi}_{T} + {\bf F}_{[1]}$, which coincides exactly with
the right-hand part of equation (\ref{1syst}) in our case. Since
the linear span of the vectors (\ref{VarDer}) coincides with the
linear span of the complete set of the regular left eigenvectors
of the operator ${\hat L}^{i}_{j[{\bf U}, \bm{\theta}_{0}]}({\bf X},T)$
with zero eigenvalues, we get that system (\ref{ConsWhitham})
is equivalent to system (\ref{RegWhithSyst}) under the constraints
(\ref{KKConstraints}).

{\hfill Lemma 2.2 is proved.}

\vspace{0.2cm}

 Let us note here that it follows from Lemma 2.2 that systems
(\ref{ConsWhitham}), obtained from different sets of conservation laws
are equivalent to each other on the submanifold given by constraints
(\ref{KKConstraints}). In other words, if system 
(\ref{EvInSyst}) has additional conservation laws 
of the form (\ref{Integrals}) then their averaging 
gives relations following from system (\ref{ConsWhitham})
under the additional constraints (\ref{KKConstraints}).

 In this paper we are going to consider multi-dimensional evolution
systems (\ref{EvInSyst}) which are Hamiltonian with respect to a 
local field-theoretic Poisson bracket 
\begin{equation}
\label{MultDimPBr}
\{ \varphi^{i} ({\bf x}) \, , \, \varphi^{j} ({\bf y}) \} \,\, = \,\,
\sum_{l_{1},\dots,l_{d}} B^{ij}_{(l_{1},\dots,l_{d})} 
(\bm{\varphi}, \bm{\varphi}_{\bf x}, \dots ) \,\,
\delta^{(l_{1})} (x^{1} - y^{1}) \, \dots \, 
\delta^{(l_{d})} (x^{d} - y^{d})
\end{equation}
$(l_{1}, \dots, l_{d} \geq 0)$, with a local Hamiltonian of the form
\begin{equation}
\label{MultDimHamFunc}
H \,\, = \,\, \int P_{H} \left(\bm{\varphi}, \bm{\varphi}_{\bf x},
\bm{\varphi}_{\bf xx}, \dots \right) \,\, d^{d} x 
\end{equation}

 As we said already, we are going to consider complete regular 
families $\Lambda$ of $m$-phase solutions of system (\ref{EvInSyst})
satisfying system (\ref{EvPhaseSyst}) with some values of
$(k^{\alpha}_{p} ({\bf U}), \, \omega^{\alpha} ({\bf U}))$. We
assume here that the solutions of the family $\Lambda$ are parametrized
by $m (d + 1) + s$ independent parameters 
$(k^{\alpha}_{p}, \omega^{\alpha}, n^{l})$, $\alpha = 1, \dots, m$,
$p = 1, \dots, d$, $l = 1, \dots, s$ excluding initial phase shifts
$\theta_{0}^{\alpha}$, and we have a set of $m (d + 1) + s$ independent
first integrals $I^{\nu}$ of the form (\ref{Integrals}) such that
their values on $\Lambda$ can be chosen as the coordinate system
$\{ U^{\nu} \}$ on $\Lambda$ (excluding initial phase shifts).
We will assume also, that the integrals $I^{\nu}$ commute with each
other and with the Hamiltonian $H$
\begin{equation}
\label{MultDimInvRel}
\{ I^{\nu} \, , \, I^{\mu} \} \, = \, 0
\,\,\,\,\, , \,\,\,\,\,
\{ I^{\nu} \, , \, H \} \, = \, 0
\end{equation}
according to bracket (\ref{MultDimPBr}).

 Let us note that according to (\ref{MultDimInvRel}) the flows
\begin{equation}
\label{Snuflows}
\varphi^{i}_{t^{\nu}} \,\, = \,\, 
S^{i\nu} (\bm{\varphi}, \bm{\varphi}_{\bf x}, \dots ) \,\, = \,\,
\{ \varphi^{i} ({\bf x}) \, , \, I^{\nu} \}
\end{equation}
generated by the functionals $I^{\nu}$ according to bracket
(\ref{MultDimPBr}) commute with the initial flow (\ref{EvInSyst}).
The flows (\ref{Snuflows}) leave invariant the full families of
$m$-phase solutions of (\ref{EvInSyst}) as well as the values
$U^{\nu} = I^{\nu}$ of the functionals $I^{\nu}$ on them. For
a complete regular family of $m$-phase solutions with independent 
parameters $(k^{1}_{p}, \dots, k^{m}_{p})$ 
it's not difficult to show that the 
flows (\ref{Snuflows}) generate linear (in time) shifts of the phases
$\theta_{0}^{\alpha}$ with some constant frequencies 
$\omega^{\alpha\nu} ({\bf U})$, such that 
\begin{equation}
\label{LinPotSnu}
S^{i\nu} \left( \bm{\Phi}, \,
k^{\beta_{1}}_{1} \, \bm{\Phi}_{\theta^{\beta_{1}}}, \dots \,
k^{\beta_{d}}_{d} \, \bm{\Phi}_{\theta^{\beta_{d}}},
\dots \right) \,\, = \,\, \omega^{\alpha\nu} ({\bf U}) \,
\Phi^{i}_{\theta^{\alpha}} (\bm{\theta}, {\bf U})
\end{equation}

 According to Lemma 2.1 we have also that the functionals
$k^{\alpha}_{p}({\bf I})$ should generate the zero flows on the 
corresponding family of $m$-phase solutions of (\ref{EvInSyst}).
For $U^{\nu}$ coinciding with the values of $I^{\nu}$ on
$\Lambda$ we get then the relations
\begin{equation}
\label{komeganulrel}
{\partial k^{\alpha}_{p} ({\bf U}) \over \partial U^{\nu}} \,\,
\omega^{\beta\nu} ({\bf U}) \,\, \equiv \,\, 0
\,\,\,\,\,\,\,\, , \,\,\,\,\, \alpha, \beta \, = \, 
1, \dots, m, \,\,\, p \, = \, 1, \dots, d
\end{equation}

 By analogy with the one-dimensional case we can try  to construct
an analogue of the Dubrovin - Novikov procedure for $d > 1$. Let us
represent the pairwise Poisson brackets of the densities
$P^{\nu} ({\bf x})$, $P^{\mu} ({\bf y})$ in the form
$$\{ P^{\nu} ({\bf x}) \, , \, P^{\mu} ({\bf y}) \} \,\, = \,\,
\sum_{l_{1},\dots,l_{d}} A^{\nu\mu}_{l_{1} \dots l_{d}}
(\bm{\varphi}, \bm{\varphi}_{\bf x}, \dots ) \,\,
\delta^{(l_{1})} (x^{1} - y^{1}) \, \dots \,
\delta^{(l_{d})} (x^{d} - y^{d}) $$
($l_{1}, \dots, l_{d} \geq 0$).

 According to relations (\ref{MultDimInvRel}) we can also write
here the relations
$$A^{\nu\mu}_{0 \dots 0} (\bm{\varphi}, \bm{\varphi}_{\bf x}, \dots ) 
\,\,\, \equiv \,\,\, \partial_{x^{1}} \, Q^{\nu\mu 1}
(\bm{\varphi}, \bm{\varphi}_{\bf x}, \dots ) \, + \, \dots \, + \,
\partial_{x^{d}} \, Q^{\nu\mu d}
(\bm{\varphi}, \bm{\varphi}_{\bf x}, \dots ) $$
for some functions $(Q^{\nu\mu 1}, \dots, Q^{\nu\mu d})$.

 In the full analogy with the Dubrovin - Novikov procedure we can 
define the expressions
$$\{ U^{\nu} ({\bf X}) \, , \, U^{\mu} ({\bf Y}) \} \,\,\, = \,\,\,
\langle A^{\nu\mu}_{10\dots0} \rangle ({\bf X}) \,\,
\delta^{\prime} (X^{1} - Y^{1}) \, \delta (X^{2} - Y^{2}) \,
\dots \, \delta (X^{d} - Y^{d}) \,\, + \,\, \dots $$
\begin{equation}
\label{MultDimDubrNovForm}
+ \,\, \langle A^{\nu\mu}_{0\dots01} \rangle ({\bf X}) \,\,
\delta (X^{1} - Y^{1}) \, \delta (X^{2} - Y^{2}) \,
\dots \, \delta^{\prime} (X^{d} - Y^{d}) \,\, +
\end{equation}
$$+ \,\, \left( \langle Q^{\nu\mu 1} \rangle_{X^{1}} 
\,\, + \dots + \,\, \langle Q^{\nu\mu d} \rangle_{X^{d}} \right) \,\,
\delta (X^{1} - Y^{1}) \, \dots \, \delta (X^{d} - Y^{d}) $$
which gives a skew-symmetric (contravariant) form on the space of
functions ${\bf U}({\bf X})$.

 The theory of the Poisson brackets having the form
(\ref{MultDimDubrNovForm}) in $d > 1$ dimensions was considered
in \cite{DubrNovDAN84,MokhovMultDimBr1,MokhovMultDimBr2}. 
As was shown in \cite{DubrNovDAN84,MokhovMultDimBr1,MokhovMultDimBr2},
the restrictions on the form of (\ref{MultDimDubrNovForm}) in 
$d > 1$ dimensions are much stronger than in the one-dimensional
case and in general should be considered by the methods of the theory
of integrable systems.

 However, the procedure described above does not give in general 
a Poisson bracket on the space of fields ${\bf U}({\bf X})$ in the
multi-dimensional situation, since
the expression (\ref{MultDimDubrNovForm}) does not satisfy the
Jacobi identity for $d > 1$ in general case. Fortunately, we don't
need a Poisson bracket on the full space of fields ${\bf U}({\bf X})$
for the Hamiltonian formulation of the Whitham approach since the
regular Whitham system is defined on the submanifold in the space of
$\{ {\bf U}({\bf X}) \}$ given by constraints (\ref{KKConstraints}).
As we will show in the next chapters, expressions 
(\ref{MultDimDubrNovForm}) define indeed a Poisson structure after
the restriction on the submanifold mentioned above which gives a
Hamiltonian structure for the regular Whitham system for $d > 1$.

 For the description of the Hamiltonian structure of the regular 
Whitham system for $d > 1$ it is convenient to introduce a
coordinate system on the submanifold corresponding to the Whitham
solutions. It is most natural to consider then the functions
$S^{\alpha} ({\bf X})$, $\alpha = 1, \dots, m$ such that
$S^{\alpha}_{X^{p}} ({\bf X}) \, \equiv \, k^{\alpha}_{p} ({\bf X})$
as a part of a coordinate system on the submanifold given by 
constraints (\ref{KKConstraints}). However, the ${\bf X}$-derivatives
of the functions $S^{\alpha} ({\bf X})$ give just $m d$ independent
parameters in the space ${\bf U}({\bf X})$ at every given ${\bf X}$.
To get the full system of coordinates on the submanifold defined by
constraints (\ref{KKConstraints}) we need additional $m + s$
parameters at every point ${\bf X}$. 
It will be convenient for us here to
take arbitrary $m + s$ parameters $U^{\gamma} ({\bf X})$,
$\gamma = 1, \dots, m + s$ from the set 
$\{ {\bf U}({\bf X}) \}$ which are 
functionally independent with the set $\{k^{\alpha}_{p} ({\bf U})\}$.

 Let us note that we put here 
$\gamma = 1, \dots, m + s$ without any loss
of generality. As a necessary condition of the functional independence
of the set $\{ k^{\alpha}_{p} ({\bf U}), U^{1}, \dots, U^{m+s} \}$
we have, in particular, that the variation derivatives (\ref{VarDer})
of the corresponding functionals $(J^{1}, \dots, J^{m+s})$ give a
linearly independent set on the family $\Lambda$. As a consequence,
we can claim that the regular left eigen-vectors 
$\bm{\kappa}^{(q)}_{[{\bf U}]}(\bm{\theta} + \bm{\theta}_{0})$ of the
operator ${\hat L}^{i}_{j[{\bf U}, \bm{\theta}_{0}]}$ can be also
expressed as linear combinations of the vectors 
$\bm{\zeta}^{(\gamma)}_{[{\bf U}]} (\bm{\theta} + \bm{\theta}_{0})$,
$\gamma = 1, \dots, m + s$, such that we have
\begin{equation}
\label{kappaexpNewParam}
\kappa^{(q)}_{i [{\bf U}]} (\bm{\theta} + \bm{\theta}_{0})
\,\, = \,\, \sum_{\gamma=1}^{m+s} \, d^{(q)}_{\gamma} ({\bf U}) \,\,
\zeta^{(\gamma)}_{i [{\bf U}]} (\bm{\theta} + \bm{\theta}_{0})
\end{equation}
with some functions $d^{(q)}_{\gamma} ({\bf U})$ on $\Lambda$.

 Representing the frequencies $\omega^{\alpha} ({\bf U})$ as 
functions of the new parameters 
$$\omega^{\alpha} \,\, = \,\, \omega^{\alpha} \, 
\left({\bf S}_{\bf X}, \, U^{1}, \dots, U^{m+s} \right)$$
we can write the second
part of the evolution system (\ref{RegWhithSyst}) together with the
constraints (\ref{InConstraints}) in the form
\begin{equation}
\label{STOmegaRel}
S^{\alpha}_{T} \,\,\, = \,\,\, \omega^{\alpha}
\left({\bf S}_{\bf X}, \, U^{1}, \dots, U^{m+s} \right)
\end{equation}

 The remaining part of the regular Whitham system can be written
as the relations
\begin{equation}
\label{AddPartWhithSyst}
U^{\gamma}_{T} \,\,\, = \,\,\, \langle Q^{\gamma 1} \rangle_{X^{1}} 
\,\, + \,\, \dots \,\, + \,\, \langle Q^{\gamma d} \rangle_{X^{d}}
\,\,\,\,\, , \,\,\,\,\,\,\,\, \gamma = 1, \dots, m + s
\end{equation}
where 
$\langle Q^{\gamma p} \rangle \, = \, \langle Q^{\gamma p} \rangle
({\bf S}_{\bf X}, \, U^{1}, \dots, U^{m+s})$.

 As follows from our considerations above the choice of the 
functionally independent parameters 
$U^{\gamma}$, $\gamma = 1, \dots, m + s$
(and the corresponding choice of the functionals $I^{\gamma}$) is
unessential for the construction.

 We will show below, that the regular Whitham system given by 
equations (\ref{STOmegaRel}) - (\ref{AddPartWhithSyst}) is 
Hamiltonian with respect to the Poisson bracket given by the 
relations
$$\{ S^{\alpha} ({\bf X}) \, , \, S^{\beta} ({\bf Y}) \} 
\, = \, 0 \,\,\, , \,\,\,\,\,
\{ S^{\alpha} ({\bf X}) \, , \, U^{\gamma} ({\bf Y}) \} \, = \, 
\omega^{\alpha\gamma} \, 
\left({\bf S}_{\bf X}, U^{1}({\bf X}), \dots, U^{m+s}({\bf X})  
\right) \, \delta ({\bf X} - {\bf Y}) $$
$\alpha, \beta =  1, \dots, m , \, \gamma = 1, \dots, m + s$, and 
relations (\ref{MultDimDubrNovForm}) for the skew-symmetric form
$\{ U^{\gamma} ({\bf X}) \, , \, U^{\rho} ({\bf Y}) \}$ restricted
to the subset $\gamma, \rho = 1, \dots, m + s$, where
$$\langle A^{\gamma\rho}_{0\dots1\dots0} \rangle \,\, = \,\,
\langle A^{\gamma\rho}_{0\dots1\dots0} \rangle \,
\left({\bf S}_{\bf X}, \, U^{1}, \dots, U^{m+s} \right)
\,\,\,\,\, , \,\,\,\,\, 
\langle Q^{\gamma\rho p} \rangle \, = \,
\langle Q^{\gamma\rho p} \rangle \,
\left({\bf S}_{\bf X}, \, U^{1}, \dots, U^{m+s} \right) $$

 The Hamiltonian function for the regular Whitham system is also 
local in this case and is equal to
$$H \,\,\, = \,\,\, \int \langle P_{H} \rangle \,
\left({\bf S}_{\bf X}, \, U^{1}({\bf X}), \dots, U^{m+s}({\bf X})
\right) \,\, d^{d} X $$

 Let us note here that although system 
(\ref{STOmegaRel}) - (\ref{AddPartWhithSyst}) as well as the 
Hamiltonian structure described above can be formally defined with
the aid of the $m + s$ additional integrals $I^{\gamma}$, the presence
of the full set $\{ I^{\nu}, \, \nu = 1, \dots, m (d + 1) + s \}$
giving the full set of parameters ${\bf U}$ on the family $\Lambda$ 
is important for the justification of the construction 
of the Hamiltonian structure in our scheme. 
The rest of equations (\ref{ConsWhitham}) gives in this
situation additional conservation laws for the regular Whitham system.
Let us note also, that the choice of the functionals
$\{ I^{\gamma}, \, \gamma = 1, \dots, m + s \}$ 
is also not important for the construction of the Hamiltonian structure 
of the regular Whitham system as well as for the Whitham system itself.
Namely, the Hamiltonian structures obtained with the aid of
different subsets 
$\{ I^{\gamma}, \, \gamma = 1, \dots, m + s \}$ transform
into each other after the corresponding change of coordinates.

 In the next chapter we start the consideration of the averaging
procedure for the bracket (\ref{MultDimPBr}) giving the Hamiltonian
structure for the regular Whitham system 
(\ref{STOmegaRel}) - (\ref{AddPartWhithSyst}).

\section{The averaging of the Poisson brackets.}
\setcounter{equation}{0}

 Let us start with the definition of a regular Hamiltonian
family of $m$-phase solutions of system (\ref{EvInSyst}) and
of a complete Hamiltonian set of the functionals (\ref{Integrals})
in the multi-dimensional case.

\vspace{0.2cm}

{\bf Definition 3.1.}

{\it We call family $\Lambda$ of $m$-phase solutions of system 
(\ref{EvInSyst}) a regular Hamiltonian family if:

1) It represents a complete regular family of $m$-phase solutions
of (\ref{EvInSyst}) in the sense of Definition 1.1;

2) The corresponding bracket (\ref{MultDimPBr}) has on $\Lambda$
constant number of annihilators $N^{1}$, $\dots$, $N^{s}$ 
with linearly independent variation derivatives
$\delta N^{l} / \delta \varphi^{i} ({\bf x})$ which
coincides with the number of independent annihilators in the 
neighborhood of $\Lambda$.
}

\vspace{0.2cm}

 According to the generalized Darboux theorem 
we can identify the number of the variation derivatives 
$\delta N^{l} / \delta \varphi^{i} ({\bf x})$ on $\Lambda$ with the
number of linearly independent quasiperiodic solutions
$v^{(l)}_{i}({\bf x})$ of the equation
$$\sum_{l_{1},\dots,l_{d}} \left.
B^{ij}_{(l_{1},\dots,l_{d})} (\bm{\varphi}, 
\bm{\varphi}_{\bf x}, \dots) 
\right|_{\Lambda} v^{(l)}_{j, \, l_{1} x^{1} \dots \, l_{d} x^{d}}
({\bf x}) \,\,\, = \,\,\, 0 $$
where $v^{(l)}_{i}({\bf x})$ have the same wave numbers as 
the corresponding functions $\bm{\varphi}({\bf x})$ on $\Lambda$.

\vspace{0.2cm}

{\bf Definition 3.2.}

{\it We call a set $(I^{1}, \dots, I^{N})$ of commuting functionals
(\ref{Integrals}) a complete Hamiltonian set on a regular Hamiltonian 
family $\Lambda$ of $m$-phase solutions of system (\ref{EvInSyst})
if:

1) The restriction of the functionals $(I^{1}, \dots, I^{N})$ 
on the quasiperiodic solutions of the family $\Lambda$ gives a complete
set of parameters $(U^{1}, \dots, U^{N})$ on this family;

2) The Hamiltonian flows generated by $(I^{1}, \dots, I^{N})$
generate on $\Lambda$ linear phase shifts of $\bm{\theta}_{0}$
with frequencies $\bm{\omega}^{\nu} ({\bf U})$, such that
$${\rm rk} \,\, || \omega^{\alpha \nu} ({\bf U}) || \,\, = \,\, m $$

3) The linear space generated by the variation derivatives
$\delta I^{\nu} / \delta \varphi^{i} ({\bf x})$ on $\Lambda$ contains
the variation derivatives of all the annihilators $N^{q}$ of the
bracket (\ref{MultDimPBr}), such that
$$\left. {\delta N^{l} \over \delta \varphi^{i} ({\bf x})} 
\right|_{\Lambda} \,\, = \,\,
\sum_{\nu=1}^{N} \gamma^{l}_{\nu} ({\bf U}) \,\, \left.
{\delta I^{\nu} \over \delta \varphi^{i} ({\bf x})} \right|_{\Lambda} $$
for some smooth functions $\gamma^{l}_{\nu} ({\bf U})$ on the
family $\Lambda$.
}

\vspace{0.2cm}

 Let us note that it follows from Definition 3.2 that if a complete
Hamiltonian set of integrals $(I^{1}, \dots, I^{N})$ exists for
a regular Hamiltonian family $\Lambda$ then the number of the additional
parameters $(n^{1}, \dots, n^{s})$ discussed above is equal to the
number of annihilators of the bracket (\ref{MultDimPBr}).
Indeed, according to Definitions 3.1 and 3.2, the number of the functionals
$I^{\nu}$ having linearly independent variation derivatives on
$\Lambda$ exactly equals to $m + s$, where $s$ is the
number of annihilators of the bracket (\ref{MultDimPBr}).
The total number of independent parameters $U^{\nu}$ on $\Lambda$ 
is then equal to $m (d + 1) + s$ due to the wave 
vectors $k^{\alpha}_{p}$, $\alpha = 1, \dots, m$, 
$p = 1, \dots, d$ which implies the above assertion.

 As follows from condition (2) of Definition 3.2
and from the invariance of the functionals $N^{l}$ and $I^{\nu}$ 
with respect to the flows (\ref{Snuflows}), the values 
$\gamma^{l}_{\nu} ({\bf U})$ can always be chosen independent
from the initial phase shifts on the family $\Lambda$. The values
$\delta I^{\nu} / \delta \varphi^{i}({\bf x}) |_{\Lambda}$ are 
linearly dependent on $\Lambda$, so it's natural to choose a
complete linearly independent subsystem. Remembering that the 
variation derivatives of $J^{\nu}$ (\ref{VarDer}) are linear 
combinations of the regular left eigenvectors 
$\kappa^{(q)}_{i[{\bf U}]} (\bm{\theta} + \bm{\theta}_{0})$,
we can write
\begin{equation}
\label{RazlozhAnn}
\left. {\delta N^{l} \over \delta \varphi^{i}({\bf x})} 
\right|_{\bm{\varphi}({\bf x}) = \bm{\Phi} 
({\bf k}_{p}({\bf U}) x^{p}  + \bm{\theta}_{0}, {\bf U})}
\,\,\,\,\, = \,\,\,\,\, \sum_{q=1}^{m+s} \, n^{l}_{q} ({\bf U}) \,\,
\kappa^{(q)}_{i[{\bf U}]} \left({\bf k}_{p}({\bf U}) \, x^{p} 
\, + \, \bm{\theta}_{0} \right) 
\end{equation}
for some smooth functions $n^{l}_{q} ({\bf U})$. The functions
$n^{l}_{q} ({\bf U})$ are then uniquely determined on
$\Lambda$ and we have
$\text{rk} \, ||n^{l}_{q} ({\bf U})|| = s$ by 
Definition 3.1.

 Let us say that the full linearly independent subsystem
of the regular left eigen-vectors of the operator
${\hat L}^{i}_{j [{\bf U}, \bm{\theta}_{0}]}$ with zero eigen-values
can be given also by the variation derivatives of the functionals
$I^{1}, \dots, I^{m+s}$ on $\Lambda$, which give exactly $m + s$
additional parameters $U^{1}, \dots, U^{m+s}$ to the parameters
$k^{\alpha}_{p}$. Relations (\ref{kappaexpNewParam}) give the 
connection between the eigenvectors 
$\kappa^{(q)}_{i[{\bf U}]} (\bm{\theta} + \bm{\theta}_{0})$ and
$\zeta^{(\gamma)}_{i[{\bf U}]} (\bm{\theta} + \bm{\theta}_{0})$,
$\gamma = 1, \dots, m + s$.

 Easy to see then that for a complete Hamiltonian set of integrals
$I^{1}, \dots, I^{N}$ we have also the relations
\begin{equation}
\label{RankOmegaAlphaGamma}
{\rm rk} \,\, ||\omega^{\alpha\gamma} ({\bf U}) || \,\, = \,\, m
\,\, , \,\,\,\,\,\,\,\, \alpha = 1, \dots, m \,\, , \,\,\,\,\,
\gamma = 1, \dots, m + s
\end{equation}
for the frequencies corresponding to the functionals
$I^{1}, \dots, I^{m+s}$.

 Let us now describe the procedure, which will be considered
in our situation. We consider now system (\ref{EvInSyst}) which is 
Hamiltonian with respect to some local bracket (\ref{MultDimPBr})
with a local Hamiltonian function of the form (\ref{MultDimHamFunc}).
We first introduce the extended space of fields
$$\bm{\varphi}({\bf x}) \,\, \rightarrow \,\, 
\bm{\varphi}(\bm{\theta}, {\bf x}) $$
where the functions $\bm{\varphi}(\bm{\theta}, {\bf x})$ 
are $2\pi$-periodic with respect to each $\theta^{\alpha}$, 
and define the extended Poisson bracket:
$$\{\varphi^{i}(\bm{\theta}, {\bf x}) \, , \, 
\varphi^{j}(\bm{\theta}^{\prime}, {\bf y})\} 
\, = \, \sum_{l_{1},\dots,l_{d}}
B^{ij}_{(l_{1},\dots,l_{d})} 
(\bm{\varphi}, \bm{\varphi}_{\bf x}, \dots)      
\,\, \delta^{(l_{1})}(x^{1} - y^{1}) \dots 
\delta^{(l_{d})}(x^{d} - y^{d}) 
\,\, \delta (\bm{\theta} - \bm{\theta}^{\prime})$$

 Let us then make the replacement 
${\bf x} \rightarrow {\bf X} = \epsilon {\bf x}$ and introduce
the Poisson bracket
\begin{multline}   
\label{EpsExtBracket}
\{\varphi^{i}(\bm{\theta}, {\bf X}) \, , \, 
\varphi^{j}(\bm{\theta}^{\prime}, {\bf Y})\}
\, = \\
= \, \sum_{l_{1},\dots,l_{d}}  \!
\epsilon^{l_{1} + \dots + l_{d}} \,
B^{ij}_{(l_{1},\dots,l_{d})} 
(\bm{\varphi}, \, \epsilon\, \bm{\varphi}_{\bf X}, \dots)
\,\, \delta^{(l_{1})}(X^{1} - Y^{1})  \dots 
\delta^{(l_{d})}(X^{d} - Y^{d})  
\, \delta (\bm{\theta} - \bm{\theta}^{\prime})
\end{multline}
on the space of fields $\bm{\varphi} (\bm{\theta}, {\bf X})$.

 We describe now the submanifold ${\cal K}$ in the space of the
functions $\bm{\varphi} (\bm{\theta}, {\bf X})$ which we are going to
consider.

 First, we will assume that the functions 
$\bm{\varphi}(\bm{\theta}, {\bf X}) \in {\cal K}$ represent
functions from the family $\Lambda$
of the $m$-phase solutions of (\ref{EvInSyst})
with some parameters ${\bf U}({\bf X})$ at every ${\bf X}$.

 Second, we impose the relations
\begin{equation}
\label{KKConstraintsPovt}
{\partial k^{\alpha}_{p} ({\bf U} ({\bf X})) \over \partial X^{l}}
\,\,\, = \,\,\,
{\partial k^{\alpha}_{l} ({\bf U} ({\bf X})) \over \partial X^{p}}
\,\,\,\,\, , \,\,\,\,\,\,\,\, \alpha = 1, \dots, m \,\,\, , \,\,\,
p, l = 1, \dots, d 
\end{equation}
for the functions ${\bf U}({\bf X})$ parametrizing the solutions
$\bm{\varphi}(\bm{\theta}, {\bf X}) \in \Lambda$ from the
submanifold ${\cal K}$.

 More precisely, let us choose for simplicity the boundary conditions
in the form ${\bf U} (X^{1}, 0, \dots , 0) \rightarrow {\bf U}_{0}$,
$X^{1} \rightarrow - \infty$, such that 
$k^{\alpha}_{1} ({\bf U}_{0}) = 0$ and define the functions 
$S^{\alpha}({\bf X})$ on ${\cal K}$ by the formula:
\begin{equation}
\label{DefinitionS}
S^{\alpha}({\bf X}) \,\,\, = \,\,\, \int_{-\infty}^{X^{1}}
k^{\alpha}_{1} (X^{1 \prime}, 0, \dots, 0) \, d X^{1 \prime} \,\,\, + 
\end{equation}
$$\,\, + \int_{0}^{X^{2}} k^{\alpha}_{2}
(X^{1}, X^{2 \prime}, 0, \dots, 0) \, d X^{2 \prime} \,\, + \,\,
\dots \,\, + \,\, \int_{0}^{X^{d}} k^{\alpha}_{d}
(X^{1}, \dots, X^{d-1}, X^{d \prime}) \, d X^{d \prime} $$

 We define then the functions 
$\bm{\varphi}(\bm{\theta}, {\bf X}) \in {\cal K}$ by the formula
\begin{equation}
\label{FormaPhinaK}
\varphi^{i} (\bm{\theta}, {\bf X}) \,\, = \,\, \Phi^{i} \left(
{{\bf S} ({\bf X}) \over \epsilon} 
\, + \, \bm{\theta}, \, {\bf S}_{\bf X}, 
U^{1}({\bf X}), \dots , U^{m+s}({\bf X}) \right)
\end{equation}
where the functions 
$k^{\alpha}_{p} ({\bf X}) \, = \, S^{\alpha}_{X^{p}} ({\bf X})$,
and the additional parameters $(U^{1}, \dots , U^{m+s})$, introduced
in the previous chapter, play now the role of parameters on the
family $\Lambda$.

 The functions 
$\{ S^{\alpha}({\bf X}), U^{1}({\bf X}), \dots , 
U^{m+s}({\bf X}) \} $ play the role of a coordinate system on the
submanifold ${\cal K}$. We have now to introduce the analogous 
coordinates in the vicinity of the submanifold ${\cal K}$.

 Let us introduce the functionals $J^{\nu}({\bf X})$ on the space 
of functions $\bm{\varphi} (\bm{\theta}, {\bf X})$ by the formula
\begin{equation}
\label{FuncJnuX}
J^{\nu}({\bf X}) \,\, = \,\, 
\int_{0}^{2\pi}\!\!\!\!\!\dots\int_{0}^{2\pi}
P^{\nu} (\bm{\varphi}, \, \epsilon \bm{\varphi}_{\bf X}, \, 
\epsilon^{2} \bm{\varphi}_{\bf XX}, \dots ) \,\, 
{d^{m} \theta \over (2\pi)^{m}} \,\,\,\,\,\,\,\, , \,\,\,\,\,
\nu = 1, \dots, N
\end{equation}
and consider their values on the functions of the family ${\cal K}$.

 We can write on ${\cal K}$:
\begin{equation}
\label{JUtransform}
J^{\nu}({\bf X}) \,\, = \,\, U^{\nu}({\bf X}) \, + \, \sum_{l\geq1}
\epsilon^{l} \, J^{\nu}_{(l)}({\bf X}) \,\,\,\,\,\,\,\, , \,\,\,\,\,
\nu = 1, \dots, N 
\end{equation}
where $J^{\nu}_{(l)}({\bf X})$ - are polynomials in the derivatives
${\bf U}_{\bf X}$, ${\bf U}_{\bf XX}$, $\dots$ 
with coefficients depending on
${\bf U}$ and have grading degree $l$ in terms of total
number of derivations with respect to ${\bf X}$.

 The higher terms in (\ref{JUtransform}) are not uniquely defined
on ${\cal K}$ due to relations (\ref{KKConstraintsPovt}). Let us
assume here that the terms $J^{\nu}_{(l)}({\bf X})$ are chosen in
some definite way in every order $l \geq 1$. The corresponding choice
will affect the definition of the functionals ${\bf U} ({\bf X})$
in the vicinity of ${\cal K}$ in the higher orders in $\epsilon$
($l \geq 1$). As we will see, this choice will not be important
in our further considerations.

 The transformation (\ref{JUtransform}) can then be inverted as 
a formal series in $\epsilon$, such that we can write
\begin{equation}
\label{UJtransform}
U^{\nu}({\bf X}) \,\, = \,\, J^{\nu}({\bf X}) \, + \, \sum_{l\geq1}
\epsilon^{l} \, U^{\nu}_{(l)}({\bf X}) \,\,\,\,\,\,\,\, , \,\,\,\,\,
\nu = 1, \dots, N  
\end{equation}
on the functions of the submanifold ${\cal K}$. 
In formula (\ref{UJtransform}) the functions $U^{\nu}_{(l)}$ 
are functions of
${\bf J}$, ${\bf J}_{\bf X}$, ${\bf J}_{\bf XX}$, $\dots$, 
polynomial in the derivatives 
${\bf J}_{\bf X}$, ${\bf J}_{\bf XX}$, $\dots$, and 
having degree $l$ in terms of the number of derivations w.r.t. 
${\bf X}$.

 We can define now the functionals ${\bf U}({\bf X})$ on the
whole functional space using the definition of the functionals
${\bf J}({\bf X})$ and relations (\ref{UJtransform}).

 Let us put now the same boundary conditions for the functionals
$k^{\alpha}_{1} ({\bf U})$ as before on the whole functional space.
We can then consider also the functionals
$S^{\alpha}[{\bf U}]({\bf X})$, given by (\ref{DefinitionS}),
as the functionals defined in the vicinity 
of the submanifold ${\cal K}$ using
the corresponding definition of the functionals ${\bf U}({\bf X})$.
On the submanifold ${\cal K}$ we will naturally have the relations
$$S^{\alpha}_{X^{p}} [{\bf U}] ({\bf X}) \,\, = \,\,
k^{\alpha}_{p} ({\bf U}({\bf X})) $$

 However, outside the submanifold ${\cal K}$ these relations are
in general not true.

 Let us introduce also the constraints
$g^{i}(\bm{\theta}, {\bf X})$
defining the submanifold ${\cal K}$ 
by the conditions $g^{i}(\bm{\theta}, {\bf X}) = 0$,
and numbered by the values of $\bm{\theta}$ and ${\bf X}$:
\begin{equation}
\label{gconstr}
g^{i} (\bm{\theta}, {\bf X}) \,\,\, = \,\,\, 
\varphi^{i} (\bm{\theta}, {\bf X}) 
\,\, - \,\, \Phi^{i} \left( 
{{\bf S} [{\bf U}[{\bf J}]]({\bf X}) \over \epsilon} 
\, + \, \bm{\theta}, \,\, S^{\alpha}_{\bf X}[{\bf U}[{\bf J}]], \,
U^{1} [{\bf J}] ({\bf X}), \, \dots , \, U^{m+s} [{\bf J}] ({\bf X}) 
\right)
\end{equation}

 The constraints 
$g^{i}(\bm{\theta}, {\bf X})$
are functionals on the whole extended space of fields
$\varphi^{i} (\bm{\theta}, {\bf X})$ by virtue of the
corresponding definition of the functionals
$J^{\nu} ({\bf X})$.

 The constraints (\ref{gconstr}) are not independent since 
the following relations hold identically for the ``gradients''
$\delta g^{i} (\bm{\theta}, {\bf X}) / \delta \varphi^{j} 
(\bm{\theta}^{\prime}, {\bf Y})$ on the submanifold ${\cal K}$:
\begin{equation}
\label{dependence}
\int \int_{0}^{2\pi}\!\!\!\!\!\dots\int_{0}^{2\pi} 
\left. {\delta S^{\alpha} ({\bf Z}) \over 
\delta \varphi^{i} (\bm{\theta}, {\bf X})}\right|_{\cal K} 
\,\, \left. {\delta g^{i} (\bm{\theta}, {\bf X}) \over  
\delta \varphi^{j} (\bm{\theta}^{\prime}, {\bf Y})}\right|_{\cal K} 
\,\, {d^{m} \theta \over (2\pi)^{m}} \,\, d^{d} X \,\, \equiv \,\, 0
\,\,\,\,\,\,\,\, , \,\,\,\,\, \alpha = 1, \dots, m 
\end{equation}
\begin{equation}
\label{dependence2}
\int \int_{0}^{2\pi}\!\!\!\!\!\dots\int_{0}^{2\pi}
\left. {\delta U^{\gamma} ({\bf Z}) \over
\delta \varphi^{i} (\bm{\theta}, {\bf X})}\right|_{\cal K}
\,\, \left. {\delta g^{i} (\bm{\theta}, {\bf X}) \over
\delta \varphi^{j} (\bm{\theta}^{\prime}, {\bf Y})}\right|_{\cal K}
\,\, {d^{m} \theta \over (2\pi)^{m}} \,\, d^{d} X \,\, \equiv \,\, 0
\,\,\,\,\,\,\,\, , \,\,\,\,\, \gamma = 1, \dots, m + s
\end{equation}

 For our purposes here it will be not necessary to choose in fact 
an independent subsystem from (\ref{gconstr}), so we keep the system of 
constraints in the form (\ref{gconstr}) keeping in mind the existence 
of identities (\ref{dependence}) - (\ref{dependence2}).

 Thus, we consider now the values of the functionals
$[S^{\alpha} ({\bf X}), U^{1}({\bf X}), \dots , U^{m+s}({\bf X}),
g^{i} (\bm{\theta}, {\bf X})]$ 
as a coordinate system
in the neighborhood of ${\cal K}$ with the relations 
(\ref{dependence}) - (\ref{dependence2}). The values of the 
functionals 
$[S^{\alpha} ({\bf X}), U^{1}({\bf X}), \dots , U^{m+s}({\bf X})]$ 
will be considered as a coordinate system on ${\cal K}$.

 For the procedure of the averaging of bracket (\ref{MultDimPBr}) 
we will need to find the pairwise brackets on ${\cal K}$
of the functionals, introduced above, according to bracket
(\ref{EpsExtBracket}).

 The pairwise Poisson brackets of the functionals
$J^{\nu} ({\bf X})$, $J^{\mu} ({\bf Y})$
have the form
\begin{multline*}
\left\{ J^{\nu} ({\bf X}) \, , \, J^{\mu} ({\bf Y}) \right\} 
\,\, = \,\, \sum_{l_{1},\dots,l_{d}} 
\int_{0}^{2\pi}\!\!\!\!\!\dots \int_{0}^{2\pi}
A^{\nu\mu}_{l_{1} \dots l_{d}} (\bm{\varphi} (\bm{\theta}, {\bf X}), 
\epsilon \bm{\varphi}_{\bf X} (\bm{\theta}, {\bf X}), \dots ) \,\,
{d^{m} \theta \over (2\pi)^{m}} \,\, \times \\
\times \,\, \epsilon^{l_{1} + \dots + l_{d}} 
\,\, \delta^{(l_{1})} (X^{1} - Y^{1}) \, \dots \,
\delta^{(l_{d})} (X^{d} - Y^{d})
\end{multline*}
where 
\begin{equation}
\label{epsAQRel}
A^{\nu\mu}_{0 \dots 0} 
(\bm{\varphi}, \epsilon \bm{\varphi}_{\bf X}, \dots)
\,\, \equiv \,\, \epsilon \, \partial_{X^{1}} \, 
Q^{\nu\mu 1} (\bm{\varphi}, \epsilon \bm{\varphi}_{\bf X}, \dots)
\, + \, \dots \, + \,
\epsilon \, \partial_{X^{d}} \, 
Q^{\nu\mu d} (\bm{\varphi}, \epsilon \bm{\varphi}_{\bf X}, \dots)
\end{equation}

 The Poisson brackets of the fields 
$\varphi^{i} (\bm{\theta}, {\bf X})$ 
with the functionals $J^{\mu}({\bf Y})$ can be written as
\begin{multline}
\label{phiJmuexp}
\{ \varphi^{i} (\bm{\theta}, {\bf X}) , J^{\mu}({\bf Y})\} \,\, = \\
= \,\, \sum_{l_{1},\dots,l_{d}} 
\epsilon^{l_{1} + \dots + l_{d}} \, C^{i\mu}_{(l_{1} \dots l_{d})}
\left( \bm{\varphi} (\bm{\theta}, {\bf X}), \epsilon \,
\bm{\varphi}_{\bf X} (\bm{\theta}, {\bf X}), \dots \right) \,\,
\delta^{(l_{1})} (X^{1} - Y^{1}) \dots \,
\delta^{(l_{d})} (X^{d} - Y^{d})
\end{multline}
with some smooth functions
$C^{i\mu}_{(l_{1} \dots l_{d})} 
(\bm{\varphi}, \epsilon \, \bm{\varphi}_{\bf X}, \dots )$.

 We also have in this case
\begin{equation}
\label{C0SmuRel}
C^{i\mu}_{(0 \dots 0)} \left( 
\bm{\varphi} (\bm{\theta}, {\bf X}), \epsilon \,
\bm{\varphi}_{\bf X} (\bm{\theta}, {\bf X}), \dots \right) 
\,\,\, \equiv \,\,\,
S^{i\mu} \left( \bm{\varphi} (\bm{\theta}, {\bf X}), \epsilon \,
\bm{\varphi}_{\bf X} (\bm{\theta}, {\bf X}), \dots \right)
\end{equation}
by virtue of (\ref{Snuflows}).

 For any function of slow variables $q ({\bf Y})$ we can write
\begin{multline}
\label{phiJq}
\left\{ \varphi^{i} (\bm{\theta}, {\bf X}) , 
\int q({\bf Y}) \, J^{\mu}({\bf Y})\, d^{d} Y \right\} \,\, = \\
= \,\, \sum_{l_{1},\dots,l_{d}} \epsilon^{l_{1} + \dots + l_{d}} \, 
C^{i\mu}_{(l_{1} \dots l_{d})} 
\left( \bm{\varphi} (\bm{\theta}, {\bf X}), \epsilon \,
\bm{\varphi}_{\bf X} (\bm{\theta}, {\bf X}), \dots \right) 
\,\, q_{l_{1}X^{1} \dots l_{d}X^{d}}
\end{multline}

 The leading term in expression (\ref{phiJq}) on ${\cal K}$ 
has the form
$$\left. \left\{ \varphi^{i} (\bm{\theta}, {\bf X}) ,
\int q({\bf Y}) \, J^{\mu}({\bf Y})\, d^{d} Y 
\right\}\right|_{{\cal K} [0]}   \, = \,\,
C^{i\mu}_{(0)} \left(  \bm{\Phi} \left( \!
{{\bf S}({\bf X}) \over \epsilon} + \bm{\theta}, {\bf X} \right),
\dots  \right)  \, q ({\bf X})  \,\, =$$
$$= \,\, q({\bf X}) \,\, S^{i\mu} \left( \bm{\Phi} \left(
{{\bf S}({\bf X}) \over \epsilon} + \bm{\theta}, {\bf X} \right),
\dots \right) $$
where $S^{i\mu} (\bm{\varphi}, \bm{\varphi}_{\bf x}, \dots)$ 
is the flow (\ref{Snuflows}) generated by the functional $I^{\mu}$.
According to (\ref{LinPotSnu}) we can write then:
\begin{equation}
\label{phiJbr}
\left. \left\{ 
\varphi^{i} (\bm{\theta}, {\bf X}) , \int q({\bf Y}) \, 
J^{\mu}({\bf Y}) \, d^{d} Y
\right\}\right|_{\cal K}
\,\, = \,\, \omega^{\alpha \mu}({\bf X}) \,\,
\Phi^{i}_{\theta^{\alpha}} \left(
{{\bf S}({\bf X}) \over \epsilon} + \bm{\theta}, 
{\bf U} ({\bf X}) \right)
\,\, q ({\bf X}) \,\, + \,\,  O (\epsilon)
\end{equation}

 Similarly, for any smooth function $Q (\bm{\theta}, {\bf X})$,
$2\pi$-periodic in each $\theta^{\alpha}$, 
we can write on the basis of (\ref{phiJmuexp})
\begin{equation}
\label{QphiJbr}
\int_{0}^{2\pi}\!\!\!\!\!\dots\int_{0}^{2\pi}
Q \left( {{\bf S}({\bf X}) \over \epsilon} + \bm{\theta}, {\bf X} 
\right) \, \left. \left\{ 
\varphi^{i}(\bm{\theta}, {\bf X}), J^{\mu} ({\bf Y}) \right\}
\right|_{\cal K} \,\, {d^{m} \theta \over (2\pi)^{m}} \,\, = 
\end{equation}
$$  = \,\, \omega^{\alpha\mu}({\bf X}) 
\int_{0}^{2\pi}\!\!\!\!\!\dots\int_{0}^{2\pi} \, 
Q (\bm{\theta}, {\bf X}) \,\, 
\Phi^{i}_{\theta^{\alpha}} (\bm{\theta}, {\bf U}({\bf X})) \,\,
{d^{m} \theta \over (2\pi)^{m}} \,\,\, \delta ({\bf X} - {\bf Y})
\,\,\, + \,\,\, O (\epsilon) $$

 In view of relations (\ref{JUtransform}) - (\ref{UJtransform}) for 
the functionals $J^{\mu}({\bf Y})$, $U^{\mu}({\bf Y})$ we can also 
write
\begin{equation}
\label{phiUbr}
\left. \left\{
\varphi^{i} (\bm{\theta}, {\bf X}) , 
\int q({\bf Y}) \, U^{\mu}({\bf Y}) \, d^{d} Y
\right\}\right|_{\cal K}
\,\, = \,\, \omega^{\alpha \mu}({\bf X}) \,\,
\Phi^{i}_{\theta^{\alpha}} \left(
{{\bf S}({\bf X}) \over \epsilon} + \bm{\theta}, 
{\bf U} ({\bf X}) \right)
\,\, q ({\bf X}) \,\, + \,\,  O (\epsilon)
\end{equation}
\begin{equation}
\label{QphiUbr}
\int_{0}^{2\pi}\!\!\!\!\!\dots\int_{0}^{2\pi}
Q \left( {{\bf S}({\bf X}) \over \epsilon} + \bm{\theta}, 
{\bf X} \right) \,
\left. \left\{ \varphi^{i}(\bm{\theta}, {\bf X}), 
U^{\mu} ({\bf Y}) \right\}
\right|_{\cal K} \,\, {d^{m} \theta \over (2\pi)^{m}} \,\, =
\end{equation}
$$  = \,\, \omega^{\alpha\mu}({\bf X})
\int_{0}^{2\pi}\!\!\!\!\!\dots\int_{0}^{2\pi} \, 
Q (\bm{\theta}, {\bf X}) \,\,
\Phi^{i}_{\theta^{\alpha}} (\bm{\theta}, {\bf U}({\bf X})) \,\,   
{d^{m} \theta \over (2\pi)^{m}} \,\,\, \delta ({\bf X} - {\bf Y})
\,\,\, + \,\,\, O (\epsilon) $$

 In this case, by virtue of (\ref{komeganulrel}) we have 
\begin{equation}
\label{phiknulbr}
\left. \left\{ 
\varphi^{i} (\bm{\theta}, {\bf X}) , 
\int q({\bf Y}) \, k^{\alpha}({\bf U} ({\bf Y})) 
\, d^{d} Y \right\}\right|_{\cal K}
\,\, = \,\, O (\epsilon)
\end{equation}
\begin{equation}
\label{QvarphikalphaKzerorel}
\int_{0}^{2\pi}\!\!\!\!\!\dots\int_{0}^{2\pi}
Q \left( {{\bf S}({\bf X}) \over \epsilon} + \bm{\theta}, 
{\bf X} \right) \,
\left. \left\{ \varphi^{i} (\bm{\theta}, {\bf X}) \, , \, 
k^{\alpha}({\bf U} ({\bf Y})) \right\}\right|_{\cal K} \,
{d^{m} \theta \over (2\pi)^{m}} \,\, = \,\, O (\epsilon)
\end{equation}
for fixed values of coordinates 
$[{\bf S}({\bf Z}), \, U^{1}({\bf Z}), \dots , U^{m + s}({\bf Z})]$.

 Let us prove now some lemmas about structure of the Poisson 
brackets on the submanifold ${\cal K}$ which we will need in the 
further consideration.

\vspace{0.2cm}

{\bf Lemma 3.1.}

{\it Let the values $U^{\nu}$ of the functionals $I^{\nu}$ on a 
complete regular family $\Lambda$ of $m$-phase solutions of
(\ref{EvInSyst}) be functionally independent
and give a complete set of parameters on $\Lambda$, 
excluding the initial phases. Then for the Poisson brackets of the 
functionals $k^{\alpha}_{p}({\bf U}({\bf X}))$ and 
$J^{\mu}({\bf Y})$ on ${\cal K}$
we have the following relations:
\begin{equation}
\label{kJmuskob}
\{ k^{\alpha}_{p} ({\bf U}({\bf X})) , 
J^{\mu}({\bf Y})\}|_{\cal K} \,\, = \,\,
\epsilon \, \left[ \omega^{\alpha \mu}({\bf U}({\bf X})) \,\, 
\delta ({\bf X} - {\bf Y}) \right]_{X^{p}} 
\,\, + \,\,  O (\epsilon^{2})
\end{equation}
}

\vspace{0.2cm}

 Proof.

The conditions of Lemma 3.1 coincide with the conditions of 
Lemmas 2.1 and 2.2. Consider the Hamiltonian flow generated by the 
functional $\int q({\bf Y}) \, J^{\mu}({\bf Y})\, d^{d} Y$ 
according to bracket 
(\ref{EpsExtBracket}) with a compactly supported function
$q({\bf Y})$ of the slow variable ${\bf Y} = \epsilon {\bf y}$.

 According to (\ref{phiJq}) we can write at the ``points''
of the submanifold ${\cal K}$:
\begin{equation}
\label{varphiqXev}
\varphi^{i}_{t} \,\, = \,\, 
q({\bf X}) \,\, \omega^{\beta\mu}({\bf U}({\bf X}))
\,\, \Phi^{i}_{\theta^{\beta}} \left(
{{\bf S}({\bf X}) \over \epsilon} + \bm{\theta}, 
{\bf U} ({\bf X}) \right)
\,\, + \,\, \epsilon \, \eta^{i}_{[q]}  \left(
{{\bf S}({\bf X}) \over \epsilon} + \bm{\theta}, 
{\bf X} \right) \,\, + \,\,
O (\epsilon^{2}) 
\end{equation}
with some ($2\pi$-periodic in each $\theta^{\alpha}$)
functions $\eta^{i}_{[q]} (\bm{\theta}, {\bf X})$. 

 Let us introduce the functionals 
$k^{\alpha}_{p} ({\bf J}({\bf X})) = 
k^{\alpha}_{p} (J^{1}({\bf X}), \dots, J^{N}({\bf X}))$
using the functions $k^{\alpha}_{p} ({\bf U})$. According
to (\ref{JUtransform}) - (\ref{UJtransform}) 
we can write
$$\{ k^{\alpha}_{p} ({\bf U}({\bf X})) , 
J^{\mu}({\bf Y})\}|_{\cal K} \,\, = \,\,
\{ k^{\alpha}_{p} ({\bf J}({\bf X})) , 
J^{\mu}({\bf Y})\}|_{\cal K} \,\, + \,\,
O (\epsilon^{2}) $$

 Consider the evolution of the functionals 
$k^{\alpha}_{p} ({\bf J}({\bf X}))$ according to the flow 
generated by
\linebreak
$\int q({\bf Y}) \, J^{\mu}({\bf Y})\, d^{d} Y$ 
on the submanifold ${\cal K}$. 
Using relations (\ref{varphiqXev}) we can write
$$k^{\alpha}_{p t} ({\bf J}({\bf X})) \,\,\, = \,\,\,
{\partial k^{\alpha}_{p} \over \partial U^{\nu}} 
\left({\bf J}({\bf X})\right) \,\,
J^{\nu}_{t}({\bf X}) \,\,\, = $$
\begin{multline*}
=
{\partial k^{\alpha}_{p} \over \partial U^{\nu}} 
\left( {\bf J}({\bf X}) \right)
\,\, q ({\bf X}) \,\, 
\omega^{\beta\mu}({\bf U}({\bf X})) \,\, \times \\
\times \, \int_{0}^{2\pi}\!\!\!\!\!\dots\!\int_{0}^{2\pi} \!
\sum_{l_{1}, \dots, l_{d}} 
\Pi^{\nu (l_{1} \dots l_{d})}_{i} \left( \bm{\Phi} \left(
{{\bf S}({\bf X}) \over \epsilon} + \bm{\theta}, 
{\bf U}({\bf X}) \right) , \,
\epsilon {\partial \over \partial {\bf X}} \bm{\Phi} \left(
{{\bf S}({\bf X}) \over \epsilon} + \bm{\theta}, 
{\bf U}({\bf X}) \right), 
\dots \right) \, \times \\
\times \,\, 
\epsilon^{l_{1} + \dots + l_{d}} \,\, 
{\partial^{l_{1}} \over \partial X^{1 \, l_{1}}} \, \dots \,
{\partial^{l_{d}} \over \partial X^{d \, l_{d}}} \,
\Phi^{i}_{\theta^{\beta}} \left(
{{\bf S}({\bf X}) \over \epsilon} + \bm{\theta}, 
{\bf U}({\bf X}) \right) \,\,
{d^{m} \theta \over (2\pi)^{m}} \,\, +
\end{multline*}
\begin{multline*}
+ \, \epsilon \,\,
{\partial k^{\alpha}_{p} \over \partial U^{\nu}} 
\left( {\bf U}({\bf X}) \right) \, \sum_{q=1}^{d} \, 
\left( q ({\bf X}) \, \omega^{\beta\mu}({\bf U}({\bf X})) 
\right)_{X^{q}} \,\, \times \\
\times \,\, \int_{0}^{2\pi}\!\!\!\!\!\dots\int_{0}^{2\pi} 
\sum_{l_{1}, \dots, l_{d}} \, l_{q} \, 
\Pi^{\nu (l_{1} \dots l_{d})}_{i} \left( \bm{\Phi} 
\left( \bm{\theta}, {\bf U}({\bf X}) \right), \, 
k^{\gamma_{1}}_{1} \, \bm{\Phi}_{\theta^{\gamma_{1}}}, \dots, \,
k^{\gamma_{d}}_{d} \, \bm{\Phi}_{\theta^{\gamma_{d}}},
\dots \right) \times \\
\times \,\, k^{\alpha^{1}_{1}}_{1} 
\dots k^{\alpha^{1}_{l_{1}}}_{1} \, \dots \,
k^{\alpha^{q}_{1}}_{q} \dots k^{\alpha^{q}_{l_{q}-1}}_{q} \, \dots \,
k^{\alpha^{d}_{1}}_{d} \dots k^{\alpha^{d}_{l_{d}}}_{d} \,\,\,
\Phi^{i}_{\theta^{\beta} \,
\theta^{\alpha^{1}_{1}}\dots\theta^{\alpha^{1}_{l_{1}}} \, \dots \,
\theta^{\alpha^{q}_{1}}\dots\theta^{\alpha^{q}_{l_{q}-1}} \, \dots \,
\theta^{\alpha^{d}_{1}}\dots\theta^{\alpha^{d}_{l_{d}}}} \,\,
{d^{m} \theta \over (2\pi)^{m}} \,\,\, +
\end{multline*}
\begin{multline*}
+ \, \epsilon \,\,
{\partial k^{\alpha}_{p} \over \partial U^{\nu}} 
\left( {\bf U}({\bf X}) \right) \,\, \times \\
\times \, \int_{0}^{2\pi}\!\!\!\!\!\dots\int_{0}^{2\pi} 
\sum_{l_{1}, \dots, l_{d}} \,
\Pi^{\nu (l_{1} \dots l_{d})}_{i} 
\left( \bm{\Phi} \left( \bm{\theta}, {\bf U}({\bf X}) \right),
\, k^{\gamma_{1}}_{1} \,
\bm{\Phi}_{\theta^{\gamma_{1}}}, \dots, \,
k^{\gamma_{d}}_{d} \, \bm{\Phi}_{\theta^{\gamma_{d}}},
\dots \right) \,\, \times \\
\times \,\, k^{\alpha^{1}_{1}}_{1}
\dots k^{\alpha^{1}_{l_{1}}}_{1} \, \dots \,
k^{\alpha^{d}_{1}}_{d} \dots k^{\alpha^{d}_{l_{d}}}_{d} \,\,
\eta^{i}_{[q] \,  
\theta^{\alpha^{1}_{1}}\dots\theta^{\alpha^{1}_{l_{1}}} \, \dots \,
\theta^{\alpha^{d}_{1}}\dots\theta^{\alpha^{d}_{l_{d}}}}
(\bm{\theta}, {\bf X})
\,\, {d^{m} \theta \over (2\pi)^{m}} \,\,\, + \,\,\, O (\epsilon^{2})
\end{multline*}

 It's not difficult to see that the first part of the above expression
contains the integrals over $\bm{\theta}$ of the expressions
$P^{\nu}_{\theta^{\beta}} (\bm{\theta}, {\bf X})$ and is equal to zero.
It is easy to see also after integration by parts that the third part 
of the above expression represents the value
$$\epsilon \,\, \int_{0}^{2\pi}\!\!\!\!\!\dots\int_{0}^{2\pi} 
{\partial k^{\alpha}_{p} \over \partial U^{\nu}} 
\left( {\bf U}({\bf X}) \right)
\,\, \zeta^{(\nu)}_{i[{\bf U}({\bf X})]} (\bm{\theta}) \,\,
\eta^{i}_{[q]} (\bm{\theta}, {\bf X}) \,\, 
{d^{m} \theta \over (2\pi)^{m}} $$
and is equal to zero according to Lemma 2.1.

 Thus, we can write in the main order the expressions
for the evolution of the functionals $k^{\alpha}_{p}({\bf J}({\bf X}))$ 
on the submanifold ${\cal K}$ in the form
\begin{multline*}  
k^{\alpha}_{p t} ({\bf J}({\bf X})) \,\, = \,\, \epsilon \,\,
{\partial k^{\alpha}_{p} \over \partial U^{\nu}}
\left( {\bf U}({\bf X}) \right) \, \sum_{q=1}^{d} \,
\left( q ({\bf X}) \, \omega^{\beta\mu}({\bf U}({\bf X}))
\right)_{X^{q}} \,\, \int_{0}^{2\pi}\!\!\!\!\!\dots\int_{0}^{2\pi}
\sum_{l_{1}, \dots, l_{d}} \, l_{q} \,
\Pi^{\nu (l_{1} \dots l_{d})}_{i \, [0]} \,\, \times \\
\times \,\, k^{\alpha^{1}_{1}}_{1}
\dots k^{\alpha^{1}_{l_{1}}}_{1} \, \dots \,
k^{\alpha^{q}_{1}}_{q} \dots k^{\alpha^{q}_{l_{q}-1}}_{q} \, \dots \,
k^{\alpha^{d}_{1}}_{d} \dots k^{\alpha^{d}_{l_{d}}}_{d} \,\,\,
\Phi^{i}_{\theta^{\beta} \,
\theta^{\alpha^{1}_{1}}\dots\theta^{\alpha^{1}_{l_{1}}} \, \dots \,
\theta^{\alpha^{q}_{1}}\dots\theta^{\alpha^{q}_{l_{q}-1}} \, \dots \,
\theta^{\alpha^{d}_{1}}\dots\theta^{\alpha^{d}_{l_{d}}}} \,\,\,
{d^{m} \theta \over (2\pi)^{m}} 
\end{multline*}

 Using relations (\ref{kalphakbetaUnu}) we get then
$$k^{\alpha}_{p t} ({\bf J}({\bf X})) \,\, = \,\, 
\epsilon \, \left[ q ({\bf X}) \,
\omega^{\alpha \mu} ({\bf U}({\bf X})) \right]_{X^{p}} \,\, + \,\,  
O (\epsilon^{2}) $$
i.e.
$$\{ k^{\alpha}_{p}({\bf J}({\bf X})) , 
J^{\mu}({\bf Y})\}|_{\cal K} \,\, = \,\,
\epsilon \, \omega^{\alpha \mu}({\bf U}({\bf X})) \,\,
\delta_{X^{p}} ({\bf X} - {\bf Y}) \,\, + \,\, \epsilon \, 
\omega^{\alpha \mu}_{X^{p}} \,\, \delta ({\bf X} - {\bf Y})
\,\, + \,\,  O (\epsilon^{2}) $$
which implies (\ref{kJmuskob}) by virtue of relations
(\ref{JUtransform}) - (\ref{UJtransform}).

{\hfill Lemma 3.1 is proved.}

\vspace{0.2cm}

 Easy to see that according to Lemma 3.1 and relations
(\ref{JUtransform}) - (\ref{UJtransform}) we can write also
\begin{equation}
\label{kUmuskob}
\{ k^{\alpha}_{p} ({\bf U}({\bf X})) \, , \, 
U^{\mu}({\bf Y})\}|_{\cal K} \,\, = \,\,
\epsilon \, \left[ \omega^{\alpha \mu}({\bf U}({\bf X})) \,\,
\delta ({\bf X} - {\bf Y}) \right]_{X^{p}} 
\,\, + \,\,  O (\epsilon^{2})
\end{equation}
on the submanifold ${\cal K}$.

 Let us formulate now some corollaries following from Lemma 3.1.

 1) Under the conditions of Lemma 3.1, we have also
\begin{equation}
\label{kknulrel}
\{ k^{\alpha}_{p} ({\bf X}) \, , \,
k^{\beta}_{l} ({\bf Y})\}|_{\cal K} \,\, = \,\,
O (\epsilon^{2})
\end{equation}
for the functionals ${\bf k}_{p} ({\bf U}({\bf X}))$.

 Indeed, by virtue of (\ref{komeganulrel}) we have
$$\{ k^{\alpha}_{p} ({\bf X}) ,  
k^{\beta}_{l} ({\bf Y})\}|_{\cal K} \, = \,
\{ k^{\alpha}_{p} ({\bf X}) , U^{\mu}({\bf Y})\}|_{\cal K} \, 
{\partial k^{\beta}_{l} \over \partial U^{\mu}} ({\bf U}({\bf Y})) 
\, = \,
\epsilon \, \omega^{\alpha \mu} ({\bf X}) \, 
k^{\beta}_{l, U^{\mu}}({\bf X})
\, \delta_{X^{p}} ({\bf X} - {\bf Y}) \, + $$
$$+ \, \epsilon \, \omega^{\alpha \mu} ({\bf X}) 
\, \left( k^{\beta}_{l, U^{\mu}}({\bf X}) \right)_{X^{p}} \, 
\delta ({\bf X} - {\bf Y}) \, +
\, \epsilon \, \left(
\omega^{\alpha \mu}({\bf X}) \right)_{X^{p}} \, 
k^{\beta}_{l, U^{\mu}}({\bf X}) \,
\delta ({\bf X} - {\bf Y}) \, + \, 
O (\epsilon^{2}) \,\, = \,\, O (\epsilon^{2}) $$

 2) Using definition (\ref{DefinitionS})
of the functionals $S^{\alpha} ({\bf X})$
we can write on ${\cal K}$ under the conditions of Lemma 3.1:
\begin{equation}
\label{SalphaJmuSkob}
\{ S^{\alpha} ({\bf X}) \, , \,
J^{\mu}({\bf Y})\}|_{\cal K} \,\,\, = \,\,\,
\epsilon \,\,  \omega^{\alpha \mu}({\bf U}({\bf X})) \,\,\,
\delta ({\bf X} - {\bf Y})   
\,\,\, + \,\,\,  O (\epsilon^{2}) 
\end{equation}
\begin{equation}
\label{SalphaUmuSkob}
\{ S^{\alpha} ({\bf X}) \, , \,
U^{\mu}({\bf Y})\}|_{\cal K} \,\,\, = \,\,\,
\epsilon \,\,  \omega^{\alpha \mu}({\bf U}({\bf X})) \,\,\,
\delta ({\bf X} - {\bf Y})   
\,\, + \,\,  O (\epsilon^{2})  
\end{equation}
\begin{equation}
\label{kalphaSalphaSbeta}
\{ k^{\alpha}_{p} ({\bf X}) \, , \,
S^{\beta} ({\bf Y}) \}|_{\cal K} \,\,\, = \,\,\, O (\epsilon^{2}) 
\,\,\,\,\,\,\,\, , \,\,\,\,\,\,\,\,
\{ S^{\alpha} ({\bf X}) \, , \,
S^{\beta} ({\bf Y}) \}|_{\cal K} \,\, = \,\, O (\epsilon^{2}) 
\end{equation}
for the functionals $S^{\alpha} ({\bf X})$.

 In the proof of Lemma 3.1, we have not used
the fact that the functionals $J^{\mu}({\bf Y})$ belong to our special 
set of functionals (\ref{FuncJnuX}) and used only the fact 
that the flow generated by the functional $I^{\mu}$
leaves invariant the family $\Lambda$ generating linear shifts of
$\theta^{\alpha}_{0}$ with constant frequencies
$\omega^{\alpha \mu} ({\bf U})$. We can therefore formulate
here the following lemma:

\vspace{0.2cm}

{\bf Lemma 3.1$^\prime$.}

{\it Let the values $U^{\nu}$ of the functionals $I^{\nu}$ on a
complete regular family $\Lambda$ of $m$-phase solutions of system
(\ref{EvInSyst}) be functionally independent
and give a complete set of parameters on $\Lambda$, excluding 
the initial phases. Let the flow generated by the functional
\begin{equation}
\label{ItildeFunc}
{\tilde I} \,\, = \,\, \int {\tilde P}
(\bm{\varphi}, \bm{\varphi}_{\bf x}, \dots ) \,\, d^{d} x 
\end{equation}
leave invariant the family $\Lambda$ generating linear shifts of
$\theta^{\alpha}_{0}$ with constant frequencies
${\tilde \omega}^{\alpha} ({\bf U})$.

 Consider the functionals
\begin{equation}
\label{JXtildeFunc}
{\tilde J} ({\bf X}) \,\, = \,\, 
\int_{0}^{2\pi} \!\!\!\!\! \dots \int_{0}^{2\pi}
{\tilde P} (\bm{\varphi}, \, 
\epsilon \bm{\varphi}_{\bf X}, \, \epsilon^{2}
\bm{\varphi}_{\bf XX}, \dots ) \,\, 
{d^{m} \theta \over (2\pi)^{m}} 
\end{equation}

Then for the Poisson brackets of the functionals
$k^{\alpha}({\bf U}({\bf X}))$ and
${\tilde J} ({\bf Y})$ on ${\cal K}$ we have the relation:
$$\{ k^{\alpha}_{p} ({\bf U}({\bf X})) , 
{\tilde J} ({\bf Y})\}|_{\cal K} \,\, = \,\,
\epsilon \, \left[ {\tilde \omega}^{\alpha}({\bf U}({\bf X})) \,\,
\delta ({\bf X} - {\bf Y}) \right]_{X^{p}} 
\,\, + \,\,  O (\epsilon^{2}) $$
}

\vspace{0.2cm}

 Proof of Lemma 3.1$^\prime$ completely repeats the proof of
Lemma 3.1.

\vspace{0.2cm}

{\bf Lemma 3.2.}

{\it Let the values $U^{\nu}$ of the functionals $I^{\nu}$ on a
complete regular family $\Lambda$ of $m$-phase solutions of system
(\ref{EvInSyst}) be functionally independent
and give a complete set of parameters on $\Lambda$,    
excluding the initial phases. Then for the constraints 
$g^{i} (\bm{\theta}, {\bf X})$ imposed by (\ref{gconstr}) 
and smooth compactly supported function
$q ({\bf X})$ as well as smooth $2\pi$-periodic
in each $\theta^{\alpha}$ function $Q (\bm{\theta}, {\bf X})$ we have 
the following relations on the submanifold ${\cal K}$:
\begin{equation}
\label{giJmunulbr}
\left. \left\{ 
g^{i} (\bm{\theta}, {\bf X}) , \int q({\bf Y}) 
\, J^{\mu}({\bf Y}) \, d^{d} Y
\right\}\right|_{\cal K} \,\, = \,\,
O (\epsilon) 
\end{equation}
\begin{equation}
\label{QRelLemma24} 
\left. \left[ \int_{0}^{2\pi}\!\!\!\!\!\dots\int_{0}^{2\pi}
Q \left({{\bf S}({\bf X}) \over \epsilon} + \bm{\theta}, 
{\bf X} \right) \,
\left\{ g^{i} (\bm{\theta}, {\bf X}) \, , \, J^{\mu}({\bf Y}) \right\} 
\,\, {d^{m} \theta \over (2\pi)^{m}} 
\right] \right|_{\cal K} \,\, = \,\, O (\epsilon) 
\end{equation}
}

\vspace{0.2cm}

 Proof.

Indeed, by (\ref {phiJbr}), (\ref{QphiJbr}), and
Lemma 3.1 we have
$$\left. \left\{
g^{i} (\bm{\theta}, {\bf X}) , \int q({\bf Y}) \, 
J^{\mu}({\bf Y}) \, d^{d} Y
\right\}\right|_{{\cal K}[0]} 
\,\, = \,\, \Phi^{i}_{\theta^{\alpha}} \left(
{{\bf S}({\bf X}) \over \epsilon} + \bm{\theta}, 
{\bf U} ({\bf X}) \right)
\,\, \omega^{\alpha \mu} ({\bf X}) \,\, q ({\bf X}) \,\, - $$
$$- \,\,\,  \Phi^{i}_{\theta^{\alpha}} \left(
{{\bf S}({\bf X}) \over \epsilon} + \bm{\theta}, 
{\bf U} ({\bf X}) \right)
\left. \left\{
S^{\alpha} ({\bf X}) \, , \,\int q({\bf Y}) \,
J^{\mu}({\bf Y}) \, d^{d} Y
\right\}\right|_{{\cal K}[0]} \,\, \equiv \,\, 0 $$

$$\left. \left[ \int_{0}^{2\pi}\!\!\!\!\!\dots\int_{0}^{2\pi}
Q \left({{\bf S}({\bf X}) \over \epsilon} + \bm{\theta}, 
{\bf X} \right) \,
\left\{ g^{i} (\bm{\theta}, {\bf X}), J^{\mu}({\bf Y}) \right\}
\,\, {d^{m} \theta \over (2\pi)^{m}} 
\right] \right|_{{\cal K}[0]} \,\, = $$
$$= \,\, \omega^{\alpha\mu}({\bf X}) 
\int_{0}^{2\pi}\!\!\!\!\!\dots\int_{0}^{2\pi}
Q (\bm{\theta}, {\bf X}) \,\,
\Phi^{i}_{\theta^{\alpha}} (\bm{\theta}, {\bf U}({\bf X}))
\,\, {d^{m} \theta \over (2\pi)^{m}} \,\,\, 
\delta ({\bf X} - {\bf Y}) \,\, -$$
$$- \, \int_{0}^{2\pi}\!\!\!\!\!\dots\int_{0}^{2\pi}
Q (\bm{\theta}, {\bf X}) \,\,
\Phi^{i}_{\theta^{\alpha}} (\bm{\theta}, {\bf U}({\bf X}))
\,\, {d^{m} \theta \over (2\pi)^{m}} \,\,\,\,\,
\left. \left\{ S^{\alpha} ({\bf X}) \, , \,
J^{\mu}({\bf Y}) \right\}\right|_{{\cal K}[0]} 
\,\,\, \equiv \,\,\, 0 $$

{\hfill Lemma 3.2 is proved.}

\vspace{0.2cm}

 Similarly to the previous case, we can formulate also
the following lemma:

\vspace{0.2cm}

{\bf Lemma 3.2$^\prime$.}

{\it Let the values $U^{\nu}$ of the functionals $I^{\nu}$ on a
complete regular family $\Lambda$ of $m$-phase solutions of system
(\ref{EvInSyst}) be functionally independent
and give a complete set of parameters on $\Lambda$,
excluding the initial phases. Let the flow generated by 
the functional (\ref{ItildeFunc})
leave invariant the family $\Lambda$ generating linear shifts of 
$\theta^{\alpha}_{0}$ with constant frequencies
${\tilde \omega}^{\alpha} ({\bf U})$. Then for the constraints
$g^{i} (\bm{\theta}, {\bf X})$ and the functionals 
${\tilde J}({\bf X})$
imposed by (\ref{JXtildeFunc}) we have the following relations
on the submanifold ${\cal K}$:
$$\left. \left\{ 
g^{i} (\bm{\theta}, {\bf X}) , 
\int q ({\bf Y}) \, {\tilde J} ({\bf Y}) \, d^{d} Y
\right\}\right|_{\cal K} \,\, = \,\,
O (\epsilon) $$
$$\left. \left[ \int_{0}^{2\pi}\!\!\!\!\!\dots\int_{0}^{2\pi}
Q \left({{\bf S}({\bf X}) \over \epsilon} + \bm{\theta}, 
{\bf X} \right) \,
\left\{ g^{i} (\bm{\theta}, {\bf X}) \, , \, 
{\tilde J} ({\bf Y}) \right\}
\,\, {d^{m} \theta \over (2\pi)^{m}} 
\right] \right|_{\cal K} \,\, = \,\, O (\epsilon) $$
under the same conditions for the functions
$q ({\bf X})$ and $Q (\bm{\theta}, {\bf X})$.
}

\vspace{0.2cm}

 Using transformations (\ref{JUtransform}) - (\ref{UJtransform})
we can also write
$$\left. \left\{
g^{i} (\bm{\theta}, {\bf X}) , \int q({\bf Y})
\, U^{\mu}({\bf Y}) \, d^{d} Y
\right\}\right|_{\cal K} \,\, = \,\,
O (\epsilon) $$
$$\left. \left[ \int_{0}^{2\pi}\!\!\!\!\!\dots\int_{0}^{2\pi}
Q \left({{\bf S}({\bf X}) \over \epsilon} + \bm{\theta},
{\bf X} \right) \,
\left\{ g^{i} (\bm{\theta}, {\bf X})\, , \, U^{\mu}({\bf Y}) \right\}
\,\, {d^{m} \theta \over (2\pi)^{m}}
\right] \right|_{\cal K} \,\, = \,\, O (\epsilon) $$
for the functionals $U^{\mu}({\bf Y})$.

 The pairwise Poisson brackets of the constraints 
$g^{i} (\bm{\theta}, {\bf X})$, 
$g^{j} (\bm{\theta}^{\prime}, {\bf Y})$
on ${\cal K}$ can be written in the form:
\begin{equation}
\label{SkobSv}
\{ g^{i} (\bm{\theta}, {\bf X}) \, , \,
g^{j} (\bm{\theta}^{\prime}, {\bf Y}) \}|_{\cal K} \, = 
\end{equation}
$$= \,\, \{ \varphi^{i} (\bm{\theta}, {\bf X}) \, , \, 
\varphi^{j} (\bm{\theta}^{\prime}, {\bf Y}) \}|_{\cal K}  \,\, - \,\,
\{ \varphi^{i} (\bm{\theta}, {\bf X}) \, , \, 
U^{\lambda}({\bf Y}) \}|_{\cal K} \,\,\,
\Phi^{j}_{U^{\lambda}} \left( {{\bf S}({\bf Y}) \over \epsilon} +
\bm{\theta}^{\prime} , {\bf U}({\bf Y}) \right)  \, - $$
$$- \, \Phi^{i}_{U^{\nu}} \left( {{\bf S}({\bf X}) \over \epsilon} +
\bm{\theta} , {\bf U}({\bf X}) \right) \, \{ U^{\nu}({\bf X}) \, , \, 
\varphi^{j} (\bm{\theta}^{\prime}, {\bf Y}) \}|_{\cal K}  \,\,\, + $$
$$+ \, \Phi^{i}_{U^{\nu}} \left( {{\bf S}({\bf X}) \over \epsilon} +
\bm{\theta} , {\bf U}({\bf X}) \right) \,
\{ U^{\nu}({\bf X}) \, , \, U^{\lambda}({\bf Y}) \}|_{\cal K} \,\,\,
\Phi^{j}_{U^{\lambda}} \left( {{\bf S}({\bf Y}) \over \epsilon} +
\bm{\theta}^{\prime} , {\bf U}({\bf Y}) \right) \, - $$
$$- \,\, {1 \over \epsilon} \,\,
\{ \varphi^{i} (\bm{\theta}, {\bf X}) \, , \, 
S^{\beta}({\bf Y}) \}|_{\cal K} \,\,\, 
\Phi^{j}_{\theta^{\prime\beta}} 
\left( {{\bf S}({\bf Y}) \over \epsilon} +
\bm{\theta}^{\prime} , {\bf U}({\bf Y}) \right)  \,\, - $$
$$- \,\, {1 \over \epsilon} \,\, \Phi^{i}_{\theta^{\alpha}} 
\left( {{\bf S}({\bf X}) \over \epsilon} +
\bm{\theta} , {\bf U}({\bf X}) \right) \,
\{  S^{\alpha}({\bf X}) \, , \,
\varphi^{j} (\bm{\theta}^{\prime}, {\bf Y}) \}|_{\cal K}
\,\,\, + $$
$$+ \,\,\, {1 \over \epsilon} \,\, 
\Phi^{i}_{\theta^{\alpha}} \left(
{{\bf S}({\bf X}) \over \epsilon} + \bm{\theta} , 
{\bf U}({\bf X}) \right) \,
\{  S^{\alpha} ({\bf X}) \, , \, U^{\lambda}({\bf Y}) \}|_{\cal K} 
\,\,\, \Phi^{j}_{U^{\lambda}} 
\left( {{\bf S}({\bf Y}) \over \epsilon} +
\bm{\theta}^{\prime} , {\bf U}({\bf Y}) \right)  \,\,\, + $$
$$+ \,\,\, {1 \over \epsilon} \,\, 
\Phi^{i}_{U^{\nu}} \left( {{\bf S}({\bf X}) \over \epsilon} +
\bm{\theta} , {\bf U}({\bf X}) \right) 
\{ U^{\nu}({\bf X}) \, , \,
S^{\beta}({\bf Y}) \}|_{\cal K} \,\,\, 
\Phi^{j}_{\theta^{\prime\beta}} 
\left( {{\bf S}({\bf Y}) \over \epsilon} 
+ \bm{\theta}^{\prime} , {\bf U}({\bf Y}) \right)  \,\,\, + $$
$$+ \,\, {1 \over \epsilon^{2}} \,\,
\Phi^{i}_{\theta^{\alpha}}  \left( 
{{\bf S}({\bf X}) \over \epsilon} + \bm{\theta} , 
{\bf U}({\bf X}) \right) \,
\{  S^{\alpha}({\bf X}) \, , \, S^{\beta}({\bf Y}) \}|_{\cal K}
\,\,\, \Phi^{j}_{\theta^{\prime\beta}}  
\left( {{\bf S}({\bf Y}) \over \epsilon} + \bm{\theta}^{\prime} , 
{\bf U}({\bf Y}) \right) $$

 Let us note that we assume that the 
parameters $U^{\nu}$, $U^{\lambda}$ 
represent here a full set of parameters ${\bf U}$
$(\nu, \lambda = 1, \dots, N)$
on $\Lambda$. The choice of the parameters $U^{\nu}$ is in fact
not important due to the invariance of the corresponding expressions
with respect to the substitutions 
$U^{\nu} \, = \, U^{\nu} ({\tilde {\bf U}})$ on $\Lambda$.
So, for the sake of brevity, we put that the set 
$(U^{1}, \dots, U^{N})$ represents the set of parameters
$(k^{\alpha}_{q}, \, U^{1}, \dots, U^{m+s})$ or the set of averaged
densities of the functionals $I^{\nu}$ introduced above.

 For convenience let us introduce the functional
$$J_{[{\bf q}]} \,\, = \,\,
\int J^{\mu} ({\bf Y}) \,\, q_{\mu} ({\bf Y}) \,\, d^{d} Y $$
for any smooth compactly supported vector-function
${\bf q} ({\bf Y}) \, = \, (q_{1} ({\bf Y}), \dots, q_{N} ({\bf Y}))$.
According to Lemma 3.2 we can write the relations
$\{ g^{i}(\bm{\theta}, {\bf X}) \, , \,   
J_{[{\bf q}]} \}|_{\cal K} \, = \, O (\epsilon)$
on the submanifold ${\cal K}$. More precisely, the first non-vanishing
term of the Poisson bracket of $g^{i} (\bm{\theta}, {\bf X})$ and
$J_{[{\bf q}]}$ on ${\cal K}$ can be written as 
$$\{ g^{i}(\bm{\theta}, {\bf X}) ,
J_{[{\bf q}]} \}|_{{\cal K}[1]} \, = \,
\{ \varphi^{i}(\bm{\theta}, {\bf X}) , 
J_{[{\bf q}]} \}|_{{\cal K}[1]} \, - \,
\Phi^{i}_{U^{\nu}} \left(
{{\bf S}({\bf X}) \over \epsilon} + \bm{\theta}, {\bf U}({\bf X}) \right)
\{ U^{\nu}({\bf X}) , J_{[{\bf q}]}  \}|_{{\cal K}[1]} \, - $$
\begin{equation}
\label{giJmu1skob}
- \,\, \Phi^{i}_{\theta^{\alpha}} \left(
{{\bf S}({\bf X}) \over \epsilon} + \bm{\theta},  
{\bf U}({\bf X}) \right)
\{ S^{\alpha}({\bf X}) \, , \, J_{[{\bf q}]}  \}|_{{\cal K}[2]}
\end{equation}

 Let us remind here that the indices $[1]$, $[2]$ mean as before the 
terms of the corresponding graded expansion on ${\cal K}$ having
degree $1$ and $2$ respectively.

 As we mentioned already, the procedure of the averaging of a Poisson
bracket is connected to some extend with the Dirac procedure of the
restriction of a Poisson bracket onto a submanifold. The more detailed
consideration of the connection of the averaging procedure with the
Dirac procedure can be found in \cite{DNMultDim}. For the justification
of the averaging procedure considered here we will need to investigate
the solubility of the system
\begin{equation}
\label{BAcond}
{\hat B}^{ij}_{[0]} ({\bf X}) \, 
B_{j[{\bf q}]} (\bm{\theta} , {\bf X} )
\,\, + \,\, A^{i}_{[1][{\bf q}]} (\bm{\theta} , {\bf X}) 
\,\, = \,\, 0
\end{equation}
where
\begin{multline}
\label{OperatorB}
{\hat B}^{ij}_{[0]} ({\bf X}) \,\, = \,\, \sum_{l_{1}, \dots, l_{d}}
B^{ij}_{(l_{1} \dots l_{d})} 
\left( \bm{\Phi} (\bm{\theta}, {\bf U}({\bf X})), \, 
k^{\gamma_{1}}_{1}({\bf X}) \, \bm{\Phi}_{\theta^{\gamma_{1}}}, 
\dots, \, k^{\gamma_{d}}_{d}({\bf X}) \, 
\bm{\Phi}_{\theta^{\gamma_{d}}}, \, \dots \right) \, \times \\ 
\times \,\, k^{\alpha^{1}_{1}}_{1}({\bf X}) \dots 
k^{\alpha^{1}_{l_{1}}}_{1}({\bf X})
\, \dots \,
k^{\alpha^{d}_{1}}_{d}({\bf X}) \dots 
k^{\alpha^{d}_{l_{d}}}_{d}({\bf X}) \,\,\,\,\,
{\partial \over \partial \theta^{\alpha^{1}_{1}}} \dots
{\partial \over \partial \theta^{\alpha^{1}_{l_{1}}}} \, \dots \,
{\partial \over \partial \theta^{\alpha^{d}_{1}}} \dots
{\partial \over \partial \theta^{\alpha^{d}_{l_{d}}}} 
\end{multline}
is the Hamiltonian operator (\ref{MultDimPBr}) on the family of
$m$-phase solutions of system (\ref{EvInSyst}) and
\begin{equation}
\label{FunktsiiA}
A^{i}_{[1][{\bf q}]} \! \left( {{\bf S}({\bf X}) \over \epsilon} + 
\bm{\theta} , {\bf X} \right) \, = \,
\{ \varphi^{i}(\bm{\theta}, {\bf X}) , 
J_{[{\bf q}]} \}|_{{\cal K}[1]} \, - \,
\Phi^{i}_{U^{\nu}} \! \left(
{{\bf S}({\bf X}) \over \epsilon} + \bm{\theta}, 
{\bf U}({\bf X}) \! \right) 
\{ U^{\nu}({\bf X}) , J_{[{\bf q}]} \}|_{{\cal K}[1]} 
\end{equation}
is the discrepancy connected with the first non-vanishing term of
the Poisson bracket of $g^{i} (\bm{\theta}, {\bf X})$ and
$J_{[{\bf q}]}$ on ${\cal K}$. Let us note also that for the sake 
of brevity we denote again by $U^{\nu}$ the full set of parameters 
${\bf U}$ on $\Lambda$ as in the expression (\ref{SkobSv}).

 System (\ref{BAcond}) represents at every ${\bf X}$ a linear system 
of partial differential equations in $\bm{\theta}$ with periodic
coefficients. For us the solubility of system (\ref{BAcond}) on
the space of $2\pi$-periodic in all $\theta^{\alpha}$ functions
will be important. Let us discuss now the properties of
system (\ref{BAcond}).

It is easy to see that the operator
${\hat B}^{ij}_{[0]} ({\bf X}) \, = \,
{\hat B}^{ij}_{[0]} ({\bf U}({\bf X}))$ 
is a differential operator on the torus.
It can be seen also that for special values of ${\bf k}_{p} ({\bf X})$
the foliation defined by the set of the vector fields
$({\bf k}_{1} ({\bf X}), \dots, {\bf k}_{d} ({\bf X}))$ can define the
tori of the lower dimensions $\mathbb{T}^{k} \subset \mathbb{T}^{m}$ 
and even one-dimensional tori embedded in $\mathbb{T}^{m}$.

 The operator ${\hat B}^{ij}_{[0]} ({\bf U})$ has in general
finite number of ``regular'' eigenvectors with zero eigenvalues
defined for all values of the parameters ${\bf U}$ and smoothly
depending on the parameters. However, for special values of ${\bf U}$
the set of eigenvectors with zero eigenvalues can be infinite
and is determined, in particular,
by the dimension of closures of the foliation leaves
in $\mathbb{T}^{m}$, defined by the vectors
${\bf k}_{p} ({\bf U})$.

 Let us note that for some special brackets
(\ref{MultDimPBr}) the differential part can be absent in the
operator ${\hat B}^{ij}_{[0]} ({\bf U})$. The operator
${\hat B}^{ij}_{[0]} ({\bf U})$ reduces then to an ultralocal   
operator acting independently at every point of $\mathbb{T}^{m}$.
As a rule, the matrix $B^{ij}_{[0]} ({\bf U})$ is non-degenerate
in this case. Easy to see that system
(\ref{BAcond}) represents a simple algebraic system in this
case and is trivially solvable. The multiphase situation is not
different then from the single-phase case. However,
for arbitrary brackets (\ref{MultDimPBr}) the operators
${\hat B}^{ij}_{[0]} ({\bf U})$ have more general form described above.

 Let us define in the space of the parameters ${\bf U}$ the set
${\cal M}$, such that for all ${\bf U} \in {\cal M}$ the dimensions
of the closures of the foliation leaves, defined by the set
$\{{\bf k}_{p} ({\bf U}) \}$ in $\mathbb{T}^{m}$ is equal to $m$.
From the condition
$${\rm rk} \,\, || \partial k^{\alpha}_{p} / \partial U^{\nu} ||
\,\, = \,\, m d $$
it follows that the set ${\cal M}$ is everywhere dense in the
parameter space ${\bf U}$ and, moreover, has the full measure.
We note also that the set ${\bf U}$ represents here the full set
of parameters ${\bf U} = (U^{1}, \dots, U^{N})$ on $\Lambda$
excluding the initial phase shifts $\bm{\theta}_{0}$.

 In the study of the solubility of (\ref{BAcond})
we must first require the orthogonality of the functions
$A^{i}_{[1][{\bf q}]} (\bm{\theta} , {\bf X})$ to the ``regular''
eigenvectors of 
${\hat B}^{ij}_{[0]} ({\bf U}({\bf X}))$ with zero eigenvalues.
Let us prove here the following lemma.

\vspace{0.2cm}

{\bf Lemma 3.3.}

{\it Let the values $U^{\nu}$ of the functionals $I^{\nu}$ on a
complete regular family $\Lambda$ of $m$-phase solutions of system
(\ref{EvInSyst}) be functionally independent
and give a complete set of parameters on $\Lambda$, excluding
the initial phases. Then we have the relations
\begin{equation}
\label{LeftVecAort}
\int_{0}^{2\pi}\!\!\!\!\!\dots\int_{0}^{2\pi}
\kappa^{(q)}_{i[{\bf U}({\bf X})]}(\bm{\theta}) \,\,
A^{i}_{[1][{\bf q}]} (\bm{\theta}, {\bf X}) \,\,
{d^{m} \theta \over (2\pi)^{m}} \,\,\, \equiv \,\,\, 0
\,\,\,\,\, , \,\,\,\,\,\,\,\, q = 1, \dots, m + s
\end{equation}
for the functions $A^{i}_{[1][{\bf q}]} (\bm{\theta}, {\bf X})$
defined by (\ref{FunktsiiA}).
}

\vspace{0.2cm}

Proof.

 Let us first prove the following statement:

The values $\zeta^{(\gamma)}_{i[{\bf U}({\bf X})]}(\bm{\theta})$,
$\gamma = 1, \dots, m + s$,
are orthogonal (for any ${\bf X}$) to the values
$A^{i}_{[1][{\bf q}]} (\bm{\theta}, {\bf X})$, i.e.
\begin{equation}
\label{kappaAort}
\int_{0}^{2\pi}\!\!\!\!\!\dots\int_{0}^{2\pi}
\zeta^{(\gamma)}_{i[{\bf U}({\bf X})]}(\bm{\theta}) \,\,
A^{i}_{[1][{\bf q}]} (\bm{\theta}, {\bf X}) \,\,
{d^{m} \theta \over (2\pi)^{m}} \,\,\, \equiv \,\,\, 0
\,\,\,\,\, , \,\,\,\,\,\,\,\, \gamma = 1, \dots, m + s
\end{equation}

 Indeed, according to (\ref{dependence2}) the convolution of the values
$\delta U^{\gamma}({\bf Z}) / \delta \varphi^{i} (\bm{\theta}, {\bf X})$
with the values 
$\{ g^{i} (\bm{\theta}, {\bf X}) \, , \, J_{[{\bf q}]} \}|_{\cal K}$
identically vanish on ${\cal K}$. According to 
(\ref{JUtransform}) - (\ref{UJtransform}) the values
$\delta U^{\gamma}({\bf Z}) / \delta \varphi^{i} (\bm{\theta}, {\bf X})$
coincide in the leading order with the values
$\delta J^{\gamma}({\bf Z}) / \delta \varphi^{i} (\bm{\theta}, {\bf X})$
on ${\cal K}$.

 Using the explicit expression for the quantities
$\delta J^{\gamma}({\bf Z}) / \delta \varphi^{i} (\bm{\theta}, {\bf X})$
on ${\cal K}$:
$$\left.
{\delta J^{\nu}({\bf Z}) \over \delta \varphi^{i} (\bm{\theta}, {\bf X})}
\right|_{\cal K} \, = \, \sum_{l_{1},\dots,l_{d}}
\Pi^{\gamma (l_{1} \dots l_{d})}_{i}
\left( \bm{\Phi} \left( {{\bf S}({\bf Z}) \over \epsilon} + \bm{\theta},
{\bf Z} \right), \dots \right) \, \epsilon^{l_{1} + \dots + l_{d}} \,
\delta_{l_{1} Z^{1} \dots l_{d} Z^{d}} ({\bf Z} - {\bf X}) $$
we can write the corresponding convolution in the form of
action of the operator
$$\int_{0}^{2\pi}\!\!\!\!\!\dots \int_{0}^{2\pi}
{d^{m} \theta \over (2\pi)^{m}} \,
\sum_{l_{1},\dots,l_{d}} \Pi^{\gamma (l_{1} \dots l_{d})}_{i}
\left( \bm{\Phi} \left( {{\bf S}({\bf Z}) \over \epsilon} + \bm{\theta},
{\bf Z} \right), \dots \right) \, \epsilon^{l_{1} + \dots + l_{d}} \,
{d^{l_{1}} \over d Z^{1 \, l_{1}}} \, \dots \,
{d^{l_{d}} \over d Z^{d \, l_{d}}} $$
on the  distributions
$\{ g^{i} (\bm{\theta}, {\bf Z}) \, , \, J_{[{\bf q}]} \}|_{\cal K}$,
which is given in the main order by the action of the operator
\begin{multline}
\label{MainOrderOpAction}
\int_{0}^{2\pi}\!\!\!\!\!\dots \int_{0}^{2\pi}
{d^{m} \theta \over (2\pi)^{m}} \,
\sum_{l_{1},\dots,l_{d}} \Pi^{\gamma (l_{1} \dots l_{d})}_{i}
\left( \bm{\Phi} \left( {{\bf S}({\bf Z}) \over \epsilon} + \bm{\theta},
{\bf Z} \right), \dots \right) \,\, \times   \\
\times \,\, k^{\alpha^{1}_{1}}_{1}({\bf Z}) \dots
k^{\alpha^{1}_{l_{1}}}_{1}({\bf Z})
\, \dots \,
k^{\alpha^{d}_{1}}_{d}({\bf Z}) \dots
k^{\alpha^{d}_{l_{d}}}_{d}({\bf Z}) \,\,\,\,\,
{\partial \over \partial \theta^{\alpha^{1}_{1}}} \dots
{\partial \over \partial \theta^{\alpha^{1}_{l_{1}}}} \, \dots \,
{\partial \over \partial \theta^{\alpha^{d}_{1}}} \dots
{\partial \over \partial \theta^{\alpha^{d}_{l_{d}}}} 
\end{multline}

 The leading order of the brackets 
$\{ g^{i} (\bm{\theta}, {\bf Z}) \, , \, J_{[{\bf q}]} \}$ on
${\cal K}$ is given by the value \linebreak
$\{ g^{i} (\bm{\theta}, {\bf Z}) \, , \, J_{[{\bf q}]} 
\}|_{{\cal K}[1]}$, defined by formula (\ref{giJmu1skob}). After
integration by parts we get that the action of the operator
(\ref{MainOrderOpAction}) is given by the convolution w.r.t.
$\bm{\theta}$ of the values 
$\{ g^{i} (\bm{\theta}, {\bf Z}) \, , \, J_{[{\bf q}]}
\}|_{{\cal K}[1]}$
with the values 
$\zeta^{(\gamma)}_{i[{\bf U}({\bf Z})]}
({\bf S}({\bf Z})/\epsilon + \bm{\theta})$.

 We know also that the values
$\zeta^{(\gamma)}_{i[{\bf U}({\bf Z})]}
({\bf S}({\bf Z})/\epsilon + \bm{\theta})$ are automatically 
orthogonal to the functions
$\Phi^{i}_{\theta^{\alpha}} 
({\bf S}({\bf Z})/\epsilon + \bm{\theta}, {\bf Z})$, so 
(after the replacement of ${\bf Z}$ to ${\bf X}$) we get relation
(\ref{kappaAort}).

 Using now relations (\ref{kappaexpNewParam}) we get the statement of 
the Lemma.

{\hfill Lemma 3.3 is proved.}

\vspace{0.2cm}

 For a regular Hamiltonian family $\Lambda$ and a complete
Hamiltonian set of integrals $(I^{1}, \dots, I^{N})$ we can also 
prove the following lemma:

\vspace{0.2cm}

{\bf Lemma 3.4.}

{\it Let the functions $B_{j[{\bf q}]} (\bm{\theta} , {\bf X})$
satisfy conditions (\ref{BAcond}). Then the functions 
$B_{j[{\bf q}]} (\bm{\theta} , {\bf X})$
automatically satisfy the conditions
\begin{equation}
\label{BetaCond}
\int_{0}^{2\pi}\!\!\!\!\!\dots\int_{0}^{2\pi}
\Phi^{j}_{\theta^{\alpha}} (\bm{\theta} , \, 
{\bf S}_{\bf X}, \, U^{1}({\bf X}), \dots, U^{m+s}({\bf X})) \,\,
B_{j[{\bf q}]} (\bm{\theta} , {\bf X}) \,\,
{d^{m} \theta \over (2\pi)^{m}} \,\,\, \equiv \,\,\, 0
\,\,\, , \,\,\,\,\,\,\,\, \alpha = 1, \dots, m
\end{equation}
}

\vspace{0.2cm}

Proof.

 Indeed, the implementation of (\ref{BAcond}) implies the
conditions
$$\int_{0}^{2\pi}\!\!\!\!\!\dots\int_{0}^{2\pi}
\zeta^{(\gamma)}_{i[{\bf U}({\bf X})]}(\bm{\theta}) \,\,
{\hat B}^{ij}_{[0]} ({\bf X}) \, B_{j[{\bf q}]} (\bm{\theta} ,
{\bf X}) \,\, {d^{m} \theta \over (2\pi)^{m}} \,\,\, = \,\,\, 0 
\,\,\,\,\, , \,\,\,\,\,\,\,\, \gamma = 1, \dots, m + s $$
which is equivalent to
$$\omega^{\alpha\gamma} ({\bf U}({\bf X})) \,
\int_{0}^{2\pi}\!\!\!\!\!\dots\int_{0}^{2\pi}
\Phi^{j}_{\theta^{\alpha}} (\bm{\theta}, {\bf U}({\bf X})) \,\,
B_{j[{\bf q}]} (\bm{\theta} , {\bf X}) \,\,
{d^{m} \theta \over (2\pi)^{m}} \,\,\, = \,\,\, 0 $$
view the skew-symmetry of ${\hat B}^{ij}_{[0]} ({\bf X})$.

 From the property (\ref{RankOmegaAlphaGamma}) of the subset
$(I^{1}, \dots, I^{m+s})$ of a complete Hamiltonian family
of commuting functionals we immediately obtain now relations 
(\ref{BetaCond}).

{\hfill Lemma 3.4 is proved.}

\vspace{0.2cm}

In the remaining
part of the article the solubility of system (\ref{BAcond})
will play the basic role for the justification of the main results.
In what follows we consider separately the single-phase $(m = 1)$
and the multiphase $(m \geq 2)$ cases. The following lemma can be 
formulated for the single-phase case $m = 1$:

\vspace{0.2cm}

{\bf Lemma 3.5.}

{\it Let $\Lambda$ be a regular Hamiltonian family of single-phase
solutions of (\ref{EvInSyst}) and $(I^{1}, \dots, I^{N})$ be a complete
Hamiltonian set of the first integrals of the form (\ref{Integrals}). 
Then the functions  $B_{j[{\bf q}]} (\theta , {\bf X})$
can be found from system (\ref{BAcond}) and can be written in 
the form 
\begin{multline}
\label{B1ApprSol}
B_{i[{\bf q}]} (\theta, {\bf X}) \,\, =  \,\, 
\beta^{\mu, \, p}_{i} (\theta, {\bf U}({\bf X})) \,\,
q_{\mu , X^{p}}({\bf X}) \,\, + \,\, 
\beta^{\mu , \, p q}_{i , \, \alpha}
(\theta, {\bf U}({\bf X})) \,
S^{\alpha}_{X^{p} X^{q}} \,\, q_{\mu}({\bf X}) \,\, + \\ 
+ \,\, \beta^{\mu , \, p}_{i , \, \gamma}
(\theta, {\bf U}({\bf X})) \,
U^{\gamma}_{X^{p}} \,\, q_{\mu}({\bf X})
\end{multline}
(summation in $\mu = 1, \dots, N$, $p, q = 1, \dots, d$,
$\gamma = 1, \dots, m + s$)
with smooth dependence on the parameters ${\bf U}({\bf X})$.}

\vspace{0.2cm}

 Proof.

 System (\ref{BAcond}) in the single-phase case is a system of ordinary 
differential equations in $\theta$ with a skew-symmetric operator
${\hat B}^{ij}_{[0]} ({\bf X})$. It is easy to see also that the 
right-hand side of system (\ref{BAcond}) has the form
\begin{multline*}
-  A^{i}_{[1][{\bf q}]} (\theta, {\bf X}) \,\, = \,\,
\xi^{i \mu, \, p} (\theta, {\bf U}({\bf X})) \,
q_{\mu , X^{p}}({\bf X}) \,\, + \,\,
\xi^{i \mu , \, p q}_{\alpha}
(\theta, {\bf U}({\bf X})) \,
S^{\alpha}_{X^{p} X^{q}} \,\, q_{\mu}({\bf X}) \,\, + \\
+ \,\, \xi^{i \mu , \, p}_{\gamma}
(\theta, {\bf U}({\bf X})) \,
U^{\gamma}_{X^{p}} \,\, q_{\mu}({\bf X}) 
\end{multline*}
with periodic in $\theta$ functions 
$\xi^{i \mu, \, p} (\theta, {\bf U}({\bf X}))$,
$\xi^{i \mu , \, p q}_{\alpha} (\theta, {\bf U}({\bf X}))$,
$\xi^{i \mu , \, p}_{\gamma} (\theta, {\bf U}({\bf X}))$.

 The orthogonality conditions (\ref{LeftVecAort}) imply then the
orthogonality of all the sets of 
functions  
$\xi^{i \mu, \, p} (\theta, {\bf U}({\bf X}))$,
$\xi^{i \mu , \, p q}_{\alpha} (\theta, {\bf U}({\bf X}))$,
$\xi^{i \mu , \, p}_{\gamma} (\theta, {\bf U}({\bf X}))$
to the functions
$\kappa^{(q)}_{i[{\bf U}({\bf X})]} (\theta)$, such that system
(\ref{BAcond}) can be split into independent inhomogeneous
systems:
\begin{equation}
\label{Bbetaxi1}
{\hat B}^{ij}_{[0]} ({\bf X}) \,\, \beta^{\mu, \, p}_{j} \left( 
\theta, {\bf U}({\bf X}) \right) \,\, = \,\, 
\xi^{i\mu, \, p} \left( \theta, {\bf U}({\bf X}) \right)
\end{equation}
\begin{equation}
\label{Bbetaxi2}
{\hat B}^{ij}_{[0]} ({\bf X}) \,\, 
\beta^{\mu, \, p q}_{j, \, \alpha} 
\left( \theta, {\bf U}({\bf X}) \right) \,\, = \,\,
\xi^{i\mu, \, p q}_{\alpha} \left( \theta, {\bf U}({\bf X}) \right)
\end{equation}
\begin{equation}
\label{Bbetaxi3}
{\hat B}^{ij}_{[0]} ({\bf X}) \,\, 
\beta^{\mu, \, p}_{j, \, \gamma}
\left( \theta, {\bf U}({\bf X}) \right) \,\, = \,\,
\xi^{i\mu, \, p}_{\gamma} \left( \theta, {\bf U}({\bf X}) \right)
\end{equation}
defining functions (\ref{B1ApprSol}).

 All the systems (\ref{Bbetaxi1}) - (\ref{Bbetaxi3})  
are systems of ordinary linear differential equations 
with periodic coefficients and a skew-symmetric operator 
${\hat B}^{ij}_{[0]} ({\bf X})$. The zero modes of the operator
${\hat B}^{ij}_{[0]} ({\bf X})$ are given by the variation
derivatives of annihilators of the bracket (\ref{MultDimPBr})
and are orthogonal to the right-hand parts of 
(\ref{Bbetaxi1}) - (\ref{Bbetaxi3}) according
to (\ref{RazlozhAnn}) and (\ref{LeftVecAort}). Eigenfunctions of 
${\hat B}^{ij}_{[0]} ({\bf X})$ form a basis in the space of
$2\pi$-periodic functions $\bm{\varphi} (\theta)$. Besides that,
the nonzero eigenvalues of ${\hat B}^{ij}_{[0]} ({\bf X})$ are 
separated from zero in this case. Thus, the $2\pi$-periodic 
functions 
$\beta^{\mu, \, p}_{i} (\theta, {\bf U}({\bf X}))$,
$\beta^{\mu , \, p q}_{i \, \alpha} (\theta, {\bf U}({\bf X}))$,
$\beta^{\mu , \, p}_{i \, \gamma} (\theta, {\bf U}({\bf X}))$
can be found from systems (\ref{Bbetaxi1}) - (\ref{Bbetaxi3}) up 
to the variation derivatives of the annihilators of the bracket
(\ref{MultDimPBr}). If we impose additional conditions of
orthogonality of 
$\beta^{\mu, \, p}_{i} (\theta, {\bf U}({\bf X}))$,
$\beta^{\mu , \, p q}_{i \, \alpha} (\theta, {\bf U}({\bf X}))$,
$\beta^{\mu , \, p}_{i \, \gamma} (\theta, {\bf U}({\bf X}))$
to the variation derivatives of the annihilators of the bracket
(\ref{MultDimPBr}) on the manifold of single-phase solutions
we can suggest a unique procedure of construction of these functions.
The functions
$\beta^{\mu, \, p}_{i} (\theta, {\bf U}({\bf X}))$,
$\beta^{\mu , \, p q}_{i \, \alpha} (\theta, {\bf U}({\bf X}))$,
$\beta^{\mu , \, p}_{i \, \gamma} (\theta, {\bf U}({\bf X}))$
then depend smoothly on the parameters ${\bf U} ({\bf X})$, 
which implies  the required properties for the functions
$B_{j[{\bf q}]} (\theta , {\bf X})$.

{\hfill Lemma 3.5 is proved.}

\vspace{0.2cm}

 Let us start now the investigation of system (\ref{BAcond}) in the
general multi-phase case.

\vspace{0.2cm}

{\bf Lemma 3.6.}

{\it Let $\Lambda$ be a regular Hamiltonian family of
$m$-phase solutions of system
(\ref{EvInSyst}) and $(I^{1}, \dots, I^{N})$ 
be a complete Hamiltonian set of commuting first integrals of
(\ref{EvInSyst}) having the form (\ref{Integrals}). Let for
${\bf U} \in {\cal M}$
$${\bf v}^{(l)}_{[{\bf U}]} (\bm{\theta}) \,\, = \,\,
\left( v^{(l)}_{1[{\bf U}]} (\bm{\theta}), \dots ,
v^{(l)}_{n[{\bf U}]} (\bm{\theta}) \right) \,\,\,\,\, , 
\,\,\,\,\,\,\,\, l = 1, \dots , s $$
be a complete set of linearly independent eigenvectors of 
the operator ${\hat B}^{ij}_{[0]} ({\bf U})$ on the torus
with zero eigenvalues, smoothly depending on 
$\bm{\theta}$. Then

\noindent
1) The number of the vectors 
${\bf v}^{(l)}_{[{\bf U}]} (\bm{\theta})$
is equal to the number of annihilators of the bracket
(\ref{MultDimPBr}) on the submanifold of 
$m$-phase solutions of (\ref{EvInSyst});

\noindent
2) The functions $A^{i}_{[1][{\bf q}]} (\bm{\theta} , {\bf X})$ 
are orthogonal to all the vectors
${\bf v}^{(l)}_{[{\bf U}({\bf X})]} (\bm{\theta})$, i.e.
$$\int_{0}^{2\pi}\!\!\!\!\!\dots\int_{0}^{2\pi}
v^{(l)}_{i[{\bf U}({\bf X})]}(\bm{\theta}) \,\,
A^{i}_{[1][{\bf q}]} (\bm{\theta} , {\bf X}) \,\,
{d^{m} \theta \over (2\pi)^{m}} \,\,\, \equiv \,\,\, 0 
\,\,\,\,\,\,\,\,\,\, , \,\,\,\,\,\,\,\,\,\,
\left( {\bf U} ({\bf X}) \, \in \, {\cal M} \right) $$

}

\vspace{0.2cm}

 Proof.

Consider the values of  ${\bf v}^{(l)}_{[{\bf U}]} (\bm{\theta})$
on any of the leaves of the foliation, defined by the set
$\{ {\bf k}_{q} ({\bf U}) \}$ on the torus $\mathbb{T}^{m}$. 
According to the definition of regular Hamiltonian
family of $m$-phase solutions of (\ref{EvInSyst}) the corresponding
functions $v^{(l)}_{i[{\bf U}]} 
({\bf k}_{1} x^{1} + \dots + {\bf k}_{d} x^{d} + \bm{\theta}_{0})$
should give the variation derivatives of some linear combination
of annihilators of the bracket (\ref{MultDimPBr}). We have then
for a fixed value of $\bm{\theta}_{0}$:
$$v^{(l)}_{i[{\bf U}]} 
({\bf k}_{j} \, x^{j} + \bm{\theta}_{0}) 
\,\, = \,\, \sum_{p=1}^{s} 
\alpha^{l}_{p} ({\bf U}, \bm{\theta}_{0}) \,\, \left. 
{\delta N^{p} \over \delta \varphi^{i}({\bf x})} 
\right|_{\bm{\varphi} = \bm{\Phi} 
({\bf k}_{j} \, x^{j} + \bm{\theta}_{0},{\bf U})}$$

 From relation (\ref{RazlozhAnn}) we have then
\begin{equation}
\label{valphankappa}
v^{(l)}_{i[{\bf U}]} ({\bf k}_{j} ({\bf U}) \, x^{j} + \bm{\theta}_{0}) 
\,\, = \,\, \sum_{p=1}^{s} \sum_{q=1}^{m+s} 
\alpha^{l}_{p} ({\bf U}, \bm{\theta}_{0}) 
\,\, n^{p}_{q} ({\bf U}) \,\,
\kappa^{(q)}_{i[{\bf U}]} \left({\bf k}_{j} ({\bf U}) \, x^{j} \, + \,
\bm{\theta}_{0} \right)   
\end{equation}

 By definition, for ${\bf U} \in {\cal M}$ the leaves of the foliation,
defined by the set $\{ {\bf k}_{q} ({\bf U}) \}$, 
are everywhere dense in $\mathbb{T}^{m}$. 
Since both the left- and the right-hand parts of
(\ref{valphankappa}) are smooth functions on $\mathbb{T}^{m}$,
we get then that they coincide on $\mathbb{T}^{m}$. We can put then
$\alpha^{l}_{p} ({\bf U}, \bm{\theta}_{0}) = \alpha^{l}_{p} ({\bf U})$
and write
\begin{equation}
\label{Sootndliavikappa}
v^{(l)}_{i[{\bf U}]} (\bm{\theta}) \,\, \equiv \,\,
\sum_{p=1}^{s} \sum_{q=1}^{m+s}
\alpha^{l}_{p} ({\bf U}) \,\, n^{p}_{q} ({\bf U}) \,\,
\kappa^{(q)}_{i[{\bf U}]} \left( \bm{\theta} \right)
\end{equation}

 It is easy to see also that any linear combination of the form
(\ref{Sootndliavikappa}) gives a regular eigenvector of the operator
${\hat B}^{ij}_{[0]} ({\bf U}({\bf X}))$ with zero eigenvalue.

 Statement (2) follows then from relation (\ref{LeftVecAort}) 
in view of the representation (\ref{Sootndliavikappa}).

{\hfill Lemma 3.6 is proved.}

\vspace{0.2cm}

 We can see, in particular, that the values
$A^{i}_{[1][{\bf q}]} (\bm{\theta}, {\bf X})$ are orthogonal
at any ${\bf X}$ to all the ``regular'' eigen-vectors
$v^{(l)}_{i[{\bf U}({\bf X})]} (\bm{\theta})$ of the operator
${\hat B}^{ij}_{[0]} ({\bf X})$ corresponding to the zero
eigen-value view the regular dependence of the values
$A^{i}_{[1][{\bf q}]} (\bm{\theta}, {\bf X})$ on the parameters
${\bf U} ({\bf X}) = ({\bf S}_{\bf X}, \, U^{1}({\bf X}), \dots,
U^{m+s}({\bf X}))$.

 However, despite the presence of Lemma 3.6, study of
system (\ref{BAcond}) is much more complicated
in the multiphase case if compared with the single-phase case.
Thus, the presence of ``resonances'' may lead to appearance 
of small eigenvalues of the operator 
${\hat B}^{ij}_{[0]} ({\bf U}({\bf X}))$
for some values of the parameters ${\bf U} ({\bf X})$. 
As a result, this circumstance may lead to insolubility of 
system (\ref{BAcond}) in the space of smooth periodic
(in all $\theta^{\alpha}$) functions for the corresponding values
of ${\bf U}({\bf X})$. The solubility of system (\ref{BAcond}) 
is thus determined by the properties of the operator 
${\hat B}^{ij}_{[0]} ({\bf U}({\bf X}))$
for the given values of ${\bf U}({\bf X})$. 

The set of the ``resonant''
values of ${\bf U}$, as a rule, has measure zero in the full
space of parameters.
Let us prove here the following Theorem which shows that the 
procedure of the bracket averaging is in fact insensitive
to the appearance of the resonant values of ${\bf U}$ and can be used
in most multi-phase cases, as well as in the single-phase case.

\vspace{0.2cm}

{\bf Theorem 3.1.}

{\it
 Let system (\ref{EvInSyst}) be a local Hamiltonian system
generated by the functional (\ref{MultDimHamFunc}) in the local
field-theoretic Hamiltonian structure (\ref{MultDimPBr}). 
Let $\Lambda$ be a regular Hamiltonian family of
$m$-phase solutions of (\ref{EvInSyst}) and $(I^{1}, \dots , I^{N})$
be a complete Hamiltonian set of commuting integrals 
(\ref{Integrals}) for this family.  

 Let the parameter space ${\bf U}$ of the family $\Lambda$
have a dense set ${\cal S} \subset {\cal M}$ on which system 
(\ref{BAcond}) is solvable in the space
of smooth $2\pi$-periodic in each $\theta^{\alpha}$ functions.
Then the bracket 
\begin{multline}
\label{AveragedBracket}
\left\{ S^{\alpha} ({\bf X})  ,  S^{\beta} ({\bf Y}) \right\}
 =  0 \, , \,\,\,\,\,
\left\{ S^{\alpha} ({\bf X})  , 
U^{\gamma} ({\bf Y}) \right\} \, = \,
\omega^{\alpha\gamma} 
\left({\bf S}_{\bf X},
U^{1}({\bf X}), \dots, U^{m+s}({\bf X}) \! \right)
\, \delta ({\bf X} - {\bf Y}) \, , \\  \\
\left\{ U^{\gamma} ({\bf X})\, , \, U^{\rho} ({\bf Y}) \right\}
\,\,\, = \,\,\, \langle A^{\gamma\rho}_{10\dots0} \rangle
\left( {\bf S}_{\bf X}, \,
U^{1}({\bf X}), \dots, U^{m+s}({\bf X}) \right) \,\,
\delta_{X^{1}} ({\bf X} - {\bf Y}) \,\,\, + \,\, \dots \,\, + \\  \\
+ \,\,\, \langle A^{\gamma\rho}_{0\dots01} \rangle
\left( {\bf S}_{\bf X}, \,
U^{1}({\bf X}), \dots, U^{m+s}({\bf X}) \right) \,\,\,
\delta_{X^{d}} ({\bf X} - {\bf Y}) \,\,\, +  \\   \\
+ \,\,\, \left[ \langle Q^{\gamma\rho \, p} \rangle
\left( {\bf S}_{\bf X}, \,
U^{1}({\bf X}), \dots, U^{m+s}({\bf X}) \right)
\right]_{X^{p}} \,\,\, \delta ({\bf X} - {\bf Y})
\,\,\,\,\,\,\,\, , \,\,\,\,\,\,\,\,\,\,\,\,\,\,\,
\gamma, \rho \, = \, 1, \dots , m + s \,\,\,\,\, ,
\end{multline}
obtained with the aid
of the functionals $(I^{1}, \dots , I^{N})$, satisfies the Jacobi
identity.
}

\vspace{0.2cm}

 Proof.

 As before, for a smooth compactly supported vector-valued function
\linebreak ${\bf q} ({\bf X}) = (q_{1}({\bf X}), \dots , 
q_{N}({\bf X}))$ 
we define the functional
$$J_{[{\bf q}]} \,\, = \,\, \int
J^{\nu} ({\bf X}) \,\, q_{\nu} ({\bf X}) \,\, d^{d} X $$
(summation in $\nu = 1, \dots, N$).

 Then, for arbitrary smooth, compactly supported in ${\bf X}$ and
$2\pi$-periodic in each $\theta^{\alpha}$ functions
${\tilde {\bf Q}} (\bm{\theta}, {\bf X}) = 
({\tilde Q}_{1}(\bm{\theta}, {\bf X}), \dots , 
{\tilde Q}_{n}(\bm{\theta}, {\bf X}))$
we define the functionals
\begin{multline*}
Q_{i}(\bm{\theta}, {\bf X}) \,\, = \\
= \,\, 
{\tilde Q}_{i}(\bm{\theta}, {\bf X}) \, - \,
\Phi^{i}_{\theta^{\beta}} (\bm{\theta}, {\bf U}({\bf X})) \,\,
M^{\beta\gamma} ({\bf U}({\bf X})) \, 
\int_{0}^{2\pi}\!\!\!\!\!\dots\int_{0}^{2\pi}
{\tilde Q}_{j}(\bm{\theta}^{\prime}, {\bf X}) \,
\Phi^{j}_{\theta^{\prime\gamma}} 
(\bm{\theta}^{\prime}, {\bf U}({\bf X})) \,\,
{d^{m} \theta^{\prime} \over (2\pi)^{m}} 
\end{multline*}
where the matrix $M^{\beta\gamma} ({\bf U})$ is the inverse of the
matrix
$$\int_{0}^{2\pi}\!\!\!\!\!\dots\int_{0}^{2\pi} \sum_{i=1}^{n} \,
\Phi^{i}_{\theta^{\beta}} (\bm{\theta}, {\bf U}) \,\,
\Phi^{i}_{\theta^{\gamma}} (\bm{\theta}, {\bf U}) \,\,
{d^{m} \theta \over (2\pi)^{m}} $$
which is always defined according to the definition of a complete
regular family of $m$-phase solutions of system (\ref{EvInSyst}).

 By definition, the functions $Q_{i}(\bm{\theta}, {\bf X})$ are local 
functionals of ${\bf U}({\bf X})$
$$Q_{i}(\bm{\theta}, {\bf X}) \,\,\, \equiv \,\,\,
Q_{i}(\bm{\theta}, {\bf X}, {\bf U}({\bf X})) $$
depending also on the arbitrary fixed functions
${\tilde {\bf Q}} (\bm{\theta}, {\bf X})$. 
Everywhere below we will assume
that ${\bf Q} (\bm{\theta}, {\bf X})$ 
is a functional of this type defined
with some function ${\tilde {\bf Q}} (\bm{\theta}, {\bf X})$.

 Easy to see that the values of $Q_{i}(\bm{\theta}, {\bf X})$ 
with arbitrary ${\tilde {\bf Q}} (\bm{\theta}, {\bf X})$ 
represent for fixed values of the
functionals ${\bf U}({\bf Z})$ all possible smooth, compactly 
supported in ${\bf X}$ and $2\pi$-periodic in each $\theta^{\alpha}$ 
functions with the only restriction
\begin{equation}
\label{OgranQ}
\int_{0}^{2\pi}\!\!\!\!\!\dots\int_{0}^{2\pi}
Q_{i} (\bm{\theta}, {\bf X}) \,\, \Phi^{i}_{\theta^{\alpha}} 
(\bm{\theta}, {\bf U}({\bf X})) \,\, {d^{m} \theta \over (2\pi)^{m}}
\,\, = \,\, 0 \,\,\,\,\,\,\,\,\,\, , \,\,\, \forall \, {\bf X} 
\,\,\, , \,\,\,\,\,\,\,\, \alpha = 1, \dots , m 
\end{equation}

 For the functionals $Q_{i}(\bm{\theta}, {\bf X})$ 
we define the functionals
$$g_{[{\bf Q}]} \,\, = \,\, \int
\int_{0}^{2\pi}\!\!\!\!\!\dots\int_{0}^{2\pi} 
g^{i} (\bm{\theta}, {\bf X}) \,\, 
Q_{i} \left( {{\bf S}({\bf X}) \over \epsilon} + \bm{\theta}, 
{\bf X} \right) \,\, {d^{m} \theta \over (2\pi)^{m}} \, d^{d} X $$

 Now, for fixed functions ${\bf q} ({\bf X})$, ${\bf p} ({\bf X})$, 
and functional ${\bf Q} (\bm{\theta}, {\bf X})$ 
consider the Jacobi identity of the form
\begin{equation}
\label{JacobigJJ}
\left\{ g_{[{\bf Q}]} \, , \, \left\{ J_{[{\bf q}]} \, , \, J_{[{\bf p}]}
\right\} \right\} \, + \, 
\left\{ J_{[{\bf p}]} \, , \, \left\{g_{[{\bf Q}]} \, , \, J_{[{\bf q}]}
\right\} \right\} \, + \,
\left\{ J_{[{\bf q}]} \, , \, \left\{ J_{[{\bf p}]} \, , \, g_{[{\bf Q}]}
\right\} \right\} \,\, \equiv \,\, 0
\end{equation}

 Expanding the values of the brackets
$\{J^{\nu}({\bf X}) \, , \, J^{\mu} ({\bf Y})\}$ 
in the neighborhood of the submanifold
${\cal K}$, we can write:
$$\left\{ J^{\nu}({\bf X}) \, , \, J^{\mu} ({\bf Y}) \right\} 
\,\, = \,\, \left. 
\left\{ J^{\nu}({\bf X}) \, , \, J^{\mu} ({\bf Y}) \right\} 
\right|_{\cal K} \,\, + $$
\begin{multline*}
+ \,\, \sum_{l_{1}, \dots, l_{d}} \left. \left[ 
{\delta \over \delta \varphi^{k} (\bm{\theta}, {\bf W})}
\int_{0}^{2\pi}\!\!\!\!\!\dots\int_{0}^{2\pi}
A^{\nu\mu}_{l_{1} \dots l_{d}} 
\left( \bm{\varphi} (\bm{\theta}^{\prime}, {\bf X}),
\epsilon \, \bm{\varphi}_{\bf X} (\bm{\theta}^{\prime}, {\bf X}), 
\dots \right) \,\, {d^{m} \theta^{\prime} \over (2\pi)^{m}} \right]
\right|_{\cal K} \, \times \\
\times \,\, g^{k} (\bm{\theta}, {\bf W}) \,\, 
{d^{m} \theta \over (2\pi)^{m}} \, d^{d} W \,\,\, 
\epsilon^{l_{1} + \dots + l_{d}} \,\,
\delta^{(l_{1})} (X^{1} - Y^{1}) \, \dots \,
\delta^{(l_{d})} (X^{d} - Y^{d})
\,\,\, + \,\,\, O ({\bf g}^{2})
\end{multline*}
where all the values on the submanifold ${\cal K}$ are calculated 
at the same values of the functionals 
$[{\bf U} ({\bf Z})]$ as for the original function
$\bm{\varphi} (\bm{\theta}, {\bf X})$.

 Let us introduce by definition
\begin{equation}
\label{JqJpVarDeriv}
\left. {\delta \{ J_{[{\bf q}]} \, , \, J_{[{\bf p}]} \} \over
\delta g^{k} (\bm{\theta}, {\bf W}) }\right|_{\cal K} \,\, \equiv 
\end{equation}
\begin{multline*}
\equiv  \sum_{l_{1}, \dots, l_{d}} \int \!\! \left. \left[
{\delta \over \delta \varphi^{k} (\bm{\theta}, {\bf W})}
\int_{0}^{2\pi}\!\!\!\!\!\dots\int_{0}^{2\pi} \!\! 
\epsilon^{l_{1} + \dots + l_{d}} \,
A^{\nu\mu}_{l_{1} \dots l_{d}}
\left( \bm{\varphi} (\bm{\theta}^{\prime}, {\bf X}),
\epsilon \bm{\varphi}_{\bf X} (\bm{\theta}^{\prime}, {\bf X}), 
\dots \right) \, {d^{m} \theta^{\prime} \over (2\pi)^{m}} \right]
\! \right|_{\cal K} \times  \\
\times \,\, q_{\nu}({\bf X}) \, 
p_{\mu, \, l_{1} X^{1} \dots l_{d} X^{d}} ({\bf X}) \,\, d^{d} X 
\end{multline*}

 Note that the notations 
$\delta / \delta g^{k} (\bm{\theta}, {\bf W})$,
generally speaking, are not natural in our situation because of
the dependence of the chosen system of constraints. Nevertheless, 
the preservation of these notations can better clarify the algebraic 
structure of the further calculations.

 The values defined by (\ref{JqJpVarDeriv}) 
can be represented in the form of the graded decompositions at 
$\epsilon \rightarrow 0$ on the submanifold ${\cal K}$.
By virtue of (\ref{epsAQRel}) it is easy to conclude that the 
expansion in $\epsilon$ of the quantities (\ref{JqJpVarDeriv}) begins 
with the first degree in $\epsilon$
\begin{equation}
\label{JqJpSkobRazl}
\left. {\delta \{ J_{[{\bf q}]} \, , \, J_{[{\bf p}]} \} \over
\delta g^{k} (\bm{\theta}, {\bf W}) }\right|_{\cal K} \,\,\, = \,\,\,
\epsilon \, \left. {\delta \{ J_{[{\bf q}]} \, , \, J_{[{\bf p}]} \} 
\over \delta g^{k} (\bm{\theta}, {\bf W}) }\right|_{{\cal K}[1]} \,\,
+ \,\, \epsilon^{2} \,
\, \left. {\delta \{ J_{[{\bf q}]} \, , \, J_{[{\bf p}]} \}
\over \delta g^{k} (\bm{\theta}, {\bf W}) }\right|_{{\cal K}[2]} \,\, 
+ \,\,\, \dots
\end{equation}

 The leading term of (\ref{JqJpSkobRazl}) can be divided into 
two parts, corresponding to the sets of the functions 
$$\left( \, 
\epsilon \, Q^{\nu\mu \, 1}_{X^{1}} (\bm{\varphi}, \epsilon \, 
\bm{\varphi}_{\bf X}, \dots ) \, , \, \dots \, , \,
\epsilon \, Q^{\nu\mu \, d}_{X^{d}} (\bm{\varphi}, \epsilon \,
\bm{\varphi}_{\bf X}, \dots ) \, \right) $$ 
and 
$$\left( \, \epsilon \, A^{\nu\mu}_{10\dots0} 
(\bm{\varphi}, \epsilon \, \bm{\varphi}_{\bf X}, \dots )
\, , \, \dots \, , \, \epsilon \, A^{\nu\mu}_{0\dots01}
(\bm{\varphi}, \epsilon \, \bm{\varphi}_{\bf X}, \dots ) \,
\right) $$
and containing the quantities $q_{\nu}({\bf W}) p_{\mu}({\bf W})$ 
and $q_{\nu}({\bf W}) p_{\mu, \, W^{l}}({\bf W})$ 
as local factors, respectively.

 The values of $\{ J_{[{\bf q}]} \, , \, J_{[{\bf p}]} \}$, are 
obviously invariant under transformations of the form
\begin{equation}
\label{DeltaSdvig}
\bm{\varphi} (\bm{\theta}, {\bf X}) \,\,\, \rightarrow \,\,\,
\bm{\varphi} (\bm{\theta} + \Delta \bm{\theta} , {\bf X}) 
\end{equation}

 From the invariance of the functionals ${\bf U}({\bf X})$ in
such transformations, we can write for the corresponding
increments of constraints (\ref{gconstr})
$$\delta g^{k} (\bm{\theta}, {\bf X}) \,\,\, = \,\,\,
\varphi^{k} (\bm{\theta} + \delta \bm{\theta} , {\bf X}) \, - \,
\varphi^{k} (\bm{\theta}, {\bf X}) $$

 As a consequence, we can write on the submanifold ${\cal K}$
\begin{equation}
\label{JqJpOrtPhi1}
\int\!\!\!\int_{0}^{2\pi}\!\!\!\!\!\dots\int_{0}^{2\pi} \! 
\left. {\delta \{ J_{[{\bf q}]} , J_{[{\bf p}]} \} \over
\delta g^{k} (\bm{\theta}, {\bf W}) }\right|_{\cal K} 
\Phi^{k}_{\theta^{\alpha}} \!
\left( { {\bf S}({\bf W}) \over \epsilon} + \bm{\theta}, \,
{\bf S}_{\bf W}, U^{1}({\bf W}), \dots, U^{m+s}({\bf W})
\!\! \right) {d^{m} \theta \over (2\pi)^{m}}
d^{d} W \, \equiv \, 0 
\end{equation}
$\alpha = 1, \dots , m$.

 The above relation is satisfied to all orders in $\epsilon$.
By the arbitrariness of the functions $q_{\nu} ({\bf X})$, relation
(\ref{JqJpOrtPhi1}) in the leading order can be strengthened.
Namely, according to the remark about the form of the leading term of 
(\ref{JqJpSkobRazl}) we can write for any ${\bf W}$
\begin{equation}
\label{JqJpVarDerOrtRel}
\int_{0}^{2\pi}\!\!\!\!\!\dots\int_{0}^{2\pi}
\left. {\delta \{ J_{[{\bf q}]} , J_{[{\bf p}]} \} \over
\delta g^{k} (\bm{\theta}, {\bf W}) }\right|_{{\cal K}[1]} \,\,
\Phi^{k}_{\theta^{\alpha}} \left( { {\bf S}({\bf W}) \over \epsilon} +
\bm{\theta}, \,
{\bf S}_{\bf W}, U^{1}({\bf W}), \dots, U^{m+s}({\bf W})
\right) \, {d^{m} \theta \over (2\pi)^{m}}
\,\, \equiv \,\, 0 
\end{equation}
$\alpha = 1, \dots , m$.

 At the same time we have the relations
$$\left. {\delta \{ J_{[{\bf q}]} , J_{[{\bf p}]} \} \over
\delta S^{\alpha} ({\bf W}) }\right|_{\cal K} \,\, = \,\,
O (\epsilon) \,\,\,\,\,\,\,\, , \,\,\,\,\,\,\,\,
\left. {\delta \{ J_{[{\bf q}]} , J_{[{\bf p}]} \} \over
\delta U^{\gamma} ({\bf W}) }\right|_{\cal K} \,\, = \,\,
O (\epsilon) $$
on the submanifold ${\cal K}$.

 Quite similarly, we have the relation
$$\left\{ g_{[{\bf Q}]} \, , \, J_{[{\bf q}]} \right\} \,\, = \,\,
\int Q_{i} \left( 
{{\bf S}({\bf X}) \over \epsilon} + \bm{\theta}, {\bf X} \right)
\,\, \left\{ g^{i} (\bm{\theta}, {\bf X}) \, , \, 
J_{[{\bf q}]} \right\} \,\,
{d^{m} \theta \over (2\pi)^{m}} \, d^{d} X \,\,\, +  $$
$$+ \,\, \int g^{i} (\bm{\theta}, {\bf X}) \,\, Q_{i,\theta^{\alpha}} 
\left( {{\bf S}({\bf X}) \over \epsilon} + \bm{\theta}, 
{\bf X} \right) \,\,
{1 \over \epsilon} \,\, 
\left\{ S^{\alpha}({\bf X}) \, , \, J_{[{\bf q}]} \right\} \,\,
{d^{m} \theta \over (2\pi)^{m}} \, d^{d} X \,\,\, + $$
$$+ \,\, \int g^{i} (\bm{\theta}, {\bf X}) \,\, Q_{i, U^{\nu}}
\left( {{\bf S}({\bf X}) \over \epsilon} + \bm{\theta}, 
{\bf X} \right) \,\,
\left\{ U^{\nu}({\bf X}) \, , \, J_{[{\bf q}]} \right\} \,\,
{d^{m} \theta \over (2\pi)^{m}} \, d^{d} X \,\, = $$

\begin{multline*}
=  \! \int \! 
Q_{i} \left( {{\bf S}({\bf X}) \over \epsilon} + \bm{\theta}, 
{\bf X} \! \right) \,\, \times \\
\times \,
\sum_{l_{1}, \dots, l_{d}} \epsilon^{l_{1} + \dots + l_{d}} \, 
C^{i\mu}_{(l_{1} \dots l_{d})} \left( 
\bm{\varphi} (\bm{\theta}, {\bf X}), \epsilon \, \bm{\varphi}_{\bf X} 
(\bm{\theta}, {\bf X}), \dots \right) \, 
{d^{m} \theta \over (2\pi)^{m}} 
\,\, q_{\mu, \, l_{1} X^{1} \dots l_{d} X^{d}}({\bf X}) 
\, d^{d} X \,\, - 
\end{multline*}
\begin{multline*}
- \, \int
Q_{i} \left( {{\bf S}({\bf X}) \over \epsilon} + \bm{\theta},
{\bf X} \right) \, \times \\
\times \,
\Phi^{i}_{k^{\alpha}_{p}} \left( {{\bf S}({\bf X}) \over \epsilon} +
\bm{\theta}, \, {\bf S}_{\bf X}, U^{1}({\bf X}), \dots, U^{m+s}({\bf X})
\right) \, \left\{ S^{\alpha}_{X^{p}}({\bf X}) \, , \, 
J_{[{\bf q}]} \right\} \, 
{d^{m} \theta \over (2\pi)^{m}} \,\, d^{d} X \,\, - 
\end{multline*}
\begin{multline*}
- \,\, \int
Q_{i} \left( {{\bf S}({\bf X}) \over \epsilon} + \bm{\theta}, 
{\bf X} \right) \, \times \\
\times \, 
\Phi^{i}_{U^{\gamma}} \left( {{\bf S}({\bf X}) \over \epsilon} +
\bm{\theta}, \, {\bf S}_{\bf X}, U^{1}({\bf X}), \dots, U^{m+s}({\bf X})
\right) \,\,
\left\{ U^{\gamma}({\bf X}) \, , \, J_{[{\bf q}]} \right\} \,\,
{d^{m} \theta \over (2\pi)^{m}} \,\, d^{d} X \,\, - 
\end{multline*}
\begin{multline*}
- \, \int \!
Q_{i} \left( {{\bf S}({\bf X}) \over \epsilon} + \bm{\theta}, 
{\bf X} \right) \, \times \\
\times \,
\Phi^{i}_{\theta^{\alpha}} \left( {{\bf S}({\bf X}) \over \epsilon} +
\bm{\theta}, \, {\bf S}_{\bf X}, U^{1}({\bf X}), \dots, U^{m+s}({\bf X})
\right) \, {1 \over \epsilon} \, 
\left\{ S^{\alpha}({\bf X}) \, , \, J_{[{\bf q}]} \right\} \,
{d^{m} \theta \over (2\pi)^{m}} \, d^{d} X \,\, + 
\end{multline*}
$$+ \, \int g^{i} (\bm{\theta}, {\bf X}) \,\, Q_{i,\theta^{\alpha}}
\left( {{\bf S}({\bf X}) \over \epsilon} + \bm{\theta}, 
{\bf X} \right) \,\, {1 \over \epsilon} \,\, 
\left\{ S^{\alpha}({\bf X}) \, , \, J_{[{\bf q}]} \right\} \,\,
{d^{m} \theta \over (2\pi)^{m}} \,\, d^{d} X \,\, + $$
$$+ \,\, \int g^{i} (\bm{\theta}, {\bf X}) \,\, Q_{i, U^{\nu}}
\left( {{\bf S}({\bf X}) \over \epsilon} + \bm{\theta}, 
{\bf X} \right) \,\,
\left\{ U^{\nu}({\bf X}) \, , \, J_{[{\bf q}]} \right\} \,\,
{d^{m} \theta \over (2\pi)^{m}} \, d^{d} X $$

 Let us note that the summation in $\gamma$ is made here for
$\gamma = 1, \dots, m + s$, while the summation in $\nu$ is made
for the full set of $U^{\nu}$, $\, \nu = 1, \dots, N$.

 According to relations (\ref{OgranQ}), we can actually see here 
that the fourth term in the above expression is identically equal to 
zero.

 According to the form of the constraints,
we have again in the neighborhood of ${\cal K}$:
$$\{ g_{[{\bf Q}]} , J_{[{\bf q}]} \} \,\,\, = \,\,\,
\left. \{ g_{[{\bf Q}]} , J_{[{\bf q}]} \}\right|_{\cal K} \, + $$
\begin{multline*}
+  \left[ \int Q_{i} \left( { {\bf S}({\bf X}) \over \epsilon}
+ \bm{\theta}^{\prime}, {\bf X} \right) \right. \,\, \times\\
\times \, \left. \left.
\sum_{l_{1}, \dots, l_{d}} \epsilon^{l_{1} + \dots + l_{d}}
\,\, {\delta C^{i\mu}_{(l_{1} \dots l_{d})} 
(\bm{\varphi}(\bm{\theta}^{\prime}, {\bf X}),
\epsilon \, \bm{\varphi}_{\bf X}(\bm{\theta}^{\prime}, {\bf X}), \dots)
\over \delta \varphi^{k} (\bm{\theta}, {\bf W})} \,
{d^{m} \theta^{\prime} \over (2\pi)^{m}} \,\,
q_{\mu, l_{1} X^{1} \dots l_{d} X^{d}}({\bf X}) \,\, d^{d} X
\right] \right|_{\cal K} \times \\
\times \,\, g^{k} (\bm{\theta}, {\bf W}) \,\,
{d^{m} \theta \over (2\pi)^{m}} \,\, d^{d} W \,\, -
\end{multline*}
\begin{multline*}
- \, \int
Q_{i} \left( {{\bf S}({\bf X}) \over \epsilon} + \bm{\theta}^{\prime}, 
{\bf X} \right) \,
\Phi^{i}_{k^{\alpha}_{p}} \left( {{\bf S}({\bf X}) \over \epsilon} +
\bm{\theta}^{\prime}, \,  
{\bf S}_{\bf X}, U^{1}({\bf X}), \dots, U^{m+s}({\bf X})
\right) \, {d^{m} \theta^{\prime} \over (2\pi)^{m}} \,\, \times \\
\times \,\, \left. {\{ S^{\alpha}_{X^{p}}({\bf X}) \, , \,
J_{[{\bf q}]} \} \over \delta \varphi^{k} (\bm{\theta}, {\bf W})}
\right|_{\cal K} \, d^{d} X \,\, 
\times \,\, g^{k} (\bm{\theta}, {\bf W}) \,\,
{d^{m} \theta \over (2\pi)^{m}} \,\, d^{d} W \,\, - 
\end{multline*}
\begin{multline*}
- \,\, \int Q_{i} (\bm{\theta}^{\prime}, {\bf X}) \,\,
\Phi^{i}_{U^{\gamma}} \left( \bm{\theta}^{\prime}, \,
{\bf S}_{\bf X}, U^{1}({\bf X}), \dots, U^{m+s}({\bf X})  \right)
\,\, {d^{m} \theta^{\prime} \over (2\pi)^{m}} \,\, \left.
{\delta \{ U^{\gamma}({\bf X}) , J_{[{\bf q}]} \}
\over \delta \varphi^{k} (\bm{\theta}, {\bf W})} \right|_{\cal K} 
\, d^{d} X \,\, \times \\
\times \,\, g^{k} (\bm{\theta}, {\bf W}) \,\,
{d^{m} \theta \over (2\pi)^{m}} \,\, d^{d} W \,\, + 
\end{multline*}
$$+ \,\, \int Q_{k, \theta^{\alpha}} 
\left({{\bf S}({\bf W}) \over \epsilon} +
\bm{\theta}, {\bf W} \right) \,\, {1 \over \epsilon} \,\,
\left. \left\{ S^{\alpha}({\bf W}) \, , \,
J_{[{\bf q}]} \right\} \right|_{\cal K} \,\,
\times \,\, g^{k} (\bm{\theta}, {\bf W}) \,\,
{d^{m} \theta \over (2\pi)^{m}} \, d^{d} W \,\, +$$
$$+ \,\, \int Q_{i, U^{\nu}}
\left( {{\bf S}({\bf W}) \over \epsilon} + \bm{\theta}, {\bf W} \right) 
\,\, \left. \left\{ U^{\nu}({\bf W}) \, , \, J_{[{\bf q}]} \right\} 
\right|_{\cal K} \,\, \times \,\, g^{k} (\bm{\theta}, {\bf W}) \,\,
{d^{m} \theta \over (2\pi)^{m}} \, d^{d} W \
\,\,\, + \,\,\, O({\bf g}^{2})$$
provided that all the values on the submanifold ${\cal K}$
are calculated at the same values of the functionals 
$[{\bf U}({\bf Z})]$.

 For the quantities $\{ g_{[{\bf Q}]} , J_{[{\bf q}]} \}$ 
we can introduce by definition
\begin{equation}
\label{ManyPhasegJvarder}
\left. {\delta \{ g_{[{\bf Q}]} , J_{[{\bf q}]} \} \over \delta 
g^{k} (\bm{\theta}, {\bf W})} \right|_{\cal K} \,\,\, \equiv 
\end{equation}
\begin{multline*}
\equiv \,\,  \left[ \int Q_{i} \left( { {\bf S}({\bf X}) \over \epsilon}
+ \bm{\theta}^{\prime}, {\bf X} \right) \right. \,\, \times\\
\times \, \left. \left.
\sum_{l_{1}, \dots, l_{d}} \epsilon^{l_{1} + \dots + l_{d}}
\,\, {\delta C^{i\mu}_{(l_{1} \dots l_{d})}
(\bm{\varphi}(\bm{\theta}^{\prime}, {\bf X}),
\epsilon \, \bm{\varphi}_{\bf X}(\bm{\theta}^{\prime}, {\bf X}), \dots)
\over \delta \varphi^{k} (\bm{\theta}, {\bf W})} \,\,
q_{\mu, l_{1} X^{1} \dots l_{d} X^{d}}({\bf X}) 
\right] \right|_{\cal K} 
{d^{m} \theta^{\prime} \over (2\pi)^{m}} \,\,d^{d} X \,\, -
\end{multline*}
\begin{multline*}
- \int \!
Q_{i} \left( {{\bf S}({\bf X}) \over \epsilon} + \bm{\theta}^{\prime},
{\bf X} \! \right) \, \times \\
\times \,
\Phi^{i}_{k^{\alpha}_{p}} \left( {{\bf S}({\bf X}) \over \epsilon} +
\bm{\theta}^{\prime}, \,
{\bf S}_{\bf X}, U^{1}({\bf X}), \dots, U^{m+s}({\bf X}) \!
\right) {d^{m} \theta^{\prime} \over (2\pi)^{m}} 
\, \left. {\{ S^{\alpha}_{X^{p}}({\bf X}) ,
J_{[{\bf q}]} \} \over \delta \varphi^{k} (\bm{\theta}, {\bf W})}
\right|_{\cal K} d^{d} X \, - 
\end{multline*}
$$- \,\, \int Q_{i} (\bm{\theta}^{\prime}, {\bf X}) \,\,
\Phi^{i}_{U^{\gamma}} \left( \bm{\theta}^{\prime}, \,
{\bf S}_{\bf X}, U^{1}({\bf X}), \dots, U^{m+s}({\bf X})  \right)
\,\, {d^{m} \theta^{\prime} \over (2\pi)^{m}} \,\, \left. 
{\delta \{ U^{\gamma}({\bf X}) , J_{[{\bf q}]} \} 
\over \delta \varphi^{k} (\bm{\theta}, {\bf W})} 
\right|_{\cal K} \, d^{d} X
\,\, + $$
$$+ \, Q_{k, \theta^{\alpha}} 
\left({{\bf S}({\bf W}) \over \epsilon} +
\bm{\theta}, {\bf W} \right) \,\, {1 \over \epsilon} \,\,
\left. \left\{ S^{\alpha}({\bf W}) \, , \,
J_{[{\bf q}]} \right\} \right|_{\cal K} \,\, +$$
$$+ \,\, Q_{i, U^{\nu}}
\left( {{\bf S}({\bf W}) \over \epsilon} + \bm{\theta}, 
{\bf W} \right) \,\,  
\left. \left\{ U^{\nu}({\bf W}) \, , \, J_{[{\bf q}]} \right\} 
\right|_{\cal K} $$

 The quantities
$\delta \{ g_{[{\bf Q}]} , J_{[{\bf q}]} \} / \delta
g^{k} (\bm{\theta}, {\bf W}) |_{\cal K}$ have the order $O (1)$ at 
$\epsilon \rightarrow 0$. We have also
$$C^{i\mu}_{(0 \dots 0)} 
\left( \bm{\varphi} (\bm{\theta}^{\prime}, {\bf X}), 
\epsilon \, \bm{\varphi}_{\bf X} (\bm{\theta}^{\prime}, {\bf X}), 
\dots \right) \,\,\, \equiv \,\,\,
S^{i\mu} \left( \bm{\varphi} (\bm{\theta}^{\prime}, {\bf X}), 
\epsilon \, \bm{\varphi}_{\bf X} (\bm{\theta}^{\prime}, {\bf X}), 
\dots \right)$$
according to relation (\ref{C0SmuRel}).

 Let us introduce the functions
$$S^{i\mu}_{k (l_{1} \dots l_{d})} 
(\bm{\varphi}, \bm{\varphi}_{\bf x}, \dots ) \,\,\, \equiv \,\,\, 
{\partial S^{i\mu} (\bm{\varphi}, \bm{\varphi}_{\bf x}, \dots )
\over \partial \varphi^{k}_{l_{1} x^{1} \dots l_{d} x^{d}}} $$

 Using now relations (\ref{SalphaJmuSkob}) and
(\ref{JqJpSkobRazl}) we can write
for the leading term of (\ref{ManyPhasegJvarder})
$$\left. {\delta \{ g_{[{\bf Q}]} , J_{[{\bf q}]} \} \over \delta
g^{k} (\bm{\theta}, {\bf W})} \right|_{{\cal K}[0]} \,\,\, = $$
\begin{multline*}
\!\!\!  = \sum_{l_{1}, \dots, l_{d}}  (-1)^{l_{1} + \dots + l_{d}} \,\,
k^{\alpha^{1}_{1}}_{1}({\bf W}) \dots
k^{\alpha^{1}_{l_{1}}}_{1}({\bf W})
\, \dots \,
k^{\alpha^{d}_{1}}_{d}({\bf W}) \dots
k^{\alpha^{d}_{l_{d}}}_{d}({\bf W}) \,\, \times \\
\times \,\, {\partial \over \partial \theta^{\alpha^{1}_{1}}} \dots
{\partial \over \partial \theta^{\alpha^{1}_{l_{1}}}} \, \dots \,
{\partial \over \partial \theta^{\alpha^{d}_{1}}} \dots
{\partial \over \partial \theta^{\alpha^{d}_{l_{d}}}}
\, \times \\
\times \, \left[ Q_{i} \left( {{\bf S}({\bf W}) \over \epsilon} +
\bm{\theta}, {\bf W} \! \right)  \left. 
S^{i\mu}_{k (l_{1} \dots l_{d})} \left(
\bm{\varphi} (\bm{\theta}, {\bf W}), \epsilon
\bm{\varphi}_{\bf W} (\bm{\theta}, {\bf W}), \dots \right) 
\right|_{{\cal K}[0]}
\right] \, q_{\mu} ({\bf W}) \,\,\, +
\end{multline*}
\begin{equation}
\label{gJVarDerNulPor}
+ \,\,\, \omega^{\alpha\mu} ({\bf W}) \,\, q_{\mu} ({\bf W}) \,\, 
Q_{k,\theta^{\alpha}} \left( {{\bf S}({\bf W}) \over \epsilon} +
\bm{\theta}, {\bf W} \right) 
\end{equation}
where
\begin{multline*}
\left. S^{i\mu}_{k (l_{1} \dots l_{d})} \left(
\bm{\varphi} (\bm{\theta}, {\bf W}), \epsilon 
\bm{\varphi}_{\bf W} (\bm{\theta}, {\bf W}), \dots \right) 
\right|_{{\cal K}[0]} \, \equiv \\ 
\equiv \, S^{i\mu}_{k (l_{1} \dots l_{d})} \left( \!
\bm{\Phi} \! \left( {{\bf S}({\bf W}) \over \epsilon} +
\bm{\theta}, {\bf S}_{\bf W}, U^{1}({\bf W}), \dots \! , 
U^{m+s}({\bf W}) \!\! \right)\! , \dots  \! \right) 
\end{multline*}

 Note one more property of the values
$\delta \{ g_{[{\bf Q}]} , J_{[{\bf q}]} \} / \delta
g^{k} (\bm{\theta}, {\bf W}) |_{{\cal K}[0]}$. As we saw earlier,
the values  $\{ g_{[{\bf Q}]} , J_{[{\bf q}]} \}$ are of the order
$O (\epsilon)$ at $\epsilon \rightarrow 0$ on the submanifold
${\cal K}$. This property is preserved also under the overall shift
of the initial phase (\ref{DeltaSdvig}).

 Indeed, for
\begin{equation}
\label{PhiDeltaTheta}
\varphi^{i} (\bm{\theta}, \, {\bf X}) \,\,\, = \,\,\,
\Phi^{i} \left( {{\bf S}({\bf X}) \over \epsilon} + \bm{\theta} +
\Delta \bm{\theta}, \,\,
{\bf S}_{\bf X}, \, U^{1}({\bf X}), \dots, U^{m+s}({\bf X}) \right)
\end{equation}
we can write
\begin{multline*}
\left\{g_{[{\bf Q}]}, J_{[{\bf q}]} \right\} \, = \,
\int Q_{i, \theta^{\alpha}} \left( {{\bf S}({\bf Z}) \over \epsilon}
+ \bm{\theta}, {\bf Z} \right) \,\, \times \\
\times \,\, {1 \over \epsilon} \,
\left\{ S^{\alpha}({\bf Z}) , J_{[{\bf q}]} \right\} \,\,
\Phi^{i} \left( {{\bf S}({\bf Z}) \over \epsilon} + \bm{\theta} +
\Delta \bm{\theta} , \,
{\bf S}_{\bf Z}, \, U^{1}({\bf Z}), \dots, U^{m+s}({\bf Z}) 
\right) \, 
{d^{m} \theta \over (2\pi)^{m}} \, d^{d} Z \,\, + 
\end{multline*}
\begin{multline*}
+ \! \int \!\! Q_{i} \left( {{\bf S}({\bf Z}) \over \epsilon}
+ \bm{\theta}, {\bf Z} \right)  C^{i\mu}_{(0 \dots 0)} \left( \!
\bm{\Phi} \! \left( {{\bf S}({\bf Z}) \over \epsilon} + \bm{\theta} +
\Delta \bm{\theta}, \,
{\bf S}_{\bf Z}, \, U^{1}({\bf Z}), \dots, U^{m+s}({\bf Z}) \right),  
\dots \! \right) \! \times \\ 
\times \,\,  q_{\mu} ({\bf Z}) \, {d^{m} \theta \over (2\pi)^{m}} 
\, d^{d} Z \,\,\, + \,\,\, O (\epsilon) 
\end{multline*}

 Because of the invariance under translations (\ref{DeltaSdvig}) 
the value of the bracket $\{ S^{\alpha}({\bf Z}), J_{[{\bf q}]} \}$ 
on the functions (\ref{PhiDeltaTheta}) is equal to its value on
${\cal K}$
$$\left\{ S^{\alpha}({\bf Z})\, , \, J_{[{\bf q}]} \right\} \,\, = \,\,
\epsilon \,\, \omega^{\alpha\nu}({\bf Z}) \, q_{\nu} ({\bf Z}) 
\,\,\, + \,\,\, O(\epsilon^{2}) $$

 Similarly, we have on the functions (\ref{PhiDeltaTheta})
\begin{multline*}
C^{i\mu}_{(0 \dots 0)} \left( 
\bm{\Phi} \left( {{\bf S}({\bf Z}) \over \epsilon} + \bm{\theta} +
\Delta \bm{\theta}, \,
{\bf S}_{\bf Z}, \, U^{1}({\bf Z}), \dots, U^{m+s}({\bf Z})
\right), \dots \right) \, = \\
= \,\, \omega^{\alpha\mu}({\bf Z}) \,\,
\Phi^{i}_{\theta^{\alpha}} \left( {{\bf S}({\bf Z}) \over \epsilon} 
+ \bm{\theta} + \Delta \bm{\theta}, \,
{\bf S}_{\bf Z}, \, U^{1}({\bf Z}), \dots, U^{m+s}({\bf Z}) \right)
\end{multline*}

 We then obtain
$$\left\{g_{[{\bf Q}]}, J_{[{\bf q}]} \right\} = \int
q_{\mu}({\bf Z}) \,\, \omega^{\alpha\mu} ({\bf Z}) \,\, 
{\partial \over \partial \theta^{\alpha}} 
\left[ Q_{i} (\bm{\theta}, {\bf Z}) \, \Phi^{i}
(\bm{\theta} + \Delta \bm{\theta}, {\bf Z}) \right] \, 
{d^{m} \theta \over (2\pi)^{m}} \, d^{d} Z \,\, + \,\, O(\epsilon) 
\,\, = \,\, O(\epsilon) $$

 As a consequence, we can write
\begin{equation}
\label{SlabVarDerRel}
\int \!\!\! \int_{0}^{2\pi}\!\!\!\!\!\dots\int_{0}^{2\pi} \!
\left. {\delta \{ g_{[{\bf Q}]} , J_{[{\bf q}]} \} \over 
\delta g^{k} (\bm{\theta}, {\bf W})} \right|_{{\cal K}[0]} \!
\Phi^{k}_{\theta^{\alpha}} \!
\left( {{\bf S}({\bf W}) \over \epsilon} 
+ \bm{\theta}, \,
{\bf S}_{\bf W}, U^{1}({\bf W}), \dots, U^{m+s}({\bf W}) \!\!
\right) \! {d^{m} \theta \over (2\pi)^{m}} d^{d} W  \equiv  0 
\end{equation}
$\alpha = 1, \dots , m$,
for the main part of $\delta \{ g_{[{\bf Q}]} , J_{[{\bf q}]} \} / 
\delta g^{k} (\bm{\theta}, {\bf W})$ on ${\cal K}$.

  Using again the fact that relation (\ref{SlabVarDerRel}) contains
arbitrary functions $q_{\mu}({\bf W})$, appearing in the integrand
expression in the form of local factors, we can rewrite
(\ref{SlabVarDerRel}) in the stronger form
\begin{equation}
\label{SilnVarDerRel}
\int_{0}^{2\pi}\!\!\!\!\!\dots\int_{0}^{2\pi}
\left. {\delta \{ g_{[{\bf Q}]} , J_{[{\bf q}]} \} \over
\delta g^{k} (\bm{\theta}, {\bf W})} \right|_{{\cal K}[0]}
\Phi^{k}_{\theta^{\alpha}} \left( {{\bf S}({\bf W}) \over \epsilon} +
\bm{\theta}, \,
{\bf S}_{\bf W}, \, U^{1}({\bf W}), \dots, U^{m+s}({\bf W})
\right) \, {d^{m} \theta \over (2\pi)^{m}} \,
\,\, \equiv \,\, 0 
\end{equation}
$\forall \, {\bf W}$, $\alpha = 1, \dots , m$.

 At the same time we have
$$\left. {\delta \{ g_{[{\bf Q}]} , J_{[{\bf q}]} \} \over
\delta S^{\alpha} ({\bf W}) }\right|_{\cal K} \,\, = \,\,
O (\epsilon) \,\,\,\,\,\,\,\, , \,\,\,\,\,\,\,\,
\left. {\delta \{ g_{[{\bf Q}]} , J_{[{\bf q}]} \} \over
\delta U^{\gamma} ({\bf W}) }\right|_{\cal K} \,\, = \,\,
O (\epsilon) $$
on the submanifold ${\cal K}$.

 We now turn back to the Jacobi identity (\ref{JacobigJJ}) for
the functionals $g_{[{\bf Q}]}$, $J_{[{\bf q}]}$, and $J_{[{\bf p}]}$.
Using expansions of the values 
$\{ J_{[{\bf q}]} , J_{[{\bf p}]} \}$,
$\{ g_{[{\bf Q}]} , J_{[{\bf q}]} \}$,
$\{ g_{[{\bf Q}]} , J_{[{\bf p}]} \}$ in terms of
$ g^{k} (\bm{\theta}, {\bf W})$, $S^{\alpha} ({\bf W})$,
$ U^{\gamma} ({\bf W})$ near ${\cal K}$, 
it is not difficult to see that after the restriction on ${\cal K}$ 
the leading term ($\sim \epsilon$) of relation (\ref{JacobigJJ}) 
can be written as
\begin{multline*}
\int \left. \left\{ g_{[{\bf Q}]} \, , \, g^{k} (\bm{\theta}, {\bf W}) 
\right\} \right|_{{\cal K}[0]} \, \left.
{\delta \{ J_{[{\bf q}]} \, , \, J_{[{\bf p}]} \} \over
\delta g^{k} (\bm{\theta}, {\bf W}) } \right|_{{\cal K}[1]} \,
{d^{m} \theta \over (2\pi)^{m}} \,\, d^{d} W \,\, + \\
+ \, \int \left. \left\{ J_{[{\bf p}]} \, , \, 
g^{k} (\bm{\theta}, {\bf W}) \right\} \right|_{{\cal K}[1]} \, \left.
{\delta \{ g_{[{\bf Q}]} \, , \, J_{[{\bf q}]} \} \over
\delta g^{k} (\bm{\theta}, {\bf W}) } \right|_{{\cal K}[0]} \,
{d^{m} \theta \over (2\pi)^{m}} \,\, d^{d} W \,\, - \\
- \, \int \left. \left\{ J_{[{\bf q}]} \, , \, 
g^{k} (\bm{\theta}, {\bf W}) \right\} \right|_{{\cal K}[1]} \, \left.
{\delta \{ g_{[{\bf Q}]} \, , \, J_{[{\bf p}]} \} \over
\delta g^{k} (\bm{\theta}, {\bf W}) } \right|_{{\cal K}[0]} \,
{d^{m} \theta \over (2\pi)^{m}} \,\, d^{d} W \,\, \equiv \,\, 0 
\end{multline*}

 The above identity can again be written in a stronger form.
Namely, making the change 
${\tilde Q}_{i}(\bm{\theta}, {\bf W}) \rightarrow 
{\tilde Q}_{i}(\bm{\theta}, {\bf W}) \, \mu_{i}({\bf W})$,
we get the corresponding change:
$Q_{i}(\bm{\theta}, {\bf W}) \rightarrow 
Q_{i}(\bm{\theta}, {\bf W}) \, \mu_{i}({\bf W})$,
where $\mu_{i}({\bf W})$ are arbitrary smooth functions of ${\bf W}$. 
From the form of the pairwise brackets of constraints given by
(\ref{SkobSv}), and (\ref{gJVarDerNulPor}), 
it is easy to see then that the integrands
are smooth functions of $\bm{\theta}$ and ${\bf W}$, 
containing $\mu_{i}({\bf W})$
in the form of local factors. 
By the arbitrariness of $\mu_{i}({\bf W})$, 
we can omit the integration over ${\bf W}$ in the above 
integrals and write for every ${\bf W}$:
\begin{multline*}
\int_{0}^{2\pi}\!\!\!\!\!\dots\int_{0}^{2\pi}
\left. \left\{ g_{[{\bf Q}]} \, , \, g^{k} (\bm{\theta}, {\bf W}) 
\right\} \right|_{{\cal K}[0]} \, \left.
{\delta \{ J_{[{\bf q}]} \, , \, J_{[{\bf p}]} \} \over
\delta g^{k} (\bm{\theta}, {\bf W}) } \right|_{{\cal K}[1]} \,
{d^{m} \theta \over (2\pi)^{m}} \,\, + \\
+ \, \int_{0}^{2\pi}\!\!\!\!\!\dots\int_{0}^{2\pi}
\left. \left\{ J_{[{\bf p}]} \, , \,
g^{k} (\bm{\theta}, {\bf W}) \right\} \right|_{{\cal K}[1]} \, \left.
{\delta \{ g_{[{\bf Q}]} \, , \, J_{[{\bf q}]} \} \over
\delta g^{k} (\bm{\theta}, {\bf W}) } \right|_{{\cal K}[0]} \,
{d^{m} \theta \over (2\pi)^{m}} \,\, - \\
- \, \int_{0}^{2\pi}\!\!\!\!\!\dots\int_{0}^{2\pi}
\left. \left\{ J_{[{\bf q}]} \, , \,
g^{k} (\bm{\theta}, {\bf W}) \right\} \right|_{{\cal K}[1]} \, \left.
{\delta \{ g_{[{\bf Q}]} \, , \, J_{[{\bf p}]} \} \over
\delta g^{k} (\bm{\theta}, {\bf W}) } \right|_{{\cal K}[0]} \,
{d^{m} \theta \over (2\pi)^{m}} \,\, \equiv \,\, 0
\end{multline*}

 Finally, using relations (\ref{giJmu1skob}) and
(\ref{SilnVarDerRel}), we can write the above identity as
\begin{multline*}
\int_{0}^{2\pi}\!\!\!\!\!\dots\int_{0}^{2\pi}
\left. \left\{ g_{[{\bf Q}]} \, , \, g^{k} (\bm{\theta}, {\bf W})
\right\} \right|_{{\cal K}[0]} \, \left.
{\delta \{ J_{[{\bf q}]} \, , \, J_{[{\bf p}]} \} \over
\delta g^{k} (\bm{\theta}, {\bf W}) } \right|_{{\cal K}[1]} \,
{d^{m} \theta \over (2\pi)^{m}} \,\, - \\
- \, \int_{0}^{2\pi}\!\!\!\!\!\dots\int_{0}^{2\pi}
A^{k}_{[1][{\bf p}]} \left( {{\bf S}({\bf W}) \over \epsilon} +
\bm{\theta}, {\bf W} \right) \,\, \left. 
{\delta \{ g_{[{\bf Q}]} \, , \, J_{[{\bf q}]} \} \over
\delta g^{k} (\bm{\theta}, {\bf W}) } \right|_{{\cal K}[0]} \,
{d^{m} \theta \over (2\pi)^{m}} \,\, + \\
+ \, \int_{0}^{2\pi}\!\!\!\!\!\dots\int_{0}^{2\pi}
A^{k}_{[1][{\bf q}]} \left( {{\bf S}({\bf W}) \over \epsilon} +
\bm{\theta}, {\bf W} \right) \,\, \left.
{\delta \{ g_{[{\bf Q}]} \, , \, J_{[{\bf p}]} \} \over
\delta g^{k} (\bm{\theta}, {\bf W}) } \right|_{{\cal K}[0]} \,
{d^{m} \theta \over (2\pi)^{m}} \,\, \equiv \,\, 0
\end{multline*}
where the functions $A^{k}_{[1][{\bf q}]} (\bm{\theta}, {\bf W})$ 
are introduced by formula (\ref{FunktsiiA}).

 Now assume that the values
$A^{k}_{[1][{\bf q}]} (\bm{\theta}, {\bf W})$ for
${\bf U}({\bf W}) \in {\cal S}$, according to (\ref{BAcond}), 
can be represented in the form
\begin{equation}
\label{ABBPredstav}
A^{k}_{[1][{\bf q}]} (\bm{\theta}, {\bf W}) \,\, = \,\, - \,
{\hat B}^{kj}_{[0]} ({\bf W}) \,\, 
B_{j[{\bf q}](1)} (\bm{\theta}, {\bf W})
\end{equation}
with some smooth in $\bm{\theta}$, $2\pi$-periodic in each
$\theta^{\alpha}$ functions 
$B_{j[{\bf q}](1)} (\bm{\theta}, {\bf W})$.

 We can then write for ${\bf U}({\bf W}) \in {\cal S}$:
$$\int_{0}^{2\pi}\!\!\!\!\!\dots\int_{0}^{2\pi}
\left. \left\{ g_{[{\bf Q}]} \, , \, g^{k} (\bm{\theta}, {\bf W})
\right\} \right|_{{\cal K}[0]} \, \left.   
{\delta \{ J_{[{\bf q}]} \, , \, J_{[{\bf p}]} \} \over
\delta g^{k} (\bm{\theta}, {\bf W}) } \right|_{{\cal K}[1]} \,
{d^{m} \theta \over (2\pi)^{m}} \,\, + $$
\begin{equation}
\label{TozhdgQJqJp}
+ \, \int_{0}^{2\pi}\!\!\!\!\!\dots\int_{0}^{2\pi}
\left[ {\hat B}^{kj}_{[0][{\bf S}]} ({\bf W}) \,\,
B_{j[{\bf p}](1)} 
\left({{\bf S}({\bf W}) \over \epsilon} + \bm{\theta}, 
{\bf W} \right) \right] \,\, \left.
{\delta \{ g_{[{\bf Q}]} \, , \, J_{[{\bf q}]} \} \over
\delta g^{k} (\bm{\theta}, {\bf W}) } \right|_{{\cal K}[0]} \,
{d^{m} \theta \over (2\pi)^{m}} \,\, -
\end{equation}
$$- \, \int_{0}^{2\pi}\!\!\!\!\!\dots\int_{0}^{2\pi}
\left[ {\hat B}^{kj}_{[0][{\bf S}]} ({\bf W}) \,\,
B_{j[{\bf q}](1)} 
\left({{\bf S}({\bf W}) \over \epsilon} + \bm{\theta},   
{\bf W} \right) \right] \,\, \left.
{\delta \{ g_{[{\bf Q}]} \, , \, J_{[{\bf p}]} \} \over
\delta g^{k} (\bm{\theta}, {\bf W}) } \right|_{{\cal K}[0]} \,
{d^{m} \theta \over (2\pi)^{m}} \,\, \equiv \,\, 0 $$
where
\begin{multline*}
{\hat B}^{ij}_{[0][{\bf S}]} ({\bf W}) 
\,\, \equiv \,\, \sum_{l_{1}, \dots, l_{d}}
B^{ij}_{(l_{1} \dots l_{d})} \left( \bm{\Phi} 
\left( {{\bf S}({\bf W}) \over \epsilon} +
\bm{\theta}, \,
{\bf S}_{\bf W}, U^{1}({\bf W}), \dots, U^{m+s}({\bf W}) \right), 
\, \dots \right) \,\, \times \\
\times \,\, 
k^{\alpha^{1}_{1}}_{1}({\bf W}) \dots
k^{\alpha^{1}_{l_{1}}}_{1}({\bf W})
\, \dots \,
k^{\alpha^{d}_{1}}_{d}({\bf W}) \dots
k^{\alpha^{d}_{l_{d}}}_{d}({\bf W}) \,\,\,\,\,
{\partial \over \partial \theta^{\alpha^{1}_{1}}} \dots
{\partial \over \partial \theta^{\alpha^{1}_{l_{1}}}} \, \dots \,
{\partial \over \partial \theta^{\alpha^{d}_{1}}} \dots
{\partial \over \partial \theta^{\alpha^{d}_{l_{d}}}}
\end{multline*}

 Using expression (\ref{SkobSv}) for the bracket of constraints
on ${\cal K}$, as well as relations (\ref{QphiUbr}), (\ref{OgranQ}),
(\ref{JqJpVarDerOrtRel}), and the skew-symmetry of the operator
${\hat B}^{ij}_{[0][{\bf S}]} ({\bf W})$,
we obtain also the following relation:
\begin{multline*}
\int_{0}^{2\pi}\!\!\!\!\!\dots\int_{0}^{2\pi}
\left. \left\{ g_{[{\bf Q}]} \, , \, g^{k} (\bm{\theta}, {\bf W})
\right\} \right|_{{\cal K}[0]} \, \left.
{\delta \{ J_{[{\bf q}]} \, , \, J_{[{\bf p}]} \} \over
\delta g^{k} (\bm{\theta}, {\bf W}) } \right|_{{\cal K}[1]} \,
{d^{m} \theta \over (2\pi)^{m}} \,\, = \\
= \,\, - \, \int_{0}^{2\pi}\!\!\!\!\!\dots\int_{0}^{2\pi}
\left[ {\hat B}^{kj}_{[0][{\bf S}]} ({\bf W}) \,\,
Q_{j} \left({{\bf S}({\bf W}) \over \epsilon} + \bm{\theta},
{\bf W} \right) \right] \, \left.
{\delta \{ J_{[{\bf q}]} \, , \, J_{[{\bf p}]} \} \over
\delta g^{k} (\bm{\theta}, {\bf W}) } \right|_{{\cal K}[1]} \,
{d^{m} \theta \over (2\pi)^{m}} 
\end{multline*}

 Expanding also the remaining terms of identity
(\ref{TozhdgQJqJp}) according to (\ref{gJVarDerNulPor}),
we get for ${\bf U}({\bf W}) \in {\cal S}$:
\begin{equation}
\label{JacobigJJmain}
\int_{0}^{2\pi}\!\!\!\!\!\dots\int_{0}^{2\pi}
\left[ {\hat B}^{kj}_{[0][{\bf S}]} ({\bf W}) \,\,
Q_{j} \left({{\bf S}({\bf W}) \over \epsilon} + \bm{\theta},
{\bf W} \right) \right] \, \left.
{\delta \{ J_{[{\bf q}]} \, , \, J_{[{\bf p}]} \} \over
\delta g^{k} (\bm{\theta}, {\bf W}) } \right|_{{\cal K}[1]} \,
{d^{m} \theta \over (2\pi)^{m}} \,\, -
\end{equation}

$$- \,\, \int_{0}^{2\pi}\!\!\!\!\!\dots\int_{0}^{2\pi}
{d^{m} \theta \over (2\pi)^{m}} 
\left[ {\hat B}^{kj}_{[0][{\bf S}]} ({\bf W}) \,\,
B_{j[{\bf p}](1)} 
\left({{\bf S}({\bf W}) \over \epsilon} + \bm{\theta},
{\bf W} \right) \right] \,\, \times $$
\begin{multline*}
\times \! \left[ 
\sum_{l_{1}, \dots, l_{d}}  (-1)^{l_{1} + \dots + l_{d}} \,\,
k^{\alpha^{1}_{1}}_{1}({\bf W}) \dots
k^{\alpha^{1}_{l_{1}}}_{1}({\bf W})
\, \dots \,
k^{\alpha^{d}_{1}}_{d}({\bf W}) \dots
k^{\alpha^{d}_{l_{d}}}_{d}({\bf W}) \right. \,\, \times \\
\times \,\,
{\partial \over \partial \theta^{\alpha^{1}_{1}}} \dots
{\partial \over \partial \theta^{\alpha^{1}_{l_{1}}}} \, \dots \,
{\partial \over \partial \theta^{\alpha^{d}_{1}}} \dots
{\partial \over \partial \theta^{\alpha^{d}_{l_{d}}}} \,\, \times \\ 
\times \, \left[ Q_{i} \left( {{\bf S}({\bf W}) \over \epsilon} +
\bm{\theta}, {\bf W} \! \right)  \left. 
S^{i\mu}_{k (l_{1} \dots l_{d})} \left(
\bm{\varphi} (\bm{\theta}, {\bf W}), \dots \right) 
\right|_{{\cal K}[0]}
\right] q_{\mu} ({\bf W}) \,\, +  \\
\left. + \,\,\, \omega^{\alpha\mu}({\bf W}) \, 
q_{\mu}({\bf W}) \, Q_{k, \theta^{\alpha}}
\left({{\bf S}({\bf W}) \over \epsilon} + \bm{\theta},
{\bf W} \right) \right] +
\end{multline*}

$$+ \,\, \int_{0}^{2\pi}\!\!\!\!\!\dots\int_{0}^{2\pi}
{d^{m} \theta \over (2\pi)^{m}}
\left[ {\hat B}^{kj}_{[0][{\bf S}]} ({\bf W}) \,\, B_{j[{\bf q}](1)} 
\left({{\bf S}({\bf W}) \over \epsilon} + \bm{\theta},
{\bf W} \right) \right] \,\, \times $$
\begin{multline*}
\times \! \left[
\sum_{l_{1}, \dots, l_{d}}  (-1)^{l_{1} + \dots + l_{d}} \,\,
k^{\alpha^{1}_{1}}_{1}({\bf W}) \dots
k^{\alpha^{1}_{l_{1}}}_{1}({\bf W})
\, \dots \,
k^{\alpha^{d}_{1}}_{d}({\bf W}) \dots
k^{\alpha^{d}_{l_{d}}}_{d}({\bf W}) \right. \,\, \times \\
\times \,\,
{\partial \over \partial \theta^{\alpha^{1}_{1}}} \dots
{\partial \over \partial \theta^{\alpha^{1}_{l_{1}}}} \, \dots \,
{\partial \over \partial \theta^{\alpha^{d}_{1}}} \dots
{\partial \over \partial \theta^{\alpha^{d}_{l_{d}}}} \,\, \times \\
\times \, \left[ Q_{i} \left( {{\bf S}({\bf W}) \over \epsilon} +
\bm{\theta}, {\bf W} \! \right)  \left. 
S^{i\mu}_{k (l_{1} \dots l_{d})} \left(
\bm{\varphi} (\bm{\theta}, {\bf W}), \dots \right)
\right|_{{\cal K}[0]}
\right] p_{\mu} ({\bf W}) \,\, +  \\
\left. + \,\,\, \omega^{\alpha\mu}({\bf W}) \,
p_{\mu}({\bf W}) \, Q_{k, \theta^{\alpha}}
\left({{\bf S}({\bf W}) \over \epsilon} + \bm{\theta},
{\bf W} \right) \right] \,\, \equiv \,\, 0
\end{multline*}

 Consider now the Jacobi identity of the form
\begin{equation}
\label{JacobiggJ}
\left\{ g_{[{\bf P}]} \, , \, \left\{ g_{[{\bf Q}]} \, , \,
J_{[{\bf q}]} \right\} \right\} \, + \,
\left\{ g_{[{\bf Q}]} \, , \, \left\{  J_{[{\bf q}]} \, , \,
g_{[{\bf P}]} \right\} \right\} \, + \,
\left\{  J_{[{\bf q}]} \, , \, \left\{ g_{[{\bf P}]} \, , \,
g_{[{\bf Q}]} \right\} \right\} \,\, \equiv \,\, 0
\end{equation}
for arbitrary fixed functions ${\bf q}({\bf X})$, and the functionals
${\bf P}(\bm{\theta}, {\bf X})$ and ${\bf Q}(\bm{\theta}, {\bf X})$
defined as before with the aid of arbitrary functions
${\tilde {\bf P}}(\bm{\theta}, {\bf X})$, 
${\tilde {\bf Q}}(\bm{\theta}, {\bf X})$.

 According to the relations
$$\left. \left\{ g_{[{\bf Q}, {\bf P}]} , 
U^{\mu} ({\bf W}) \right\} \right|_{\cal K} \, = \, O (\epsilon)
\,\,\, , \,\,\,
\left. \left\{ J_{[{\bf q}]} , \, g^{k} (\bm{\theta}, {\bf W})
\right\} \right|_{\cal K} \, = \, O (\epsilon)
\,\,\, , \,\,\,
\left. \left\{ J_{[{\bf q}]} , 
U^{\mu} ({\bf W}) \right\} \right|_{\cal K} 
\, = \, O (\epsilon) $$
it's not difficult to see that after the restriction on ${\cal K}$
the major term (in $\epsilon$) of (\ref{JacobiggJ}) will be written 
as 
\begin{multline*}
\int \left. \left\{ g_{[{\bf P}]} \, , \, g^{k} (\bm{\theta}, {\bf W})
\right\} \right|_{{\cal K}[0]} \, \left.
{\delta \{ g_{[{\bf Q}]} \, , \, J_{[{\bf q}]} \} \over
\delta g^{k} (\bm{\theta}, {\bf W}) } \right|_{{\cal K}[0]} \,
{d^{m} \theta \over (2\pi)^{m}} \, d^{d} W \,\, - \\
- \, \int \left. \left\{ g_{[{\bf Q}]} \, , \, 
g^{k} (\bm{\theta}, {\bf W}) \right\} \right|_{{\cal K}[0]} \, \left.
{\delta \{ g_{[{\bf P}]} \, , \, J_{[{\bf q}]} \} \over
\delta g^{k} (\bm{\theta}, {\bf W}) } \right|_{{\cal K}[0]} \,
{d^{m} \theta \over (2\pi)^{m}} \, d^{d} W \,\,\, \equiv \,\,\, 0
\end{multline*}

 Again recalling that $q_{\nu}({\bf W})$ are arbitrary functions of 
${\bf W}$ appearing in the integrand in the form of local factors, 
we can write the above relation in a stronger form.
That is, for every ${\bf W}$
\begin{equation}
\label{JacobiggJmain}
\int_{0}^{2\pi}\!\!\!\!\!\dots\int_{0}^{2\pi}
\left. \left\{ g_{[{\bf P}]} \, , \, g^{k} (\bm{\theta}, {\bf W})
\right\} \right|_{{\cal K}[0]} \, \left.
{\delta \{ g_{[{\bf Q}]} \, , \, J_{[{\bf q}]} \} \over 
\delta g^{k} (\bm{\theta}, {\bf W}) } \right|_{{\cal K}[0]} \,
{d^{m} \theta \over (2\pi)^{m}} \,\,\,\,\, - 
\end{equation}
$$- \,\, \int_{0}^{2\pi}\!\!\!\!\!\dots\int_{0}^{2\pi}
\left. \left\{ g_{[{\bf Q}]} \, , \,
g^{k} (\bm{\theta}, {\bf W}) \right\} \right|_{{\cal K}[0]} \, \left.
{\delta \{ g_{[{\bf P}]} \, , \, J_{[{\bf q}]} \} \over
\delta g^{k} (\bm{\theta}, {\bf W}) } \right|_{{\cal K}[0]} \,
{d^{m} \theta \over (2\pi)^{m}} \,\,\, \equiv \,\,\, 0 $$

 As well as in the case of identity (\ref{JacobigJJmain}),
using relations (\ref{SkobSv}), (\ref{QphiUbr}), (\ref{OgranQ}),
(\ref{SilnVarDerRel}), we can write identity
(\ref{JacobiggJmain}) in the form:
\begin{equation}
\label{ggJRasp}
\int_{0}^{2\pi}\!\!\!\!\!\dots\int_{0}^{2\pi}
{d^{m} \theta \over (2\pi)^{m}}
\left[ {\hat B}^{kj}_{[0][{\bf S}]} ({\bf W}) \,\,
P_{j} \left({{\bf S}({\bf W}) \over \epsilon} + \bm{\theta},
{\bf W} \right) \right] \,\, \times 
\end{equation}
\begin{multline*}
\times \! \left[
\sum_{l_{1}, \dots, l_{d}}  (-1)^{l_{1} + \dots + l_{d}} \,\,
k^{\alpha^{1}_{1}}_{1}({\bf W}) \dots
k^{\alpha^{1}_{l_{1}}}_{1}({\bf W})
\, \dots \,
k^{\alpha^{d}_{1}}_{d}({\bf W}) \dots
k^{\alpha^{d}_{l_{d}}}_{d}({\bf W}) \right. \,\, \times \\
\times \,\,
{\partial \over \partial \theta^{\alpha^{1}_{1}}} \dots
{\partial \over \partial \theta^{\alpha^{1}_{l_{1}}}} \, \dots \,
{\partial \over \partial \theta^{\alpha^{d}_{1}}} \dots
{\partial \over \partial \theta^{\alpha^{d}_{l_{d}}}} \,\, \times \\
\times \, \left[ Q_{i} \left( {{\bf S}({\bf W}) \over \epsilon} +
\bm{\theta}, {\bf W} \! \right)  \left.
S^{i\mu}_{k (l_{1} \dots l_{d})} \left(
\bm{\varphi} (\bm{\theta}, {\bf W}), \dots \right)
\right|_{{\cal K}[0]}
\right] q_{\mu} ({\bf W}) \,\, +  \\   
\left. + \,\,\, \omega^{\alpha\mu}({\bf W}) \,
q_{\mu}({\bf W}) \, Q_{k, \theta^{\alpha}}
\left({{\bf S}({\bf W}) \over \epsilon} + \bm{\theta},
{\bf W} \right) \right] -
\end{multline*}

$$- \,\, \int_{0}^{2\pi}\!\!\!\!\!\dots\int_{0}^{2\pi}
{d^{m} \theta \over (2\pi)^{m}}
\left[ {\hat B}^{kj}_{[0][{\bf S}]} ({\bf W}) \,\,
Q_{j} \left({{\bf S}({\bf W}) \over \epsilon} + \bm{\theta},
{\bf W} \right) \right] \,\, \times $$
\begin{multline*}
\times \! \left[
\sum_{l_{1}, \dots, l_{d}}  (-1)^{l_{1} + \dots + l_{d}} \,\,
k^{\alpha^{1}_{1}}_{1}({\bf W}) \dots
k^{\alpha^{1}_{l_{1}}}_{1}({\bf W})
\, \dots \,   
k^{\alpha^{d}_{1}}_{d}({\bf W}) \dots
k^{\alpha^{d}_{l_{d}}}_{d}({\bf W}) \right. \,\, \times \\
\times \,\,
{\partial \over \partial \theta^{\alpha^{1}_{1}}} \dots
{\partial \over \partial \theta^{\alpha^{1}_{l_{1}}}} \, \dots \,
{\partial \over \partial \theta^{\alpha^{d}_{1}}} \dots  
{\partial \over \partial \theta^{\alpha^{d}_{l_{d}}}} \,\, \times \\
\times \, \left[ P_{i} \left( {{\bf S}({\bf W}) \over \epsilon} +
\bm{\theta}, {\bf W} \! \right)  \left.
S^{i\mu}_{k (l_{1} \dots l_{d})} \left(
\bm{\varphi} (\bm{\theta}, {\bf W}), \dots \right)
\right|_{{\cal K}[0]}
\right] q_{\mu} ({\bf W}) \,\, +  \\
\left. + \,\,\, \omega^{\alpha\mu}({\bf W}) \,
q_{\mu}({\bf W}) \, P_{k, \theta^{\alpha}}
\left({{\bf S}({\bf W}) \over \epsilon} + \bm{\theta},
{\bf W} \right) \right] \,\, \equiv \,\, 0
\end{multline*}

 Note now that the values of
${\bf Q} (\bm{\theta}, {\bf X})$ and 
${\bf P} (\bm{\theta}, {\bf X})$
are arbitrary $2\pi$-periodic functions of $\bm{\theta}$, 
satisfying conditions (\ref{OgranQ}).
In particular, we can put in (\ref{ggJRasp}) at 
${\bf U} ({\bf W}) \in {\cal S}$
\begin{equation}
\label{PBPB}
{\bf P} (\bm{\theta}, {\bf W}) \,\, = \,\, 
{\bf B}_{[{\bf p}](1)} (\bm{\theta}, {\bf W}) \,\,\,\,\,\,\,\, 
\text{or}
\,\,\,\,\,\,\,\, {\bf P} (\bm{\theta}, {\bf W}) \,\, = \,\,
{\bf B}_{[{\bf q}](1)} (\bm{\theta}, {\bf W})
\end{equation}

 By analogy with (\ref{gJVarDerNulPor}) we introduce for convenience
the notation for ${\bf U} ({\bf W}) \in {\cal S}$:
\begin{equation}
\label{gBJqVarDerDefinition}
\left. {\delta \{ g_{[\epsilon {\bf B}_{[{\bf p}](1)}]} \, , \,
J_{[{\bf q}]} \} \over \delta g^{k} (\bm{\theta}, {\bf W})}
\right|_{{\cal K}[1]} \,\,\, \equiv
\end{equation}
\begin{multline*}
\equiv \left[ 
\sum_{l_{1}, \dots, l_{d}}  (-1)^{l_{1} + \dots + l_{d}} \,\,
k^{\alpha^{1}_{1}}_{1}({\bf W}) \dots
k^{\alpha^{1}_{l_{1}}}_{1}({\bf W})  
\, \dots \,
k^{\alpha^{d}_{1}}_{d}({\bf W}) \dots
k^{\alpha^{d}_{l_{d}}}_{d}({\bf W}) \right. \,\, \times \\
\times \,\,
{\partial \over \partial \theta^{\alpha^{1}_{1}}} \dots
{\partial \over \partial \theta^{\alpha^{1}_{l_{1}}}} \, \dots \,
{\partial \over \partial \theta^{\alpha^{d}_{1}}} \dots
{\partial \over \partial \theta^{\alpha^{d}_{l_{d}}}} \,\, \times \\
\times \,
\left[ B_{i[{\bf p}](1)} \left( {{\bf S}({\bf W}) \over \epsilon} +
\bm{\theta}, {\bf W} \! \right)  \left.
S^{i\mu}_{k (l_{1} \dots l_{d})} \left(
\bm{\varphi} (\bm{\theta}, {\bf W}), \dots \right)
\right|_{{\cal K}[0]}
\right] q_{\mu} ({\bf W}) \,\, +  \\
\left. + \,\,\, \omega^{\alpha\mu}({\bf W}) \,
q_{\mu}({\bf W}) \, B_{k[{\bf p}](1), \theta^{\alpha}}
\left({{\bf S}({\bf W}) \over \epsilon} + \bm{\theta},
{\bf W} \right) \right] 
\end{multline*}
for arbitrary smooth functions ${\bf q}({\bf X})$, ${\bf p}({\bf X})$.

 Note that the functional $g_{[\epsilon {\bf B}_{[{\bf p}](1)}]}$ is
not defined on the whole functional space, so relation
(\ref{gBJqVarDerDefinition}) plays just a role of a formal notation 
for ${\bf U} ({\bf W}) \in {\cal S}$.

 Using now (\ref{ggJRasp}) for the functions (\ref{PBPB}),
we can rewrite (\ref{JacobigJJmain}) in the form
\begin{multline*}
\int_{0}^{2\pi}\!\!\!\!\!\dots\int_{0}^{2\pi}
{d^{m} \theta \over (2\pi)^{m}} \, 
\left[ {\hat B}^{kj}_{[0][{\bf S}]} ({\bf W}) \,\,
Q_{j} \left({{\bf S}({\bf W}) \over \epsilon} + \bm{\theta},
{\bf W} \right) \right] \,\, \times 
\cr
\cr
\times \,\, \left( \left.
{\delta \{ J_{[{\bf q}]} \, , \, J_{[{\bf p}]} \} \over
\delta g^{k} (\bm{\theta}, {\bf W}) } \right|_{{\cal K}[1]} \, - \,
\left. {\delta \{ g_{[\epsilon {\bf B}_{[{\bf p}](1)}]} \, , \,
J_{[{\bf q}]} \} \over \delta g^{k} (\bm{\theta}, {\bf W}) } 
\right|_{{\cal K}[1]} \, + \, 
\left. {\delta \{ g_{[\epsilon {\bf B}_{[{\bf q}](1)}]} \, , \,
J_{[{\bf p}]} \} \over \delta g^{k} (\bm{\theta}, {\bf W}) }
\right|_{{\cal K}[1]} \right) \,\,\, \equiv \,\,\, 0
\end{multline*}
provided that ${\bf U} ({\bf W}) \in {\cal S}$.

 Using the skew-symmetry of the Hamiltonian operator
${\hat B}^{kj}_{[0][{\bf S}]} ({\bf W})$
on $\Lambda$ we can then write
\begin{multline*}
\int_{0}^{2\pi}\!\!\!\!\!\dots\int_{0}^{2\pi}
{d^{m} \theta \over (2\pi)^{m}} \,
Q_{j} \left( {{\bf S}({\bf W}) \over \epsilon} 
+ \bm{\theta}, {\bf W} \right)
\,\,\times \\
\times \left[ {\hat B}^{jk}_{[0][{\bf S}]} ({\bf W}) 
\left( \left.
{\delta \{ J_{[{\bf q}]} , J_{[{\bf p}]} \} \over
\delta g^{k} (\bm{\theta}, {\bf W}) } \right|_{{\cal K}[1]} \, - \,
\left. {\delta \{ g_{[\epsilon {\bf B}_{[{\bf p}](1)}]} , 
J_{[{\bf q}]} \} \over \delta g^{k} (\bm{\theta}, {\bf W}) }
\right|_{{\cal K}[1]} \, + \,
\left. {\delta \{ g_{[\epsilon {\bf B}_{[{\bf q}](1)}]} ,
J_{[{\bf p}]} \} \over \delta g^{k} (\bm{\theta}, {\bf W}) }
\right|_{{\cal K}[1]} \right) \! \right] \, \equiv \, 0
\end{multline*}

 The values $Q_{j} (\bm{\theta}, {\bf W} )$ are arbitrary smooth
$2\pi$-periodic functions of $\bm{\theta}$ with the only
restriction (\ref{OgranQ}). We know also that the value in the
brackets is a smooth $2\pi$-periodic in each $\theta^{\alpha}$
function of $\bm{\theta}$ for ${\bf U} ({\bf W}) \in {\cal S}$.
As a consequence, we can write for ${\bf U} ({\bf W}) \in {\cal S}$
\begin{multline*}
{\hat B}^{jk}_{[0][{\bf S}]} ({\bf W}) \, \left( \left.
{\delta \{ J_{[{\bf q}]} , J_{[{\bf p}]} \} \over
\delta g^{k} (\bm{\theta}, {\bf W}) } \right|_{{\cal K}[1]} \!\! - 
\left. {\delta \{ g_{[\epsilon {\bf B}_{[{\bf p}](1)}]}  , 
J_{[{\bf q}]} \} \over \delta g^{k} (\bm{\theta}, {\bf W}) }
\right|_{{\cal K}[1]}  \!\! + 
\left. {\delta \{ g_{[\epsilon {\bf B}_{[{\bf q}](1)}]}  , 
J_{[{\bf p}]} \} \over \delta g^{k} (\bm{\theta}, {\bf W}) }
\right|_{{\cal K}[1]} \right)  \,\, \equiv \\
\equiv \, \sum_{\alpha=1}^{m} 
a^{\alpha}_{[{\bf q}, {\bf p}]} 
({\bf U}({\bf W}), {\bf U}_{\bf W}({\bf W})) \,\,
\Phi^{j}_{\theta^{\alpha}} \left( 
{{\bf S}({\bf W}) \over \epsilon} + \bm{\theta}, 
{\bf U}({\bf W}) \right) 
\end{multline*}
with some coefficients 
$a^{\alpha}_{[{\bf q}, {\bf p}]} 
({\bf U}({\bf W}), {\bf U}_{\bf W}({\bf W}))$.

 The values in parentheses are smooth $2\pi$-periodic
functions of $\bm{\theta}$ at ${\bf U} ({\bf W}) \in {\cal S}$. 
According to Lemma 3.6 we can therefore say 
that up to a linear combination of
the regular eigenvectors 
${\bf v}^{(l)}_{[{\bf U}({\bf W})]} 
({\bf S}({\bf W})/\epsilon + \bm{\theta})$
of ${\hat B}^{jk}_{[0][{\bf S}]} ({\bf W})
= {\hat B}^{jk}_{[0][{\bf S}]} ({\bf U}({\bf W}))$,
corresponding to zero eigenvalues,
the value in parentheses is a linear combination of the
variation derivatives (\ref{VarDer}), generating linear shifts
of the phases on $\Lambda$. For a complete Hamiltonian set of the
integrals $(I^{1}, \dots , I^{N})$ we can then write according to
(\ref{SviazZetaKappa}) and (\ref{Sootndliavikappa})
$$\left. {\delta \{ J_{[{\bf q}]} , J_{[{\bf p}]} \} \over
\delta g^{k} (\bm{\theta}, {\bf W}) } \right|_{{\cal K}[1]} \!\! -
\left. {\delta \{ g_{[\epsilon {\bf B}_{[{\bf p}](1)}]}  ,
J_{[{\bf q}]} \} \over \delta g^{k} (\bm{\theta}, {\bf W}) }
\right|_{{\cal K}[1]}  \!\! +
\left. {\delta \{ g_{[\epsilon {\bf B}_{[{\bf q}](1)}]}  ,
J_{[{\bf p}]} \} \over \delta g^{k} (\bm{\theta}, {\bf W}) }
\right|_{{\cal K}[1]}  \,\, \equiv $$
\begin{equation}
\label{PredstavVarDer} 
\equiv \,\, \sum_{q=1}^{m+s} 
b_{q [{\bf q}, {\bf p}]} 
({\bf U}({\bf W}), {\bf U}_{\bf W}({\bf W})) 
\,\, \kappa^{(q)}_{k[{\bf U}({\bf W})]}
\left( {{\bf S}({\bf W}) \over \epsilon} + \bm{\theta}\right)
\end{equation}
with some coefficients 
$b_{q [{\bf q}, {\bf p}]} 
({\bf U}({\bf W}), {\bf U}_{\bf W}({\bf W}))$ at
${\bf U} ({\bf W}) \in {\cal S}$.

 Consider now the Jacobi identity of the form
$$\left\{ \left\{ J_{[{\bf q}]} , J_{[{\bf p}]} \right\} ,
J_{[{\bf r}]} \right\} \,\, + \,\, 
\left\{ \left\{ J_{[{\bf p}]} , J_{[{\bf r}]} \right\} ,
J_{[{\bf q}]} \right\} \,\, + \,\,
\left\{ \left\{ J_{[{\bf r}]} , J_{[{\bf q}]} \right\} ,
J_{[{\bf p}]} \right\} \,\, \equiv \,\, 0$$
with arbitrary smooth functions 
${\bf q}({\bf X})$, ${\bf p}({\bf X})$, ${\bf r}({\bf X})$.

 In the main ($\sim \epsilon^{2}$) order on ${\cal K}$ the given
identity leads to the relations
$$\int \left. {\delta \{ J_{[{\bf q}]} , J_{[{\bf p}]} \} \over
\delta S^{\alpha} ({\bf W}) } \right|_{{\cal K}[1]}
\left. \left\{ S^{\alpha}({\bf W}) \, , \, J_{[{\bf r}]} \right\}
\right|_{{\cal K}[1]} \, d^{d} W \,\,\,\,\, + \,\,\,\,\, c.p.
\,\,\,\,\, +  $$
\begin{equation}
\label{JJJskobrazl}
+ \,\,\, \int \left. 
{\delta \{ J_{[{\bf q}]} , J_{[{\bf p}]} \} \over
\delta U^{\gamma} ({\bf W}) } \right|_{{\cal K}[1]} 
\left. \left\{ U^{\gamma}({\bf W}) \, , \, J_{[{\bf r}]} \right\} 
\right|_{{\cal K}[1]} \, d^{d} W \,\,\,\,\, + \,\,\,\,\, c.p.
\,\,\,\,\, + 
\end{equation}
$$+ \,\, \int \!\! \int_{0}^{2\pi}\!\!\!\!\!\dots\int_{0}^{2\pi}
\left. {\delta \{ J_{[{\bf q}]} , J_{[{\bf p}]} \} \over
\delta g^{k} (\bm{\theta}, {\bf W}) } \right|_{{\cal K}[1]} 
\left. \left\{ g^{k} (\bm{\theta}, {\bf W}) \, , \, 
J_{[{\bf r}]} \right\}
\right|_{{\cal K}[1]} \, {d^{m} \theta \over (2\pi)^{m}} \,\, d^{d} W
\,\,\, + \,\,\, c.p. \,\,\, \equiv \,\,\, 0 $$
($\alpha = 1, \dots, m$, $\, \gamma = 1, \dots, m + s$).

 Again, using relations (\ref{giJmu1skob}), (\ref{FunktsiiA}) and
(\ref{JqJpVarDerOrtRel}), we can replace identity (\ref{JJJskobrazl})
by the following relation
$$\int \left. {\delta \{ J_{[{\bf q}]} , J_{[{\bf p}]} \} \over
\delta S^{\alpha} ({\bf W}) } \right|_{{\cal K}[1]}
\left. \left\{ S^{\alpha}({\bf W}) \, , \, J_{[{\bf r}]} \right\}
\right|_{{\cal K}[1]} \, d^{d} W \,\,\,\,\, + \,\,\,\,\, c.p.
\,\,\,\,\, +  $$
$$+ \,\,\, \int \left. 
{\delta \{ J_{[{\bf q}]} , J_{[{\bf p}]} \} \over
\delta U^{\gamma} ({\bf W}) } \right|_{{\cal K}[1]}
\left. \left\{ U^{\gamma}({\bf W}) \, , \, J_{[{\bf r}]} \right\}
\right|_{{\cal K}[1]} \, d^{d} W \,\,\,\,\, + \,\,\,\,\, c.p.
\,\,\,\,\, + $$
$$+ \,\, \int \!\! \int_{0}^{2\pi}\!\!\!\!\!\dots\int_{0}^{2\pi}
\left. {\delta \{ J_{[{\bf q}]} , J_{[{\bf p}]} \} \over
\delta g^{k} (\bm{\theta}, {\bf W}) } \right|_{{\cal K}[1]}
A^{k}_{[1][{\bf r}]} \left( 
{{\bf S}({\bf W}) \over \epsilon} + \bm{\theta}, {\bf W} \right) 
\,\, {d^{m} \theta \over (2\pi)^{m}} \,\, d^{d} W
\,\,\, + \,\,\, c.p. \,\,\, \equiv \,\,\, 0 $$

 Using relations (\ref{LeftVecAort}) and the representations
(\ref{ABBPredstav}) and (\ref{PredstavVarDer}), we can write for 
${\bf U} ({\bf W}) \in {\cal S}$:
\begin{equation}
\label{gchastTozhdJnak}
\int_{0}^{2\pi}\!\!\!\!\!\dots\int_{0}^{2\pi}
\left. {\delta \{ J_{[{\bf q}]} , J_{[{\bf p}]} \} \over
\delta g^{k} (\bm{\theta}, {\bf W}) } \right|_{{\cal K}[1]}
A^{k}_{[1][{\bf r}]} \left( 
{{\bf S}({\bf W}) \over \epsilon} + \bm{\theta}, {\bf W} \right)    
\,\, {d^{m} \theta \over (2\pi)^{m}}
\,\,\, + \,\,\, c.p. \,\,\, =
\end{equation}
\begin{multline*}
= \int_{0}^{2\pi}\!\!\!\!\!\!\!\dots\!\int_{0}^{2\pi} \! 
\left[ {\hat B}^{kj}_{[0][{\bf S}]} ({\bf W}) \, B_{j[{\bf r}](1)} \! 
\left( {{\bf S}({\bf W}) \over \epsilon} + \bm{\theta},
{\bf W} \right) \! \right] \, \times \\
\times \, \left[ 
\left. {\delta \{ g_{[\epsilon {\bf B}_{[{\bf q}](1)}]}  ,
J_{[{\bf p}]} \} \over \delta g^{k} (\bm{\theta}, {\bf W}) }
\right|_{{\cal K}[1]} \!\! - 
\left. {\delta \{ g_{[\epsilon {\bf B}_{[{\bf p}](1)}]}  ,
J_{[{\bf q}]} \} \over \delta g^{k} (\bm{\theta}, {\bf W}) }
\right|_{{\cal K}[1]} \! \right] \! {d^{m} \theta \over (2\pi)^{m}}
\,\, + 
\end{multline*}
\begin{multline*}
+ \int_{0}^{2\pi}\!\!\!\!\!\!\!\dots\!\int_{0}^{2\pi} \! 
\left[ {\hat B}^{kj}_{[0][{\bf S}]} ({\bf W}) \, B_{j[{\bf q}](1)} \! 
\left( {{\bf S}({\bf W}) \over \epsilon} + \bm{\theta},
{\bf W} \right) \! \right] \, \times \\
\times \, \left[
\left. {\delta \{ g_{[\epsilon {\bf B}_{[{\bf p}](1)}]}  ,
J_{[{\bf r}]} \} \over \delta g^{k} (\bm{\theta}, {\bf W}) }
\right|_{{\cal K}[1]} \!\! - 
\left. {\delta \{ g_{[\epsilon {\bf B}_{[{\bf r}](1)}]}  ,
J_{[{\bf p}]} \} \over \delta g^{k} (\bm{\theta}, {\bf W}) }  
\right|_{{\cal K}[1]} \! \right] {d^{m} \theta \over (2\pi)^{m}}
\,\, + 
\end{multline*}
\begin{multline*}
+ \int_{0}^{2\pi}\!\!\!\!\!\!\!\dots\!\int_{0}^{2\pi} \! 
\left[ {\hat B}^{kj}_{[0][{\bf S}]} ({\bf W}) \, B_{j[{\bf p}](1)} \! 
\left( {{\bf S}({\bf W}) \over \epsilon} + \bm{\theta},
{\bf W} \right) \! \right] \, \times \\
\times \, \left[
\left. {\delta \{ g_{[\epsilon {\bf B}_{[{\bf r}](1)}]}  ,
J_{[{\bf q}]} \} \over \delta g^{k} (\bm{\theta}, {\bf W}) }
\right|_{{\cal K}[1]} \!\! - 
\left. {\delta \{ g_{[\epsilon {\bf B}_{[{\bf q}](1)}]}  ,
J_{[{\bf r}]} \} \over \delta g^{k} (\bm{\theta}, {\bf W}) }
\right|_{{\cal K}[1]} \! \right] {d^{m} \theta \over (2\pi)^{m}} 
\end{multline*}

 Similarly to earlier arguments, substituting now in identity
(\ref{ggJRasp}):
$${\bf Q} (\bm{\theta}, {\bf W}) \,\, = \,\,
{\bf B}_{[{\bf r}](1)} (\bm{\theta}, {\bf W}) \,\,\,\,\,\,\,\, ,
\,\,\,\,\,\,\,\, {\bf P} (\bm{\theta}, {\bf W}) \,\, = \,\,
{\bf B}_{[{\bf p}](1)} (\bm{\theta}, {\bf W}) $$
for ${\bf U} ({\bf W}) \in {\cal S}$ we obtain the identities
\begin{equation}
\label{ParnTozhd}
\int_{0}^{2\pi}\!\!\!\!\!\dots\int_{0}^{2\pi}
\left[ {\hat B}^{kj}_{[0][{\bf S}]} ({\bf W}) \,\, B_{j[{\bf p}](1)} 
\left( {{\bf S}({\bf W}) \over \epsilon} + \bm{\theta},
{\bf W} \right) \right] \,\, \left. 
{\delta \{ g_{[\epsilon {\bf B}_{[{\bf r}](1)}]}  ,
J_{[{\bf q}]} \} \over \delta g^{k} (\bm{\theta}, {\bf W}) }
\right|_{{\cal K}[1]} {d^{m} \theta \over (2\pi)^{m}} \,\,\, -
\end{equation}
$$- \,\,\, \int_{0}^{2\pi}\!\!\!\!\!\dots\int_{0}^{2\pi}
\left[ {\hat B}^{kj}_{[0][{\bf S}]} ({\bf W}) \,\, B_{j[{\bf r}](1)}   
\left( {{\bf S}({\bf W}) \over \epsilon} + \bm{\theta},
{\bf W} \right) \right] \,\, \left.
{\delta \{ g_{[\epsilon {\bf B}_{[{\bf p}](1)}]}  ,
J_{[{\bf q}]} \} \over \delta g^{k} (\bm{\theta}, {\bf W}) }
\right|_{{\cal K}[1]} {d^{m} \theta \over (2\pi)^{m}} \,\,\, 
= \,\,\, 0 $$

 Using the cyclic permutations of the functions ${\bf q} ({\bf X})$,
${\bf p}({\bf X})$, and ${\bf r}({\bf X})$ 
in identity (\ref{ParnTozhd}),
it's not difficult to see that the right-hand part of relation
(\ref{gchastTozhdJnak}) is identically equal to zero at 
${\bf U} ({\bf W}) \in {\cal S}$. It's not difficult to see also that
the left-hand side of the relation (\ref{gchastTozhdJnak}) 
is a smooth regular function of the parameters 
${\bf U} ({\bf W}) = ({\bf S}_{\bf W}, \, U^{1} ({\bf W}), \dots,
U^{m+s} ({\bf W}))$ and their derivatives. 
Using the fact that the set
${\cal S}$ is everywhere dense in the parameter space
${\bf U}$, we can conclude that the left-hand side of equation
(\ref{gchastTozhdJnak}) is identically equal to zero under the
conditions of the theorem.

We have, therefore, that under the conditions of the theorem,
the identity (\ref{JJJskobrazl}) implies the relations
$$\int \left. {\delta \{ J_{[{\bf q}]} , J_{[{\bf p}]} \} \over
\delta S^{\alpha} ({\bf W}) } \right|_{{\cal K}[1]}
\left. \left\{ S^{\alpha}({\bf W}) \, , \, J_{[{\bf r}]} \right\}
\right|_{{\cal K}[1]} \, d^{d} W \,\,\,\,\, + \,\,\,\,\, c.p.
\,\,\,\,\, +  $$
$$+ \,\,\, \int \left.
{\delta \{ J_{[{\bf q}]} , J_{[{\bf p}]} \} \over
\delta U^{\gamma} ({\bf W}) } \right|_{{\cal K}[1]}
\left. \left\{ U^{\gamma}({\bf W}) \, , \, J_{[{\bf r}]} \right\}
\right|_{{\cal K}[1]} \, d^{d} W \,\,\,\,\, + \,\,\,\,\, c.p.
\,\,\,\,\, \equiv \,\,\,\,\, 0 $$

 Using the relations
$$\left. {\delta \{ J_{[{\bf q}]} , J_{[{\bf p}]} \}
\over \delta S^{\alpha} ({\bf W}) } \right|_{{\cal K}[1]} \,\, = \,\,
{\delta \{ U_{[{\bf q}]} , U_{[{\bf p}]} \}_{DN}
\over \delta S^{\alpha} ({\bf W}) } \,\,\,\,\, , \,\,\,\,\,
\left. {\delta \{ J_{[{\bf q}]} , J_{[{\bf p}]} \}
\over \delta U^{\gamma} ({\bf W}) } \right|_{{\cal K}[1]} \,\, = \,\,
{\delta \{ U_{[{\bf q}]} , U_{[{\bf p}]} \}_{DN} 
\over \delta U^{\gamma} ({\bf W}) } $$
$$\left. \left\{ U^{\gamma}({\bf W}) , J_{[{\bf r}]} \right\}
\right|_{{\cal K}[1]} \,\, = \,\,
\left\{ U^{\gamma}({\bf W}) , U_{[{\bf r}]} \right\}_{DN} $$
we get then the identities
\begin{multline}
\label{DubrNovformIden}
\int {\delta \{ U_{[{\bf q}]} , U_{[{\bf p}]} \}_{DN}
\over \delta S^{\alpha} ({\bf W}) } \,\,
\left\{ S^{\alpha}({\bf W}) , U_{[{\bf r}]} \right\}_{DN}
\,\, d^{d} W \,\,\,\,\, + \,\,\,\,\, c. p. \,\,\, + \\
+ \,\,\, \int {\delta \{ U_{[{\bf q}]} , U_{[{\bf p}]} \}_{DN}
\over \delta U^{\gamma} ({\bf W}) } \,\,
\left\{ U^{\gamma}({\bf W}) , U_{[{\bf r}]} \right\}_{DN}
\,\, d^{d} W \,\,\,\,\, + \,\,\,\,\, c. p. \,\,\, \equiv \,\,\, 0
\end{multline}
(summation in $\alpha = 1, \dots, m$, $\gamma = 1, \dots, m + s$).

 Here the notation 
$\{ S^{\alpha}({\bf W}) , U_{[{\bf r}]} \}_{DN}$ means the pairing
of $S^{\alpha}({\bf W})$ as the functional of ${\bf U}({\bf Z})$
with $U_{[{\bf r}]}$ given by the Dubrovin - Novikov
skew-symmetric form and coinciding with the corresponding
value for the bracket (\ref{AveragedBracket}). Easy to see then
that relations (\ref{DubrNovformIden}) give in particular the
Jacobi identities for the bracket (\ref{AveragedBracket}) on the
space of fields 
$(S^{\alpha} ({\bf X}), U^{1} ({\bf X}), \dots, U^{m+s} ({\bf X}))$.

\hfill{ Theorem 3.1 is proved.}

\vspace{0.2cm}

 We will not consider here applications of Theorem 3.1 in details.
The example of the application of Theorem 3.1 in one-dimensional
case can be found in \cite{DNMultDim}. In general, we can say that
Theorem 3.1 gives justification of the procedure of the bracket
averaging for a wide class of systems having multi-phase solutions.

 Using Lemma 3.5 we can formulate now the theorem that gives the 
justification of the procedure of the bracket averaging for a regular 
Hamiltonian family of single-phase solutions of system (\ref{EvInSyst}).

\vspace{0.2cm}

{\bf Theorem 3.1$^{\prime}$.}

{\it Let $\Lambda$ be a regular Hamiltonian family of
single-phase solutions of system (\ref{EvInSyst}). Let
$(I^{1}, \dots, I^{N})$ be a complete Hamiltonian set of
commuting first integrals of system (\ref{EvInSyst}) on $\Lambda$
having the form (\ref{Integrals}).
Then the bracket
\begin{multline}
\label{AvBracketSinglePhase}
\left\{ S ({\bf X}) , S ({\bf Y}) \right\}
\, = \, 0 \,\,\, , \,\,\,\,\,
\left\{ S ({\bf X}) , 
U^{\gamma} ({\bf Y}) \right\} \,\, = \,\,
\omega^{\gamma} \,
\left( S_{\bf X}, \,
U^{1}({\bf X}), \dots, U^{s+1}({\bf X}) \right)
\,\, \delta ({\bf X} - {\bf Y}) \,\, , \\  \\
\left\{ U^{\gamma} ({\bf X})\, , \, U^{\rho} ({\bf Y}) \right\}
\,\,\, = \,\,\, \langle A^{\gamma\rho}_{10\dots0} \rangle
\left( S_{\bf X}, \,
U^{1}({\bf X}), \dots, U^{s+1}({\bf X}) \right) \,\,
\delta_{X^{1}} ({\bf X} - {\bf Y}) \,\,\, + \,\, \dots \,\, + \\  \\
+ \,\,\, \langle A^{\gamma\rho}_{0\dots01} \rangle
\left( S_{\bf X}, \,
U^{1}({\bf X}), \dots, U^{s+1}({\bf X}) \right) \,\,\,
\delta_{X^{d}} ({\bf X} - {\bf Y}) \,\,\, +  \\   \\
+ \,\,\, \left[ \langle Q^{\gamma\rho \, p} \rangle
\left( S_{\bf X}, \,
U^{1}({\bf X}), \dots, U^{s+1}({\bf X}) \right)
\right]_{X^{p}} \,\,\, \delta ({\bf X} - {\bf Y})
\,\,\,\,\,\,\,\, , \,\,\,\,\,\,\,\,\,\,\,\,\,\,\,
\gamma, \rho \, = \, 1, \dots , s + 1
\end{multline}
on the space of fields
$(S ({\bf X}) \, , \, U^{\gamma} ({\bf X}))$,
$\gamma \, = \, 1, \dots, s + 1$,
satisfies the Jacobi identity.
}

\vspace{0.2cm}

 Let us prove now the second theorem justifying the invariance of
the bracket averaging procedure.

\vspace{0.2cm}

{\bf Theorem 3.2.}

{\it Let $\Lambda$ be a regular Hamiltonian family of $m$-phase
solutions of (\ref{EvInSyst}). Let $(I^{1}, \dots, I^{N})$ and
$(I^{\prime 1}, \dots, I^{\prime N})$ be two different complete
Hamiltonian sets of commuting first integrals of
(\ref{EvInSyst}) having the form (\ref{Integrals}). Then the
brackets (\ref{AveragedBracket}) obtained using the sets
$(I^{1}, \dots, I^{N})$ and $(I^{\prime 1}, \dots, I^{\prime N})$
coincide with each other.
}

\vspace{0.2cm}

 Proof.

Note that the sets $(I^{1}, \dots, I^{N})$,
$(I^{\prime 1}, \dots, I^{\prime N})$ correspond to two different 
systems of coordinates $(U^{1}, \dots, U^{N})$,
$(U^{\prime 1}, \dots, U^{\prime N})$ on the family $\Lambda$,
given by the averages of the functionals ${\bf I}$ and
${\bf I}^{\prime}$. Let us first prove that the Dubrovin - Novikov
form obtained using the set 
$(I^{\prime 1}, \dots, I^{\prime N})$ coincides with the form,
obtained using the set $(I^{1}, \dots, I^{N})$, after
the corresponding change of coordinates.
$$U^{\prime\nu} \,\, = \,\, U^{\prime\nu} ({\bf U}) $$

 Consider the functionals
$$J^{\nu} ({\bf X}) \,\, = \,\, 
\int_{0}^{2\pi}\!\!\!\!\!\dots\int_{0}^{2\pi}
P^{\nu} (\bm{\varphi}, \epsilon \bm{\varphi}_{\bf X}, \dots ) \,\,
{d^{m} \theta \over (2\pi)^{m}} \,\,\,\,\,\,\,\, , \,\,\,\,\,
\nu = 1, \dots, N $$

 Consider now the values of the functionals $S^{\alpha}({\bf X})$,
$U^{\gamma}({\bf X})$,
and the constraints $g^{i} (\bm {\theta}, {\bf X})$, 
introduced with the aid of the functionals 
${\bf J} ({\bf X})$, as a coordinate system in the neighborhood of the 
submanifold ${\cal K}$.

 We obviously have the relations
$$\left. J^{\nu} ({\bf X}) \right|_{\cal K} \,\, = \,\,
U^{\nu} ({\bf X}) \, + \, O (\epsilon) \,\,\,\,\,\,\,\, , 
\,\,\,\,\,\,\,\,
\left. J^{\prime\nu} ({\bf X}) \right|_{\cal K} \,\, = \,\,
U^{\prime\nu} ({\bf X}) \, + \, O (\epsilon)$$
on the submanifold ${\cal K}$.

 For values of the functionals $J^{\prime\nu}({\bf X})$ 
on the submanifold ${\cal K}$ we can then write the relations 
$$\left. J^{\prime\nu}({\bf X}) \right|_{\cal K} \,\, = \,\,
 U^{\prime\nu} \left( {\bf J} ({\bf X}) \right) \,\, + \,\,
\sum_{l\geq1} \epsilon^{l} \, j^{\prime\nu}_{(l)}
\left( {\bf S}_{\bf X}, U^{1}({\bf X}), \dots , U^{m+s}({\bf X}),
\dots \right) $$
where $j^{\prime\nu}_{(l)}$ are smooth functions of
$({\bf S}_{\bf X}, U^{1}({\bf X}), \dots , U^{m+s}({\bf X}), \dots)$ 
and their derivatives, polynomial in the derivatives 
$({\bf S}_{\bf XX}, {\bf U}_{\bf X}, \dots )$
and having degree $l$ according to our previous definition.

 Expanding the values of $J^{\prime\nu}({\bf X})$ 
in the neighborhood of the submanifold ${\cal K}$, we can write
$$J^{\prime\nu}({\bf X}) \,\, = \,\,
U^{\prime\nu} \left( {\bf J} ({\bf X}) \right) \,\, + \,\,
\sum_{l\geq1} \epsilon^{l} \, j^{\prime\nu}_{(l)}
\left( {\bf S}_{\bf X}, U^{1}({\bf X}), \dots , U^{m+s}({\bf X}),
\dots \right) \,\, + $$  
$$+ \,\, \int \int_{0}^{2\pi}\!\!\!\dots\int_{0}^{2\pi}
T^{\prime\nu}_{i} ({\bf X}, \bm{\theta}, {\bf Y}, \epsilon) \,\,
g^{i} (\bm{\theta}, {\bf Y}) \,\, 
{d^{m} \theta \over (2\pi)^{m}} \, d^{d} Y 
\,\, + \,\, O ({\bf g}^{2}) $$
where, according to the form of constraints (\ref{gconstr}),
we can put
$$T^{\prime\nu}_{i} ({\bf X}, \bm{\theta}, {\bf Y}, \epsilon) 
 =  \sum_{l_{1}, \dots, l_{d}} 
\left. \Pi^{\prime\nu (l_{1} \dots l_{d})}_{i} \left(
\bm{\varphi} (\bm{\theta}, {\bf X}),
\epsilon \bm{\varphi}_{\bf X} (\bm{\theta}, {\bf X}), \dots \right)
\right|_{\cal K} \epsilon^{l_{1} + \dots + l_{d}} \,
\delta_{l_{1} X^{1} \dots l_{d} X^{d}} ({\bf X} - {\bf Y}) 
\equiv $$
$$\equiv \,\, \sum_{l_{1}, \dots, l_{d}} \left.
{\partial P^{\prime\nu} \over 
\partial \varphi^{i}_{l_{1} x^{1} \dots l_{d} x^{d}}}
\left( \bm{\varphi} (\bm{\theta}, {\bf X}),
\epsilon \bm{\varphi}_{\bf X} (\bm{\theta}, {\bf X}), \dots \right)
\right|_{\cal K} \, \epsilon^{l_{1} + \dots + l_{d}} \,\,\, 
\delta_{l_{1} X^{1} \dots l_{d} X^{d}} ({\bf X} - {\bf Y}) $$

 Considering the functionals
$J^{\prime}_{[{\bf q}]} = 
\int q_{\nu} ({\bf X}) \, J^{\prime\nu} ({\bf X}) \, d^{d} X$
with arbitrary smooth (compactly supported) functions 
$q_{\nu} ({\bf X})$, we can write in the vicinity of ${\cal K}$:
\begin{equation}
\label{qnuJprimenu}
J^{\prime}_{[{\bf q}]} \,\, = \,\, \int q_{\nu} ({\bf X}) \left[
U^{\prime\nu} \left( {\bf J} ({\bf X}) \right) \, + \,
\sum_{l\geq1} \epsilon^{l} \, j^{\prime\nu}_{(l)}
\left( {\bf S}_{\bf X}, U^{1}({\bf X}), \dots , U^{m+s}({\bf X}),
\dots \right) \right] \, d^{d} X \,\, +
\end{equation}
\begin{multline*}
+ \int \! \sum_{l_{1}, \dots, l_{d}} (-1)^{l_{1} + \dots + l_{d}}
\,\, \epsilon^{l_{1} + \dots + l_{d}} \,\, \times \\
\times \, \left[ {d^{l_{1}} \over d Y_{1}^{l_{1}}} \dots 
{d^{l_{d}} \over d Y_{d}^{l_{d}}} \, q_{\nu} ({\bf Y})  \left.
\Pi^{\prime\nu (l_{1} \dots l_{d})}_{i} 
\left( \bm{\varphi} (\bm{\theta}, {\bf Y}),
\epsilon \bm{\varphi}_{\bf Y} (\bm{\theta}, {\bf Y}), \dots \right)
\right|_{\cal K} \right]
g^{i} (\bm{\theta}, {\bf Y}) \,\,
{d^{m} \theta \over (2\pi)^{m}} \, d^{d} Y \,\, + \\ 
+ \,\,\,\,\, O ({\bf g}^{2}) 
\end{multline*}

 The leading term (in $\epsilon$) in the second part of
expression (\ref{qnuJprimenu}) is given by the expression
\begin{multline*}
\int q_{\nu} ({\bf Y}) \sum_{l_{1}, \dots, l_{d}}
(-1)^{l_{1} + \dots + l_{d}} \,\,
k^{\alpha^{1}_{1}}_{1}({\bf Y}) \dots
k^{\alpha^{1}_{l_{1}}}_{1}({\bf Y})
\, \dots \,
k^{\alpha^{d}_{1}}_{d}({\bf Y}) \dots
k^{\alpha^{d}_{l_{d}}}_{d}({\bf Y}) \,\, \times \\
\times \,\, \Pi^{\prime\nu (l_{1} \dots l_{d})}_{i, \,
\theta^{\alpha^{1}_{1}}\dots\theta^{\alpha^{1}_{l_{1}}}
\dots \theta^{\alpha^{d}_{1}}\dots\theta^{\alpha^{d}_{l_{d}}}}
\left( \bm{\Phi} \left(
{{\bf S}({\bf Y}) \over \epsilon} + \bm{\theta}, {\bf Y} \right),
\dots \right) \, g^{i} (\bm{\theta}, {\bf Y}) \,\,
{d^{m} \theta \over (2\pi)^{m}} \,\, d^{d} Y 
\end{multline*}
and coincides with the value
$$\int\!\int_{0}^{2\pi}\!\!\!\!\!\dots\int_{0}^{2\pi}
q_{\nu} ({\bf Y}) \, \zeta^{\prime(\nu)}_{i[{\bf U}({\bf Y})]}
\left( {{\bf S}({\bf Y}) \over \epsilon} + \bm{\theta} \right) \,
g^{i} (\bm{\theta}, {\bf Y}) \,\,
{d^{m} \theta \over (2\pi)^{m}} \, d^{d} Y $$
where
$$\zeta^{\prime(\nu)}_{i[{\bf U}]} (\bm{\theta}) \,\, = \,\,
\left. \left[ {\delta \over \delta \varphi^{i}(\bm{\theta})}
\int_{0}^{2\pi}\!\!\!\!\!\dots\int_{0}^{2\pi} P^{\prime\nu}
(\bm{\varphi}, \, k_{1}^{\beta_{1}} 
\bm{\varphi}_{\theta^{\beta_{1}}}, \dots,
k_{d}^{\beta_{d}} \bm{\varphi}_{\theta^{\beta_{d}}}, \dots ) \,\,
{d^{m} \theta \over (2\pi)^{m}} \right]
\right|_{\bm{\varphi}(\bm{\theta}) = \bm{\Phi} (\bm{\theta}, {\bf U})}$$

 As the values $\zeta^{(\nu)}_{i[{\bf U}]} (\bm{\theta})$, the values
$\zeta^{\prime(\nu)}_{i[{\bf U}]} (\bm{\theta})$ represent regular left
eigenvectors of the operator ${\hat L}^{i}_{j[{\bf U}]}$ corresponding
to the zero eigenvalue. In the case of a complete regular
family of $m$-phase solutions, we have therefore
$$\zeta^{\prime(\nu)}_{i[{\bf U}]} (\bm{\theta}) \,\, = \,\,
\sum_{q} \Gamma^{\nu}_{q} ({\bf U}) \,\,
\kappa^{(q)}_{i[{\bf U}]} (\bm{\theta}) $$
for some functions $\Gamma^{\nu}_{q} ({\bf U})$.  

 We can write, therefore, up to quadratic
terms in ${\bf g} (\bm{\theta}, {\bf X})$:
\begin{multline*}
\int q_{\nu} ({\bf X}) \, J^{\prime\nu} ({\bf X}) \, d^{d} X
\,\, = \\
= \,\, \int q_{\nu} ({\bf X}) \left[
U^{\prime\nu} \left( {\bf J} ({\bf X}) \right) \, + \,
\sum_{l\geq1} \epsilon^{l} \, j^{\prime\nu}_{(l)}
\left( {\bf S}_{\bf X}, U^{1}({\bf X}), \dots , U^{m+s}({\bf X}),
\dots \right) \right] \, d^{d} X \,\, + \\
+ \, \int q_{\nu} ({\bf X}) \left[ \sum_{q}
\Gamma^{\nu}_{q} ({\bf U}) \,\, \kappa^{(q)}_{i[{\bf U}({\bf X})]}
\left( {{\bf S}({\bf X}) \over \epsilon} + \bm{\theta} \right) 
\, + \, O (\epsilon) \right] g^{i} (\bm{\theta}, {\bf X}) \,\,
{d^{m} \theta \over (2\pi)^{m}} \, d^{d} X
\,\,\, + \,\,\, O ({\bf g}^{2}) 
\end{multline*}

 Consider the Poisson brackets:
$$\left. \left\{ J^{\prime}_{[{\bf q}]} , J^{\prime}_{[{\bf p}]}
\right\} \right|_{\cal K} \,\, = \,\,
\left. \left\{ \int q_{\nu} ({\bf X}) \, 
J^{\prime\nu} ({\bf X}) \, d^{d} X
\,\, , \,\, \int p_{\mu} ({\bf Y}) \, 
J^{\prime\mu} ({\bf Y}) \, d^{d} Y \right\}
\right|_{\cal K} \,\, = $$
$$= \,\, \int q_{\nu} ({\bf X}) \,
{\partial U^{\prime\nu} \over \partial U^{\lambda}}({\bf X}) \, 
\left[ \left. \left\{ J^{\lambda}({\bf X}) \,\, , \,\,
\int p_{\mu} ({\bf Y}) \, J^{\prime\mu} ({\bf Y}) \, d^{d} Y \right\}
\right|_{\cal K} \, + \, O (\epsilon^{2}) \right] \, d^{d} X \,\, + $$
\begin{multline*}
+ \,\, \int
q_{\nu} ({\bf X}) \sum_{q} \Gamma^{\nu}_{q} ({\bf U}({\bf X})) \left[
\kappa^{(q)}_{i[{\bf U}({\bf X})]}
\left( {{\bf S}({\bf X}) \over \epsilon} + \bm{\theta} \right) \, + \,
O (\epsilon) \right] \times \\
\times \left. \left\{  g^{i} (\bm{\theta}, {\bf X}) \, , \,
J^{\prime\mu} ({\bf Y}) \right\} \right|_{\cal K} p_{\mu} ({\bf Y}) \,\,
{d^{m} \theta \over (2\pi)^{m}} \, d^{d} X \, d^{d} Y
\end{multline*}

 By Lemma 3.2$^{\prime}$ we have the relation
$\{ g^{i} (\bm{\theta}, {\bf X}) \, , \, 
J^{\prime}_{[{\bf p}]} \}|_{\cal K}
= O (\epsilon)$ on the submanifold ${\cal K}$. In addition,
completely analogous to relation (\ref{LeftVecAort}) holds
the relation
$$\int_{0}^{2\pi}\!\!\!\!\!\dots\int_{0}^{2\pi}
\kappa^{(q)}_{i[{\bf U}({\bf X})]}
\left( {{\bf S}({\bf X}) \over \epsilon} + \bm{\theta} \right) \,
\left\{ g^{i} (\bm{\theta}, {\bf X}) \, , \,
J^{\prime}_{[{\bf p}]} \right\}|_{{\cal K}[1]} \,
{d^{m} \theta \over (2\pi)^{m}} \,\, \equiv \,\, 0 $$
by virtue of the original dependence of the constraints
$g^{i} (\bm{\theta}, {\bf X})$. We thus obtain:
$$\int\!\!\int q_{\nu} ({\bf X}) \, \left. \left\{ 
J^{\prime\nu} ({\bf X}) \, , \, J^{\prime\mu} ({\bf Y}) \right\} 
\right|_{\cal K} p_{\mu} ({\bf Y}) \,
d^{d} X \, d^{d} Y \,\, = $$
$$= \,\, \int q_{\nu} ({\bf X}) \,
{\partial U^{\prime\nu} \over \partial U^{\lambda}} ({\bf X}) \,
\left. \left\{ J^{\lambda}({\bf X}) \, , \,
\int p_{\mu} ({\bf Y}) \, 
J^{\prime\mu} ({\bf Y}) \, d^{d} Y \right\} \right|_{\cal K}
\, d^{d} X \,\, + \,\, O (\epsilon^{2}) $$

 Repeating the arguments for the functional
$\int p_{\mu} ({\bf Y}) \, J^{\prime\mu} ({\bf Y}) \, d^{d} Y$ 
we finally obtain
\begin{equation}
\label{JprimeJrel}
\int\!\!\int q_{\nu} ({\bf X}) \, 
\left. \left\{ J^{\prime\nu} ({\bf X})
\, , \, J^{\prime\mu} ({\bf Y}) \right\} 
\right|_{\cal K} \, p_{\mu} ({\bf Y}) \,\,
d^{d} X \, d^{d} Y \,\, =
\end{equation}
$$= \,\, \int\!\!\int q_{\nu} ({\bf X}) \,\,
{\partial U^{\prime\nu} \over \partial U^{\lambda}} ({\bf X}) \,
\left. \left\{ J^{\lambda}({\bf X}) \, , \,
J^{\sigma} ({\bf Y}) \right\} \right|_{\cal K} \,
{\partial U^{\prime\mu} \over \partial U^{\sigma}} ({\bf Y}) \,\,
p_{\mu} ({\bf Y}) \,\, d^{d} X \, d^{d} Y 
\,\, + \,\, O (\epsilon^{2}) $$

 Given that the principal (in $\epsilon$) terms in the expressions
$\{ J^{\lambda}({\bf X}) \, , \, J^{\sigma} ({\bf Y}) \} |_{\cal K}$ 
\linebreak and
$\{ J^{\prime\nu} ({\bf X}) \, , \, J^{\prime\mu} ({\bf Y}) \} |_{\cal K}$
coincide with the Dubrovin - Novikov forms, obtained with the aid of
the sets $(I^{1}, \dots, I^{N})$ and $(I^{\prime 1}, \dots, I^{\prime N})$
respectively, we conclude from (\ref{JprimeJrel}):
$$\left\{ U^{\prime\nu} ({\bf X}) \, ,
\, U^{\prime\mu} ({\bf Y}) \right\}^{\prime}_{DN}
\,\, = \,\, 
{\partial U^{\prime\nu} \over \partial U^{\lambda}} ({\bf X}) \,
\left\{ U^{\lambda}({\bf X}) \, , \, U^{\sigma} ({\bf Y}) \right\}_{DN} \,
{\partial U^{\prime\mu} \over \partial U^{\sigma}} ({\bf Y}) $$
which means the coinciding of the forms $\{ \dots , \dots \}_{DN}$ 
and $\{ \dots , \dots \}^{\prime}_{DN}$.

 As a result, we can claim that the relations
$$\left\{ k^{\alpha}_{p} ({\bf U}({\bf X})) \, , \,
k^{\beta}_{q} ({\bf U}({\bf Y})) \right\} \, = \, 0
\,\,\, , \,\,\,\,\,
\left\{ k^{\alpha}_{p} ({\bf U}({\bf X})) \, , \,
U^{\gamma}({\bf Y}) \right\} \, = \,
\left[ \omega^{\alpha\gamma} ({\bf U}({\bf X})) \,\,
\delta ({\bf X} - {\bf Y}) \right]_{X^{p}} $$
($\alpha, \beta = 1, \dots, m$, $\gamma = 1, \dots, m + s$)
and the expressions for
$$\left\{ U^{\gamma}({\bf X}) \, , \, U^{\rho}({\bf Y})
\right\}_{DN} \,\,\,\,\,\,\,\, , \,\,\,\,\,\,\,\,
\gamma , \rho \, = \, 1, \dots, m + s $$
transform into the expressions
$$\left\{ k^{\alpha}_{p} ({\bf U}^{\prime}({\bf X})) ,
k^{\beta}_{q} ({\bf U}^{\prime}({\bf Y})) \right\} \, = \, 0
\,\,\, , \,\,\,\,\,
\left\{ k^{\alpha}_{p} ({\bf U}^{\prime}({\bf X})) , 
U^{\prime \gamma}({\bf Y}) \right\} \, = \,
\left[ \omega^{\prime \alpha\gamma} ({\bf U}^{\prime}({\bf X})) 
\,\, \delta ({\bf X} - {\bf Y}) \right]_{X^{p}} $$
and
$$\left\{ U^{\prime \gamma}({\bf X}) \, , \, 
U^{\prime \rho}({\bf Y}) \right\}^{\prime}_{DN}$$
after the change of coordinates
$$\left( U^{1}, \dots, U^{N} \right) \,\, \rightarrow \,\,
\left( U^{\prime 1}, \dots, U^{\prime N} \right) $$
on $\Lambda$. We can claim then that brackets 
(\ref{AveragedBracket}) obtained with the aid of the sets
$(I^{1}, \dots, I^{N})$ and $(I^{\prime 1}, \dots, I^{\prime N})$
transform into each other after the change of coordinates
$$\left( S^{1} ({\bf X}), \! \dots, S^{m} ({\bf X}), 
U^{1} ({\bf X}), \dots, U^{m+s} ({\bf X}) \right) \rightarrow
\left( S^{1} ({\bf X}), \! \dots, S^{m} ({\bf X}), 
U^{\prime 1} ({\bf X}), \dots, U^{\prime m+s} ({\bf X}) \right) $$
where 
$U^{\prime \gamma} ({\bf X}) = U^{\prime \gamma} \, ({\bf S}_{\bf X}, 
\, U^{1} ({\bf X}), \dots, U^{m+s} ({\bf X}))$. 

{\hfill Theorem 3.2 is proved.}

\vspace{0.2cm}

 Let us note that Theorem 3.2 means, in particular, that bracket
(\ref{AveragedBracket}) is also invariant with respect to the
choice of the functionals $(I^{1}, \dots, I^{m+s})$ among the full
set $\{I^{\nu}, \,\, \nu = 1, \dots, N \}$.

 Finally, we prove the theorem about the Hamiltonian properties of the 
Whitham system (\ref{STOmegaRel}) - (\ref{AddPartWhithSyst})
under the same conditions as before.

\vspace{0.2cm}

{\bf Theorem 3.3.}

{\it Let $\Lambda$ be a regular Hamiltonian family of $m$-phase
solutions of (\ref{EvInSyst}). Let $(I^{1}, \dots, I^{N})$ be a
complete Hamiltonian set of commuting first integrals of
(\ref{EvInSyst}) having the form (\ref{Integrals}) and $H$ be the
Hamiltonian function for system (\ref{EvInSyst})
having the form (\ref{MultDimHamFunc}). Then the Whitham system
(\ref{STOmegaRel}) - (\ref{AddPartWhithSyst})
is Hamiltonian with respect to the
corresponding bracket (\ref{AveragedBracket})
with the Hamiltonian function
$$H^{av} \,\, = \,\, \int \langle P_{H} \rangle
\left( {\bf S}_{\bf X}, \, U^{1} ({\bf X}), \dots, U^{m+s} ({\bf X})
\right) \,\, d^{d} X $$

}

\vspace{0.2cm}

 Proof.

 By Theorem 3.2, without loss of generality we can assume that
the Hamiltonian functional $H$ belongs to the set 
$(I^{1}, \dots, I^{m + s})$,
$H = I^{\mu_{0}}$. It is easy to verify then that the corresponding
Hamiltonian $H^{av}$ generates in this case the system
$$S^{\alpha}_{T} \,\,\, = \,\,\, \omega^{\alpha \mu_{0}}
\left({\bf S}_{\bf X}, \, U^{1}, \dots, U^{m+s} \right)
\,\,\,\,\,\,\,\, , \,\,\,\,\,\,\,\,\,\, \alpha = 1, \dots, m $$
$$U^{\gamma}_{T} \,\,\, = \,\,\, 
\langle Q^{\gamma \mu_{0} 1} \rangle_{X^{1}}
\,\, + \,\, \dots \,\, + \,\, 
\langle Q^{\gamma \mu_{0} d} \rangle_{X^{d}}
\,\,\,\,\,\,\,\, , \,\,\,\,\,\,\,\,\,\, \gamma = 1, \dots, m + s $$
i.e. exactly system (\ref{STOmegaRel}) - (\ref{AddPartWhithSyst}).

{\hfill Theorem 3.3 is proved.}

\vspace{0.2cm}

\section{On the canonical forms of the averaged brackets.}
\setcounter{equation}{0}

 Let us consider now ``canonical forms'' acquired by the averaged 
brackets in some special coordinates. We first prove here the
following lemma:

\vspace{0.2cm}

{\bf Lemma 4.1.}

{\it Consider bracket (\ref{AveragedBracket}) in the
coordinates 
$(S^{1}({\bf X}), \dots, S^{m}({\bf X}), \, U^{1}({\bf X}), \dots, 
U^{m+s}({\bf X}))$. There exists locally an invertible change of
coordinates
\begin{multline*}
\left( S^{1}({\bf X}), \dots, S^{m}({\bf X}), \, U^{1}({\bf X}), 
\dots, U^{m+s}({\bf X}) \right) \,\, \rightarrow \\
\rightarrow \,\, 
\left( S^{1}({\bf X}), \dots, S^{m}({\bf X}), 
\, Q_{1}({\bf X}), \dots, Q_{m}({\bf X}),
\, N^{1}({\bf X}), \dots, N^{s}({\bf X}) \right) 
\end{multline*}
where 
$$Q_{\alpha} \,\, = \,\, Q_{\alpha} \left( {\bf S}_{\bf X}, \,
U^{1}, \dots , U^{m+s} \right) \,\,\,\,\, , \,\,\,\,\,
N^{l} \,\, = \,\,  N^{l}  \left( {\bf S}_{\bf X}, \,
U^{1}, \dots , U^{m+s} \right) $$
such that the bracket (\ref{AveragedBracket}) has in the new
coordinates the form:
\begin{equation}
\label{FirstPartBracket}
\left\{ S^{\alpha} ({\bf X}) \, , \, S^{\beta} ({\bf Y}) \right\}
\, = \, 0 \,\,\, , \,\,\,
\left\{ S^{\alpha} ({\bf X}) \, , \, Q_{\beta} ({\bf Y}) \right\}
\, = \, \delta^{\alpha}_{\beta} \,\, 
\delta ({\bf X} - {\bf Y}) \,\,\, , \,\,\,
\left\{ S^{\alpha} ({\bf X}) \, , \, N^{l}  ({\bf Y}) \right\}
\, = \, 0
\end{equation}
$$\left\{ Q_{\alpha} ({\bf X}) \, , \, Q_{\beta} ({\bf Y}) \right\}
\,\, = \,\, J_{\alpha\beta} \left[ {\bf S}, {\bf N} \right] 
({\bf X} - {\bf Y}) \,\,\,\,\, , \,\,\,\,\,
\left\{ Q_{\alpha} ({\bf X}) \, , \, N^{l}  ({\bf Y}) \right\}
\,\, = \,\, J_{\alpha}^{l} \left[ {\bf S}, {\bf N} \right]
({\bf X} - {\bf Y}) $$
\begin{equation}
\label{SecondPartBracket}
\left\{  N^{l}  ({\bf X}) \, , \, N^{q}  ({\bf Y}) \right\}
\,\, = \,\, J^{lq} \left[ {\bf S}, {\bf N} \right]
({\bf X} - {\bf Y}) 
\end{equation}
($\alpha , \beta = 1, \dots, m \, , \, l , q = 1, \dots, s$),
where $J_{\alpha\beta}$, $J_{\alpha}^{l}$, $J^{lq}$ are local 
distributions of the gradation degree 1 given by linear combinations 
of the function $\delta ({\bf X} - {\bf Y})$ and its first derivatives
with local coefficients, depending on the values
$({\bf S}_{\bf X}, \, {\bf N} ({\bf X}), \, {\bf S}_{\bf XX}, \,
{\bf N}_{\bf X} ({\bf X}))$.

}

\vspace{0.2cm}

Proof.

 Let us fix the parameters $k^{\alpha}_{p} = S^{\alpha}_{X^{p}}$ on
the family $\Lambda$ and consider the coordinates \linebreak
${\bf U} = (U^{1}, \, \dots, \, U^{m+s})$ on the $(m + s)$-dimensional
submanifolds $k^{\alpha}_{p} = {\rm const}$ (for all $\alpha, p$). 
Let us consider the vector fields
$${\vec \xi}_{(\alpha)} \,\, = \,\,
\left( \, \omega^{\alpha \, 1} ({\bf S}_{\bf X}, \,  {\bf U}) , \dots,
\, \omega^{\alpha \, m + s} ({\bf S}_{\bf X}, \,  {\bf U}) \,
\right)^{t} $$
on the submanifolds $k^{\alpha}_{p} = {\rm const}$.

 From the Jacobi identities 
$$\left\{ \left\{ U^{\nu} ({\bf X}) \, , \, S^{\alpha} ({\bf Y})
\right\} \, , \, S^{\beta} ({\bf Z}) \right\} \,\, - \,\,
\left\{ \left\{ U^{\nu} ({\bf X}) \, , \, S^{\beta} ({\bf Z})
\right\} \, , \, S^{\alpha} ({\bf Y}) \right\} 
\,\,\, \equiv \,\,\, 0 $$
for the bracket (\ref{AveragedBracket}) it is easy to get the
relations
$$\left[ \, {\vec \xi}_{(\alpha)} \, , \, {\vec \xi}_{(\beta)} \,
\right] \,\,\, \equiv \,\,\, 0 \,\,\,\,\, , \,\,\,\,\,
\alpha, \beta = 1, \dots, m $$ 
for the commutators of the vectors fields ${\vec \xi}_{(\alpha)}$
on the submanifolds $k^{\alpha}_{p} = {\rm const}$.

 According to relations (\ref{RankOmegaAlphaGamma}) for the
frequencies generated by the functionals \linebreak
$(I^{1}, \dots, I^{m+s})$
we can state also that the set $\{ {\vec \xi}_{(\alpha)} \}$ is
linearly independent at every point.

 We can claim then that on every submanifold 
$$\left\{ k^{\alpha}_{p} = {\rm const} \,\,\, , \,\,\,
\alpha = 1, \dots, m \,\, , \,\, p = 1, \dots, d \right\} $$ 
there exists locally an invertible change of coordinates
$$ \left( U^{1}, \dots, U^{m+s} \right) 
\,\, \rightarrow \,\,
\left( \, Q_{1}({\bf S}_{\bf X}, {\bf U}), \dots, \, 
Q_{m}({\bf S}_{\bf X}, {\bf U}), \,\, 
N^{1}({\bf S}_{\bf X}, {\bf U}), \dots, 
N^{s}({\bf S}_{\bf X}, {\bf U}) \, \right) $$
depending
smoothly on the parameters $k^{\alpha}_{p} = S^{\alpha}_{X^{p}}$,
such that the coordinate representation of the vector fields
${\vec \xi}_{(\alpha)}$ on these submanifolds has the form
$${\vec \xi}_{(1)} \, = \, ( 1, 0, \dots, 0 )^{t} \,\, , \,\,\,\,\,
\dots \,\,\,\,\, , \,\, {\vec \xi}_{(m)} \, = \, 
( 0, \dots, 0, 1, 0, \dots, 0  )^{t} $$

 It is not difficult to see then that the corresponding change of
coordinates
$$\left( \, {\bf S} ({\bf X}) \, , \, {\bf U} ({\bf X}) \, \right)
\,\,\, \rightarrow \,\,\, \left( \, {\bf S} ({\bf X}) \, , \,
 {\bf Q} ({\bf X}) \, , \,  {\bf N} ({\bf X}) \, \right) $$
provides relations (\ref{FirstPartBracket}) for bracket
(\ref{AveragedBracket}). 

 The general form of the distributions
$J_{\alpha\beta}$, $J_{\alpha}^{l}$, $J^{lq}$ follows then from the
form of bracket (\ref{AveragedBracket}) and for the proof of the
Lemma we have just to prove the relations
\begin{equation}
\label{ZeroDepRel}
\delta J_{\alpha\beta} \, / \, \delta Q_{\gamma} ({\bf Z}) \, \equiv \, 0
\,\,\,\,\, , \,\,\,\,\,
\delta J_{\alpha}^{l} \, / \, \delta Q_{\gamma} ({\bf Z}) \, \equiv \, 0
\,\,\,\,\, , \,\,\,\,\,
\delta J^{lq} \, / \, \delta Q_{\gamma} ({\bf Z}) \, \equiv \, 0
\,\,\,\,\,\,\,\, , \,\,\,\,\, \gamma = 1, \dots, m
\end{equation}
for the functionals $J_{\alpha\beta}$, $J_{\alpha}^{l}$, $J^{lq}$.

 Using the Jacobi identities
$$\left\{ \left\{ Q_{\alpha} ({\bf X}) ,  Q_{\beta} ({\bf Y})
\right\} ,  S^{\gamma} ({\bf Z}) \right\} \,\, + \,\,
c.p. \,\,\, \equiv \,\,\, 0 \,\,\,\,\, , \,\,\,\,\,
\left\{ \left\{ Q_{\alpha} ({\bf X}) \, , \, N^{l} ({\bf Y})
\right\} \, , \, S^{\gamma} ({\bf Z}) \right\} \,\, + \,\,
c.p. \,\, \equiv \,\, 0 $$
$$\left\{ \left\{ N^{l} ({\bf X}) , N^{q} ({\bf Y})
\right\} , S^{\gamma} ({\bf Z}) \right\} \,\, + \,\,
c.p. \,\, \equiv \,\, 0 $$
we then easily get relations (\ref{ZeroDepRel}).

{\hfill Lemma 4.1 is proved.}

\vspace{0.2cm}

 Let us note here that according to their definition the 
coordinates $( Q_{\alpha} \, , \, N^{l} )$ are defined modulo
the transformations
$$Q_{\alpha} \,\, \rightarrow \,\, Q_{\alpha} \, + \,
q_{\alpha} ({\bf S}_{\bf X}, \, {\bf N}) \,\,\,\,\, , \,\,\,\,\,
N^{l} \,\, \rightarrow \,\, {\tilde N}^{l} 
({\bf S}_{\bf X}, \, {\bf N}) $$
where the transformation
$N^{l} \,\, \rightarrow \,\, {\tilde N}^{l}
({\bf S}_{\bf X}, \, {\bf N})$ is invertible for every fixed
${\bf S}_{\bf X}$.

 Let us consider now a special case, when the additional parameters
$(n^{1}, \dots, n^{s})$ are absent on the family $\Lambda$ and the 
full regular family of $m$-phase solutions of system (\ref{EvInSyst})
can be parametrized by the $m (d + 1)$ parameters 
$(k^{\alpha}_{p}, \omega^{\alpha})$, $\alpha = 1, \dots, m$,
$p = 1, \dots, d$.

 In this situation we can prove the following theorem about the
canonical form of the averaged bracket (\ref{AveragedBracket}).

\vspace{0.2cm}

{\bf Theorem 4.1.}

{\it
 Let system (\ref{EvInSyst}) be a local Hamiltonian system
generated by the functional (\ref{MultDimHamFunc}) in the local
field-theoretic Hamiltonian structure (\ref{MultDimPBr}).
Let $\Lambda$ be a regular Hamiltonian family of
$m$-phase solutions of (\ref{EvInSyst}) and 
$(I^{1}, \dots , I^{m(d+1)})$
be a complete Hamiltonian set of commuting integrals
(\ref{Integrals}) for this family.

 Let relations (\ref{AveragedBracket}) represent a Poisson bracket
on the space of fields 
$$(S^{1} ({\bf X}), \dots, S^{m} ({\bf X}), \, 
U^{1} ({\bf X}), \dots, U^{m} ({\bf X}))$$

 Then locally there exists 
an invertible change of coordinates
$$\left( S^{1}({\bf X}), \dots, S^{m}({\bf X}), \, U^{1}({\bf X}),
\dots, U^{m}({\bf X}) \right) \,\, \rightarrow  \,\,
\left( S^{1}({\bf X}), \dots, S^{m}({\bf X}),
\, Q_{1}({\bf X}), \dots, Q_{m}({\bf X}) \right) $$
where
$Q_{\alpha} \,\, = \,\, Q_{\alpha} ( {\bf S}_{\bf X}, \, {\bf U})$ 
such that the bracket (\ref{AveragedBracket}) acquires in the 
coordinates \linebreak $({\bf S} ({\bf X}) \, , \, {\bf Q} ({\bf X}))$ 
the following non-degenerate canonical form:
$$ \left\{ S^{\alpha} ({\bf X}) \, , \, S^{\beta} ({\bf Y}) \right\}
\, = \, 0 \,\,\, , \,\,\,\,\,
\left\{ S^{\alpha} ({\bf X}) \, , \, Q_{\beta} ({\bf Y}) \right\}
\, = \, \delta^{\alpha}_{\beta} \,\,
\delta ({\bf X} - {\bf Y}) \,\,\, , \,\,\,\,\,
\left\{ Q_{\alpha} ({\bf X}) \, , \, Q_{\beta}  ({\bf Y}) \right\}
\, = \, 0 $$

}

\vspace{0.2cm}

Proof.

 Let us give here just a proof in abstract field-theoretical form.
The same considerations can be made also in the differential-geometric
form after concrete substitutions for the coordinate transformations. 
However, we would like to skip here the corresponding calculations
to avoid rather bulky expressions in the text.

 In accordance with Lemma 4.1 let us write down the averaged bracket
(\ref{AveragedBracket}) in coordinates \linebreak
$(S^{\alpha} ({\bf X}) \, , \, {\tilde Q}_{\alpha} ({\bf X}))$
such that
$$ \left\{ S^{\alpha} ({\bf X}) \, , \, S^{\beta} ({\bf Y}) \right\}
\,\, = \,\, 0 \,\,\,\,\, , \,\,\,\,\,
\left\{ S^{\alpha} ({\bf X}) \, , \, 
{\tilde Q}_{\beta} ({\bf Y}) \right\}
\,\, = \,\, \delta^{\alpha}_{\beta} \,\,
\delta ({\bf X} - {\bf Y}) $$
$$\left\{ {\tilde Q}_{\alpha} ({\bf X}) \, , \, 
{\tilde Q}_{\beta}  ({\bf Y}) \right\}
\,\, = \,\, {\tilde J}_{\alpha\beta} \left[ {\bf S} \right]
({\bf X} - {\bf Y}) $$

 From the Jacobi identities
$$\left\{ \left\{ {\tilde Q}_{\alpha} ({\bf X}) \, , \,  
{\tilde Q}_{\beta} ({\bf Y}) \right\} \, , \, 
{\tilde Q}_{\gamma} ({\bf Z}) \right\} \,\,\, + \,\,\,
c.p. \,\,\, \equiv \,\,\, 0 $$
we then easily get the relations
\begin{equation}
\label{ClosedJRelations}
{ \delta {\tilde J}_{\alpha\beta} [{\bf S}] ({\bf X}, {\bf Y}) 
\over \delta S^{\gamma} ({\bf Z})} \,\, + \,\,
{ \delta {\tilde J}_{\beta\gamma} [{\bf S}] ({\bf Y}, {\bf Z})
\over \delta S^{\alpha} ({\bf X})} \,\, + \,\,
{ \delta {\tilde J}_{\gamma\alpha} [{\bf S}] ({\bf Z}, {\bf X})
\over \delta S^{\beta} ({\bf Y})} \,\,\, \equiv \,\,\, 0
\end{equation}
for the form 
${\tilde J}_{\alpha\beta} [{\bf S}] ({\bf X}, {\bf Y})
\, \equiv \, {\tilde J}_{\alpha\beta} [{\bf S}] ({\bf X} - {\bf Y})$.

 Relations (\ref{ClosedJRelations}) mean that the 2-form
${\tilde J}_{\alpha\beta} [{\bf S}] ({\bf X}, {\bf Y})$ on the space
of fields \linebreak
$(S^{1} ({\bf X}), \dots, S^{m} ({\bf X}))$ is closed and
can be locally represented as
$${\tilde J}_{\alpha\beta} [{\bf S}] ({\bf X}, {\bf Y}) 
\,\,\, = \,\,\,
{ \delta q_{\alpha} [{\bf S}] ({\bf X})
\over \delta S^{\beta} ({\bf Y})} \,\, - \,\,
{ \delta q_{\beta} [{\bf S}] ({\bf Y})
\over \delta S^{\alpha} ({\bf X})} $$
for some 1-form $q_{\alpha} [{\bf S}] ({\bf X})$ on the same phase 
space. Easy to see now that the change
$$Q_{\alpha} ({\bf X}) \,\,\, = \,\,\, 
{\tilde Q}_{\alpha} ({\bf X}) \,\, - \,\, 
q_{\alpha} [{\bf S}] ({\bf X})$$
gives the necessary coordinates $Q_{\alpha} ({\bf X})$.

 From the form of the functionals
${\tilde J}_{\alpha\beta} [{\bf S}] ({\bf X}, {\bf Y})$
it's not difficult to show also that the 1-form
$q_{\alpha} [{\bf S}] ({\bf X})$ can be chosen in the form
$q_{\alpha} [{\bf S}] ({\bf X}) \, 
\equiv \, q_{\alpha} ({\bf S}_{\bf X})$
with some smooth functions $q_{\alpha} ({\bf S}_{\bf X})$,
which gives the necessary form of the corresponding coordinate
transformation.

{\hfill Theorem 4.1 is proved.}

\vspace{0.2cm}

 Let us note that in more general case of the presence of additional
parameters \linebreak
$(n^{1}, \dots, n^{s})$ we can expect in fact, that the 
bracket (\ref{FirstPartBracket}) - (\ref{SecondPartBracket}) can have
more complicated, maybe even degenerate, form. 
 
 We can note also that the bracket 
$\{ N^{l} ({\bf X}) \, , \, N^{q}  ({\bf Y}) \}$ given by
(\ref{SecondPartBracket}) represents a Poisson bracket 
on the space of fields ${\bf N} ({\bf X})$ for any
fixed values of ${\bf S}({\bf X})$ as follows from the form of
the bracket (\ref{FirstPartBracket}) - (\ref{SecondPartBracket}).
This actually gives rather strong restrictions on the form of the
bracket (\ref{SecondPartBracket}).

 At last we consider just a very simple example of the averaging of
a non-degenerate bracket in the single-phase case. Let us consider
the nonlinear wave equation in $d$ spatial dimensions having the form
\begin{equation}
\label{NonLinWaveEq}
\varphi_{tt} \,\, - \,\, \Delta \, \varphi \,\, = \,\, 
V^{\prime} (\varphi)
\end{equation}
with some potential function $V (\varphi)$.

 The periodic one-phase solutions of (\ref{NonLinWaveEq}) are defined
by the equation
\begin{equation}
\label{OnePhaseNLWE}
\left( \omega^{2} - k_{1}^{2} - \dots - k_{d}^{2} \right) \,
\varphi_{\theta\theta} \,\, = \,\, V^{\prime} (\varphi)
\end{equation}
which implies
$${\omega^{2} - k_{1}^{2} - \dots - k_{d}^{2} \over 2} \,\,
\varphi_{\theta}^{2} \,\, = \,\, V (\varphi) \, + \, C $$
with some integration constant $C$. We can assume
for example that we have here $\omega^{2} \, > \, {\bf k}^{2}$
though it does not affect the final conclusions.

 The solutions $\varphi (\theta, \, \omega, k_{1}, \dots, k_{d})$
are then given by the quadrature
$$\pm \, \sqrt{{\omega^{2} - k_{1}^{2} - \dots - k_{d}^{2} \over 2}} 
\,\, \int {d \, \varphi \over \sqrt{V (\varphi) + C}} \,\,\, = \,\,\,
\theta \, + \, \theta_{0} $$
where the value of $\varphi$ oscillates between two subsequent zeros
of the expression $V (\varphi) + C$, while the condition
$$\sqrt{{\omega^{2} - k_{1}^{2} - \dots - k_{d}^{2} \over 2}}
\,\, \oint {d \, \varphi \over \sqrt{V (\varphi) + C}} \,\,\, = \,\,\,
2 \pi $$
fixes the value of the integration constant $C$. 

 Let us denote the corresponding function
$\varphi (\theta, \, \omega, k_{1}, \dots, k_{d})$ as
$\Phi (\theta + \theta_{0}, \, \omega, k_{1}, \dots, k_{d})$.

 Equation (\ref{NonLinWaveEq}) can be written in the Hamiltonian form
\begin{equation}
\label{NonLinWEHamForm}
\varphi_{t} \, = \, \psi \,\,\,\,\,\,\,\,\,\, ,
\,\,\,\,\,\,\,\,\,\,
\psi_{t} \, = \, \Delta \, \varphi \, + \, V^{\prime} (\varphi) 
\end{equation}
with the Hamiltonian function
\begin{equation}
\label{NonLinWEHamFunc}
H \,\, \equiv \,\, \int P_{H} (\varphi, \, \psi, \dots ) \, d^{d} x
\,\, \equiv \,\, \int \left( {\psi^{2} \over 2} 
\, + \, {(\nabla \varphi)^{2} \over 2} \, - \,
V (\varphi) \right) \, d^{d} x 
\end{equation}
and the Poisson bracket
$$\{ \varphi ({\bf x}) \, , \, \varphi ({\bf y}) \} \, = \, 0 
\,\,\,\,\, , \,\,\,\,\,
\{ \varphi ({\bf x}) \, , \, \psi ({\bf y}) \} \, = \,
\delta ({\bf x} - {\bf y}) \,\,\,\,\, , \,\,\,\,\,
\{ \psi ({\bf x}) \, , \,  \psi ({\bf y}) \} \, = \, 0 $$

 The family $\Lambda$ is represented then by the family of the 
functions
$$\left(
\begin{array}{c}
\varphi \cr \psi
\end{array} \right) \,\, = \,\,
\left(
\begin{array}{c}
\Phi (\theta + \theta_{0}, \, \omega, k_{1}, \dots, k_{d}) \cr
\omega \, 
\Phi_{\theta} (\theta + \theta_{0}, \, \omega, k_{1}, \dots, k_{d}) 
\end{array} \right) $$
parametrized by the values $(\omega, k_{1}, \dots, k_{d})$ and the
initial phase $\theta_{0}$.

 Using equation (\ref{OnePhaseNLWE}) it is not difficult to show
that the family $\Lambda$ is a complete regular family of single-phase 
solutions of (\ref{NonLinWEHamForm}). The set of the commuting 
functionals $(I^{1}, \dots, I^{d+1})$ is given here by the
momentum functionals
$$I_{q} \,\, \equiv \,\, 
\int P_{q} (\varphi, \, \psi, \dots ) \, d^{d} x
\,\, \equiv \,\, \int \varphi_{x^{q}} \, \psi \,\, d^{d} x 
\,\,\,\,\, , \,\,\,\,\,\,\,\, q = 1, \dots, d $$
and the Hamiltonian functional (\ref{NonLinWEHamFunc}), which give
a complete Hamiltonian set of the first integrals for the family
$\Lambda$. The values of the averaged densities 
$\langle P_{q} \rangle$ on $\Lambda$ are equal to
$$\langle P_{q} \rangle 
\,\,\, =  \,\,\, k_{q} \, \omega \, \int_{0}^{2\pi} 
\Phi_{\theta}^{2} \,\, {d \theta \over 2 \pi} \,\,\, = \,\,\,
k_{q} \, Q $$
where
$$Q \,\,\, \equiv \,\,\, \omega \, \int_{0}^{2\pi}
\Phi_{\theta}^{2} \,\, {d \theta \over 2 \pi} $$
while the functionals $I_{q}$ generate on $\Lambda$ linear shifts
of $\theta_{0}$ with the frequencies $\omega_{q} \, = \, k_{q}$.

 The pairwise Poisson brackets of the densities $P_{q} ({\bf x})$,
$P_{H} ({\bf x})$ can be written in the form
\begin{equation}
\label{PqPlSkobka}
\left\{ P_{q} ({\bf x}) \, , \, P_{l} ({\bf y}) \right\} 
\,\,\, = \,\,\,
P_{q} ({\bf x}) \,\, \delta_{x^{l}} ({\bf x} - {\bf y}) \,\, + \,\,
P_{l} ({\bf x}) \,\, \delta_{x^{q}} ({\bf x} - {\bf y}) \,\, + \,\,
\left[  P_{q} ({\bf x}) \right]_{x^{l}} \, \delta ({\bf x} - {\bf y})
\end{equation}
\begin{multline}
\label{PqPHSkobka}
\left\{ P_{q} ({\bf x}) \, , \, P_{H} ({\bf y}) \right\} 
\,\,\, = \,\,\,
\psi^{2} ({\bf x}) \,\, \delta_{x^{q}} ({\bf x} - {\bf y}) \,\, + \,\,
\varphi_{x^{q}} \, \sum_{l} \, \varphi_{x^{l}} \, 
\delta_{x^{l}} ({\bf x} - {\bf y}) \,\, + \\
+ \,\, \left[ {1 \over 2} \, \psi^{2} \, - \, {1 \over 2} 
\left( \nabla \varphi \right)^{2} \, + \, V(\varphi) \right]_{x^{q}} \,
\delta ({\bf x} - {\bf y}) \,\, + \,\, \sum_{l} \, \left(
\varphi_{x^{q}} \varphi_{x^{l}} \right)_{x^{l}} \,
\delta ({\bf x} - {\bf y}) \hspace{1cm}
\end{multline}
\begin{equation}
\label{PHPHSkobka}
\left\{ P_{H} ({\bf x}) \, , \, P_{H} ({\bf y}) \right\}
\,\,\, = \,\,\, 2 \, \sum_{l} \, P_{l} ({\bf x}) \,\, 
\delta_{x^{l}} ({\bf x} - {\bf y}) \,\, + \,\,
\sum_{l} \, \left[  P_{l} ({\bf x})  \right]_{x^{l}} \,
\delta ({\bf x} - {\bf y}) 
\end{equation}

 Thus, if we choose the coordinates $S ({\bf X})$,
$U_{(q)} ({\bf X}) \, = \, \langle P_{q} \rangle ({\bf X})$ for some
$q = 1, \dots, d$, the corresponding bracket (\ref{AvBracketSinglePhase})
will be written in the form:
\begin{equation}
\label{NumberqAvBr}
\begin{array}{c}
\left\{ S ({\bf X}) \, , \, S ({\bf Y}) \right\} \,\, = \,\, 0
\,\,\,\,\,\,\,\, , \,\,\,\,\,\,\,\,
\left\{ S ({\bf X}) \, , \, U_{(q)} ({\bf Y}) \right\} \,\, = \,\,
S_{X^{q}} \,\, \delta ({\bf X} - {\bf Y}) \cr \cr
\left\{ U_{(q)} ({\bf X}) \, , \, U_{(q)} ({\bf Y}) \right\} \,\, = \,\,
2 \, U_{(q)} ({\bf X}) \,\,  \delta_{X^{q}} ({\bf X} - {\bf Y}) 
\,\, + \,\,
U_{(q) \, X^{q}} \,\, \delta ({\bf X} - {\bf Y}) 
\end{array}
\end{equation}

 It's not difficult to check that in the coordinates
$$ S ({\bf X}) \,\,\,\,\, , \,\,\,\,\,
Q  ({\bf X}) \,\, = \,\, \langle P_{1} \rangle / S_{X^{1}}
\,\, =  \,\, \dots \,\, = \,\,  \langle P_{d} \rangle / S_{X^{d}} $$
any of the brackets (\ref{NumberqAvBr}) takes the form
\begin{equation}
\label{CanOnePhaseAvBr}
\left\{ S ({\bf X}) \, , \, S ({\bf Y}) \right\} \,\, = \,\, 0
\,\,\,\,\,\,\,\, , \,\,\,\,\,\,\,\,
\left\{ S ({\bf X}) \, , \, Q ({\bf Y}) \right\} \,\, = \,\,
\delta ({\bf X} - {\bf Y}) 
\,\,\,\,\,\,\,\, , \,\,\,\,\,\,\,\,
\left\{ Q ({\bf X}) \, , \, Q ({\bf Y}) \right\} \,\, = \,\, 0
\end{equation}

 We can see then that all the brackets (\ref{NumberqAvBr})
represent in fact the same bracket (\ref{CanOnePhaseAvBr}) in different
coordinates. Thus, we can claim that all the brackets (\ref{NumberqAvBr})
transform into each other after the corresponding change of coordinates
$ U_{(q)} ({\bf X}) \, = \, S_{X^{q}} \, U_{(l)} ({\bf X}) / S_{X^{l}}$.

 The averaged density $\langle P_{H} \rangle$ on $\Lambda$ can be 
represented in the form
\begin{multline*}
\langle P_{H} \rangle \,\, = \,\, \int_{0}^{2\pi} \left[
{\omega^{2} + k_{1}^{2} + \dots +  k_{d}^{2} \over 2} \,\,\,
\Phi_{\theta}^{2} \,\, - \,\, V  \left( \Phi \right) \right] \,\,
{d \theta \over 2\pi} \,\, = \\
= \,\, \omega^{2} \, \int_{0}^{2\pi} \Phi_{\theta}^{2} \,
{d \theta \over 2\pi} \,\, - \,\, \int_{0}^{2\pi} \left[
{\omega^{2} - k_{1}^{2} - \dots -  k_{d}^{2} \over 2} \,\,\,
\Phi_{\theta}^{2} \,\, + \,\, V  \left( \Phi \right) \right] \,\,
{d \theta \over 2\pi}
\end{multline*}

 Using the fact that the variation derivative of the second integral 
is equal to zero on $\Lambda$ according to (\ref{OnePhaseNLWE}) it is
not difficult to obtain the relation
\begin{equation}
\label{DifferentialPH}
d \, \langle P_{H} \rangle \,\, = \,\, {\omega} \, d Q
\,\, + \,\, Q \sum_{l} {k_{l} \over \omega} \,\, d k_{l} 
\end{equation}

 From (\ref{DifferentialPH}) it is easy to get the following
relations in the bracket (\ref{CanOnePhaseAvBr}):
$$\left\{ S ({\bf X}) \, , \, \langle P_{H} \rangle ({\bf Y}) \right\} 
\,\, = \,\, \omega \, 
(S_{\bf X}, \, Q) \,\, \delta ({\bf X} - {\bf Y})$$
$$\left\{ \langle P_{H} \rangle ({\bf X}) \, , \,
\langle P_{H} \rangle ({\bf Y}) \right\} \,\, = \,\, 2 \, \sum_{l} \,
\langle P_{l} \rangle ({\bf X}) \,\, 
\delta_{X^{l}} ({\bf X} - {\bf Y}) \,\, + \,\, \sum_{l} \,
\langle P_{l} \rangle_{X^{l}} \,\, \delta ({\bf X} - {\bf Y}) $$

 From expression (\ref{PHPHSkobka})
for the pairwise Poisson brackets of the 
functionals $P_{H} ({\bf x})$, $P_{H} ({\bf y})$ we can see then that
the bracket (\ref{AvBracketSinglePhase}) obtained with the aid of the
functional $H$ coincides with the bracket (\ref{CanOnePhaseAvBr})
(and all (\ref{NumberqAvBr})) after the corresponding change of
coordinates.

 The Whitham equations can be easily written using the averaged
Poisson bracket and the functional 
$H^{av} = \int \langle P_{H} \rangle \, d^{d} X$. 
Thus, in the coordinates
$( S({\bf X}), \, Q({\bf X}) )$ the Whitham system takes the form
$$S_{T} \,\, = \,\, \omega \, (S_{\bf X}, \, Q) 
\,\,\,\,\,\,\,\, , \,\,\,\,\,\,\,\,
Q_{T} \,\, = \,\, \sum_{l} \, \left( 
{S_{X^{l}}  \over \omega} \,\, Q \right)_{X^{l}} $$

 In conclusion the author expresses his gratitude to Prof. M.V. Pavlov 
for the interest to this work and fruitful discussions.

 This work was financially supported by the Russian Federation Government 
Grant No. 2010-220-01-077, Grant of the President of Russian Federation
NSh-4995.2012.1, and Grant RFBR No. 11-01-12067-ofi-m-2011.

\end{document}